\documentclass[12pt]{article}
\usepackage{slashed}
\usepackage{mathrsfs}
\makeatletter \@addtoreset{equation}{section} \makeatother
\renewcommand{\theequation}{\thesection.\arabic{equation}}
\newcommand{\ba}{\begin{array}}
\newcommand{\ea}{\end{array}}
\newcommand{\beq}{\begin{equation}}
\newcommand{\eeq}{\end{equation}}
\newcommand{\bea}{\begin{eqnarray}}
\newcommand{\eea}{\end{eqnarray}}

\def\bce{\begin{center}}
\def\ece{\end{center}}

\def\nonu{\nonumber}

\def\pa{\partial}
\def\al{\alpha}
\def\be{\beta}
\def\ga{\gamma}

\def\de{\delta}

\def\ep{\epsilon}

\def\la{\lambda}

\def\eps6{{\displaystyle \mathop{\epsilon}^{6}}{}}
\def\g6{{\displaystyle \mathop{g}^{6}}{}}

\def\nab6{{\displaystyle \mathop{\nabla}^{6}}{}}

\def\0{{\sst{(0)}}}
\def\1{{\sst{(1)}}}
\def\2{{\sst{(2)}}}
\def\3{{\sst{(3)}}}
\def\4{{\sst{(4)}}}
\def\5{{\sst{(5)}}}
\def\6{{\sst{(6)}}}
\def\7{{\sst{(7)}}}
\def\8{{\sst{(8)}}}

\def\ba{\begin{array}}
\def\ea{\end{array}}
\def\beq{\begin{equation}}
\def\eeq{\end{equation}}
\def\be{\begin{equation}}
\def\ee{\end{equation}}

\def\la{\lambda}
\def\eps{\epsilon}

\def\ba{\begin{array}}
\def\ea{\end{array}}
\def\beq{\begin{equation}}
\def\eeq{\end{equation}}
\def\be{\begin{equation}}
\def\ee{\end{equation}}

\def\la{\lambda}
\def\eps{\epsilon}

\def\eps6{{\displaystyle \mathop{\epsilon}^{6}}{}}

\def\nab6{{\displaystyle \mathop{\nabla}^{6}}{}}

\newcommand{\bean}{\begin{eqnarray*}}
\newcommand{\eean}{\end{eqnarray*}}

\usepackage{amsmath}

\usepackage{kotex}

\usepackage{amsmath}
\usepackage{graphicx}
\usepackage{amssymb}
\usepackage{amsthm}
\usepackage{grffile}
\usepackage{simplewick}
\usepackage{tikz}

\usepackage{nccmath}
\usepackage{physics}
\usepackage{wrapfig}
\usepackage{mathtools}
\usepackage{stackengine}

\usetikzlibrary{intersections,decorations.markings}

\allowdisplaybreaks

\begin{document}
\thispagestyle{empty} \addtocounter{page}{-1}
   \begin{flushright}
\end{flushright}

\vspace*{1.3cm}
  
\centerline{ \Large \bf
  A Supersymmetric Extension of $w_{1+\infty}$ Algebra
}
\vspace*{0.5cm}
\centerline{ \Large \bf
in the Celestial Holography}
\vspace*{1.5cm}
\centerline{ {\bf  Changhyun Ahn$^\dagger$}
  and {\bf Man Hea Kim$^{\ast}$}
} 
\vspace*{1.0cm} 
\centerline{\it 
$\dagger$ Department of Physics, Kyungpook National University, Taegu
41566, Korea} 
\vspace*{0.3cm}
\centerline{\it 
  $\ast$
  Department of Physics Education,  Sunchon National University,
  Sunchon 57922, Korea }
\vspace*{0.8cm} 
\centerline{\tt ahn@knu.ac.kr,
manhea.kim10000@gmail.com
} 
\vskip2cm

\centerline{\bf Abstract}
\vspace*{0.5cm}

We determine the ${\cal N}=1$ supersymmetric
topological $W_{\infty}
$ algebra
by using the $\la $ deformed bosons $(\beta,\gamma)$ and
fermions $(b,c)$
ghost system.
By considering the real  bosons
and the real  fermions at $\la=0$ (or $\la=\frac{1}{2}$),
the ${\cal N}=1$ supersymmetric $W_{\frac{\infty}{2}}$ algebra
is obtained.
At $\la=\frac{1}{4}$, other ${\cal N}=1$ supersymmetric
$W_{1+\infty}[\la=\frac{1}{4}]$ algebra is determined.
We also obtain the extension of Lie superalgebra $PSU(2,2|{\cal N}=4)$
appearing in the worldsheet theory
by using the symplectic bosons and the fermions.
We identify the soft current algebra between the graviton, the gravitino,
the photon (the gluon), the photino (the gluino) or
the scalars, equivalent to
${\cal N}=1$ supersymmetric
$W_{1+\infty}[\la]$ algebra,
in two dimensions
with the ${\cal N}=1$ supergravity theory in four dimensions discovered by
Freedman, van Nieuwenhuizen and Ferrara
in 1976 and its matter coupled theories, via celestial holography.

\vspace*{2cm}

\baselineskip=18pt
\newpage
\renewcommand{\theequation}
{\arabic{section}\mbox{.}\arabic{equation}}

\tableofcontents

\section{ Introduction
}

A celestial holography connects the gravitational scattering amplitudes
in the 
asymptotically flat four dimensional spacetimes and the conformal
field theory living on the two dimensional celestial sphere
\cite{Strominger,Raclariu,Pasterski,PPR,Donnay}.

In \cite{Strominger2105}, a soft current algebra
between the gravitons and the gluons in the Einstein Yang-Mills theory
is obtained.
The wedge subalgebra of $w_{1+\infty}$ algebra \cite{Bakas}
appears.
See also the relevant work in
\cite{GHPS} where the previous works in
\cite{FFT,PRSY} are used.
In \cite{FSTZ}, by considering the fermionic partners,
gravitinos and gluinos as well as the above bosonic ones in the
${\cal N}=1$
supersymmetric Einstein Yang-Mills theory, the operator product
expansions(OPEs)
of
gravitinos and gluinos with gravitons and gluons
are determined, along the lines of \cite{FFT,PRSY}.
The corresponding ${\cal N}=1$ supersymmetric soft current algebra
between the above soft particles
is obtained in \cite{Ahn2111,Ahn2202}
by generalizing the works of \cite{Strominger2105,GHPS}.
Note that the anticommutators between the fermionic
operators are vanishing.

On the other hand, in \cite{Prabhu},
the ${\cal N}=1$ superconformal algebra \cite{FQS,BKT}
where there exists a nontrivial anticommutator
for the modes of the fermionic current as well as
the usual two commutators
is reproduced from the Lie superalgebra based on the BMS symmetries
\cite{BvM,Sachs}. It is natural to ask how we can construct
the supersymmetric soft current algebra where
the anticommutators between the fermionic
operators are nonvanishing.

In \cite{AK2309}, by considering the deformation parameter $\la$,
the generalization of \cite{PRS} is obtained.
The bosonic subalgebras are given by
$W^K_{1+\infty}[\la] \times W^K_{1+\infty}[\la+\frac{1}{2}]$.
The first factor is realized by $K$ values of
$(b,c)$ fermions and the
second factor is realized by $K$ values of $(\beta,\gamma)$ bosons.
The above deformation parameter $\la$ appears in the weights
of above fields nontrivially. 
By construction, there exist several anticommutators between the
modes for the fermionic currents \cite{AK2309}.


In this paper, we would like to construct the following
things as follows.

i) One ${\cal N}=1$ supersymmetric extension of $w_{1+\infty}$ algebra
is found in the work of \cite{PRSS}. Because  the ${\cal N}=2$
supersymmetric $W_{1+\infty}[\la]$ algebra for generic $\la$ parameter
exists in \cite{AK2309}, 
the corresponding ${\cal N}=1$ supersymmetric
topological $W_{\infty}^{\text{top}}$ algebra
with $\la$ deformation, which is a generalization of
\cite{PRSS}, can be determined explicitly.
ii) Other type of ${\cal N}=1$ supersymmetric $W_{\frac{\infty}{2}}$
algebra is studied \cite{BdVnpb,BdVplb} without any explicit structure
constants. By introducing the real free bosons and fermions
with $\la=0$ (or $\la =\frac{1}{2}$), half of the bosonic currents
and half of the fermionic currents can survive.
The explicit (anti)commutator relations (or OPEs)
for ${\cal N}=1$ supersymmetric $W_{\frac{\infty}{2}}$
algebra
can be obtained.
iii) Furthermore, at the particular value of $\la=\frac{1}{4}$,
the ${\cal N}=1$ supersymmetry remains in \cite{BdVnpb,BdVplb,Ahn2203}.
It is difficult to extract the structure constants from
\cite{BdVnpb,BdVplb}.
We determine the complete structure of this
${\cal N}=1$ $W_{1+\infty}[\la=\frac{1}{4}]$
algebra
by providing the explicit structure constants.
iv) We can examine the particular limit of \cite{AK2309} in order to
check whether there exist the nontrivial anticommutator relations
between the modes for the fermionic currents with a
general $\la$ dependence in the context of ${\cal N}=1$ supersymmetric
$W_{1+\infty}[\la]$ algebra.
v) In \cite{GG2104,GG2105,Ahn2109}, the free field worldsheet realization
provides the weight $\frac{1}{2}$ conjugate pairs of symplectic
boson fields and four weight $\frac{1}{2}$ complex fermion fields.
From our experience in \cite{AK2309}, we can determine the extension of
Lie superalgebra $PSU(2,2|4)$ by considering the derivatives
acting on the quadratic free fields. 
vi) The asymptotic symmetry algebra of ${\cal N}=8$ supergravity
theory in four dimensions is studied in \cite{BRS1,BRS2}. It seems that
there should appear the soft current algebra in two dimensions. 
As a first step, we try to describe and interpret
the algebras obtained above by  taking the particular
limit of the parameter and
inserting the
helicities appropriately 
in the context of the (conformal)
supergravity theory via celestial holography.

In section $2$, we apply the procedure of \cite{PRSS} to the algebra
studied in \cite{AK2309}.
In section $3$, at the particular value $\la = 0$,
by using the free field construction in \cite{BdVnpb,BdVplb},
the new algebra is obtained. For the $\la=\frac{1}{4}$,
other new algebra is determined.
Finally, for generic $\la$, we take the particular limit for the
rescaling of the modes for the currents and obtain
the various (anti)commutators. 
In section $4$,
by using the free field construction in \cite{GG2104,GG2105},
the new algebra is obtained.
In section $5$, the corresponding dual bulk theories
are described in the celestial holography.
In section $6$, we summarize what we have obtained in this paper
and present some open problems.
In Appendices, some details appearing in the previous sections
are described.

The Thielemans package \cite{Thielemans} is used
  together with a mathematica \cite{mathematica} all
  the times in this paper.
  We list some relevant works
\cite{AGS,33,34,35,36,37,38,39,40,41,42,43,44,45,46,47,48,49,50,BHP,Tropper}
on the supersymmetric construction
in the context of \cite{Strominger}.

We present the different kinds of bosonic $W$ algebras
in the Table 1.
Historically, the so-called `GMP'
algebra in terms of Fourier transformed density operators
appeared in the quantum Hall effect \cite{GMP} \footnote{
  We thank the referee for pointing this out. }:
the equation $(4.13)$ of \cite{GMP}.
For example, in \cite{Cappelli2103,Cappelli1512},
the $W_{1+\infty}$ algebra
(or $w_{1+\infty}$ algebra) can be written in terms of above GMP algebra
by changing the polynomial basis into  the Fourier basis.
See also earlier works given by \cite{CTZ} and \cite{IKS}.
It would be interesting to apply our results in this paper to the
quantum Hall effect in three dimensional spacetimes.

\begin{table}[]
    \centering
\renewcommand{\arraystretch}{1.7}
\begin{tabular}{|c|c| }
\hline
The bosonic $W$ algebra & The references
\\
\hline
$w_{\infty}$ & \cite{GMP,Bakas}
\\
\hline
$W_{\infty}$ & \cite{PRS1990PLB,Pope9112}
\\
\hline
$W_{1+\infty}$ & \cite{PRS,Pope9112}
\\
\hline
$w_{1+\infty}$ & \cite{Sezgin}
\\
\hline
$W_{\frac{\infty}{2}}$ & \cite{BPRSS,BdVnpb,BdVplb}
\\
\hline
$W_{\frac{1+\infty}{2}}$ & \cite{BPRSS,BdVnpb,BdVplb}
\\
\hline
$w_{\frac{\infty}{2}}$ & \mbox{the subsection $3.1$ of this paper}
\\
\hline
$w_{\frac{1+\infty}{2}}$ & \mbox{the subsection $3.1$ of this paper}
\\
\hline
$W_{\infty}^K$
&  \cite{BK1990MPL,Odake}
\\
\hline
$W_{1+\infty}^L$ &
\cite{OS,Odake}
\\
\hline
$w_{\infty}^K$
&  \cite{Ahn2202}
\\
\hline
$w_{1+\infty}^L$ &
\cite{Ahn2202}
\\
\hline
$W_{\frac{\infty}{2}}^K$
&  \mbox{the subsection $3.1$ of this paper}
\\
\hline
$W_{\frac{1+\infty}{2}}^L$ &
\mbox{the subsection $3.1$ of this paper}
\\
\hline
$w_{\frac{\infty}{2}}^K$
&  \mbox{the subsection $3.1$ of this paper}
\\
\hline
$w_{\frac{1+\infty}{2}}^L$ &
\mbox{the subsection $3.1$ of this paper}
\\
\hline
$W_{\infty}[\la]$ & \cite{BdVnpb,BdVplb}
 \\
\hline 
$W_{1+\infty}[\la]$
& \cite{BdVnpb,BdVplb} \\
\hline
$w_{\infty}[\la]$ & \mbox{the subsection $3.4$ of this paper}
 \\
\hline 
$w_{1+\infty}[\la]$
& \mbox{the subsection $3.4$ of this paper}
\\
\hline
$W_{\infty}^K[\la]$
&  \cite{Ahn2203,AK2309}
\\
\hline
$W_{1+\infty}^L[\la]$ &
\cite{Ahn2203,AK2309}
\\
\hline
$w_{\infty}^K[\la]$
&  \mbox{the subsection $3.4$ of this paper}
\\
\hline
$w_{1+\infty}^L[\la]$ &
\mbox{the subsection $3.4$ of this paper}
\\
\hline
\end{tabular}
\caption{ The various types of bosonic $W$ algebras with corresponding
references.}
\end{table}

\section{ A $\la$ deformation of the ${\cal N}=1$
  supersymmetric topological $W_{\infty}^{\text{top}}$ algebra  }

The operator product expansions
of the $(\beta , \ga)$ and $(b , c)$ systems are given by \cite{CHR}
\bea
\ga^{i,\bar{a}}(z)\, \beta^{\bar{j},b}(w) =
\frac{1}{(z-w)}\, \de^{i \bar{j}}\, \de^{\bar{a} b} + \cdots\, ,
\qquad
c^{i, \bar{a}}(z) \, b^{\bar{j},b}(w) =
\frac{1}{(z-w)}\, \de^{i \bar{j}}\, \de^{\bar{a} b} + \cdots\, .
\label{fundOPE}
\eea
The $(\beta, \ga)$ fields are bosonic operators while
the $(b, c)$ fields are fermionic operators.
The
conformal weights of $(\beta , \ga)$ fields are given by
$(\la,1-\la)$ while
those of $(b , c)$ fields are given by
$(\frac{1}{2}+\la,\frac{1}{2}-\la)$.
There are fundamental indices $a, b $ and
antifundamental indices $\bar{a}, \bar{b}$
of $SU(K)$ and 
there are  fundamental indices $i, j $
and antifundamental indices $\bar{i}, \bar{j}$ of
$SU(N)$.

\subsection{The bosonic currents}

From Appendix $(A.6)$ or $(D.1)$ of \cite{AK2309},
we can calculate
the modes for the bosonic currents in the $\la$ deformed
${\cal N}=1$
supersymmetric  topological
 $W_{\infty}^{\text{top}}$ algebra by using
 the anticommutator $(6)$ of \cite{PRSS} as follows:
\bea
\acomm{(\bar{Q}^{\la}_{\frac{3}{2}})_{-\frac{1}{2}}}{(Q^{\la}_{h+\frac{1}{2}})_{m+\frac{1}{2}}}&=&
\sum_{h_3=1}^{h+1\,\, }\sum_{k=0}^{h-h_3+1}
\,
2(-1)^{h_3+1}\,(4q)^{h-h_3+1}\,(2h_3-1)! \nonu \\
&\times & 
\bigg[\,\,U_F^{\,\,h,1,h_3,k}(\la)\,[m+h]_{h-h_3-k+1}\,[0]_k\,(W^{\la}_{F,\,h_3})_{m}
\nonu \\
&
+ &
U_B^{\,\,h,1,h_3,k}(\la)\,[m+h]_{k}\,[0]_{h-h_3-k+1}\,(W^{\la}_{B,\,h_3})_{m}
\bigg]\,
\nonu \\
& = & 2(\tilde{W}^{\la}_{F,\,h+1})_m+2(\tilde{W}^{\la}_{B,\,h+1})_m
   \equiv 2(\tilde{W}^{\la}_{\,h+1})_m\,,
\label{bosonic}
\eea
where the $\la$ deformed modes on the right hand side of
(\ref{bosonic}) are
\bea
    (\tilde{W}_{F,\,h}^{\lambda})_m
  &
    =& (W_{F,\,h}^{\lambda})_m-
    q\frac{2(h-1)-4\lambda}{2(h-2)+1}(m+h-1)(W_{F,\,h-1}^{\lambda})_{m}\,,
    \nonu \\
     (\tilde{W}_{B,\,h}^{\lambda})_m
    &
    =& (W_{B,\,h}^{\lambda})_m+
    q\frac{2(h-2)+4\lambda}{2(h-2)+1}(m+h-1)(W_{B,\,h-1}^{\lambda})_{m}\,.
    \label{twobosonic}
\eea
It is obvious that
these lead to the previous results in \cite{Ahn2202,PRSS} for $\la =0$.
The $\la$ dependent terms of the coefficients on the right hand side
in (\ref{bosonic}) (that is, the second terms in (\ref{twobosonic})) 
come from the case of $h_3=h$ while the $\la$ independent terms of
those (the first terms in (\ref{twobosonic}))
appear when $h_3=h+1$. For the $h_3 < h$, then there are
no contributions in the summation of (\ref{bosonic}).
There is no central term in the above anticommutator
due to the particular combinations of the modes on the left hand side.
The falling
Pochhammer symbol $[a]_n \equiv a(a-1) \cdots (a-n+1)$ is used.
The structure constants in (\ref{bosonic}) are given by Appendix $C$.

In terms of currents, by using the standard
procedure in the conformal field theory,
we obtain the following results
corresponding to (\ref{twobosonic})
\footnote{By construction, the weight $h$ is greater than or equal to
$2$.}
\bea
   \tilde{W}_{F, \,h}^{\lambda}
&=&
W_{F,\,h}^{\lambda}
+ q\frac{2(h-1)- 4 \lambda}{2(h-2)+1}\partial W_{F,\,h-1}^{\lambda}\,,\
\nonu
\\  
\tilde{W}_{B, \,h}^{\lambda}
& = & W_{B,\,h}^{\lambda}
- q \frac{2(h-2)  + 4\lambda}{2(h-2)+1}\partial W_{B,\,h-1}^{\lambda}.
\label{deformedexp}
\eea
Of course, when we take $\la \rightarrow 0$ limit,
the results in
(\ref{deformedexp})
become those in \cite{Ahn2202,PRSS}.
In principle, these currents can be obtained
from the OPE between the fermionic currents
corresponding to the left hand side of (\ref{bosonic}).

In terms of free fields appearing in (\ref{fundOPE}),
the above currents in (\ref{deformedexp})
take the form
\bea
  \tilde{W}_{F,\,h}^{\lambda}(z)
  &= & (-4q)^{h-2}
\sum_{i=0}^{h-1} \Big[2a^{i}\big(h,\lambda+\tfrac{1}{2}\big)-\alpha^i(h,\lambda)\Big]\,
\partial_z^{h-i-1}
  \Big((\partial^{i}b^{\bar{l} b})\, \delta_{b \bar{a}}\delta_{\bar{l} l}\,c^{l \bar{a}}\Big)(z),
\nonu \\
\tilde{W}_{B,\,h}^{\lambda}(z)
&=& (-4q)^{h-2}
\sum_{i=0}^{h-1}  \Big[2a^{i}(h,\lambda)-\alpha^i(h,\lambda)\Big]\,
\partial_z^{h-i-1}
  \Big((\partial^{i}\beta^{\bar{l} b})\, \delta_{b \bar{a}}\delta_{\bar{l} l}\,\gamma^{l \bar{a}}\Big)(z),
  \nonu 
  \\
\tilde{W}^{\la}_h(z)  & \equiv &
\tilde{W}^{\la}_{F,\,h}(z) + \tilde{W}^{\la}_{B,\,h}(z),
\nonu \\
    \tilde{W}_{F,\,h}^{\lambda,\hat{A}}(z)
  &=& (-4q)^{h-2}
\sum_{i=0}^{h-1} \Big[2a^{i}\big(h,\lambda+\tfrac{1}{2}\big)-\alpha^i(h,\lambda)\Big]\,
\partial_z^{h-i-1}
  \Big((\partial^{i}b^{\bar{l} b})\, t^{\hat{A}}_{b \bar{a}}\delta_{\bar{l} l}\,c^{l \bar{a}}\Big)(z),
\nonu \\
\tilde{W}_{B,\,h}^{\lambda,\hat{A}}(z)
&=& (-4q)^{h-2}
\sum_{i=0}^{h-1}  \Big[2a^{i}(h,\lambda)-\alpha^i(h,\lambda)\Big]\,
\partial_z^{h-i-1}
\Big((\partial^{i}\beta^{\bar{l} b})\, t^{\hat{A}}_{b \bar{a}}\delta_{\bar{l} l}\,
\gamma^{l \bar{a}}\Big)(z),\nonu \\
\tilde{W}^{\la, \hat{A}}_h(z)  & \equiv &
\tilde{W}^{\la, \hat{A}}_{F,\,h}(z) + \tilde{W}^{\la, \hat{A}}_{B,\,h}(z),
\label{WFBA}
\eea
as well as the fermionic currents in \cite{AK2309}
\bea
Q_{h+\frac{1}{2}}^{\lambda}(z)
& = & \sqrt{2}\,(-4q)^{h-1}\sum_{i=0}^{h-1} \beta^{i}(h+1,\lambda)\,
\partial_z^{h-i-1}
((\partial_z^{i}b^{\bar{l} b})\,\delta_{b\bar{a}} \delta_{\bar{l} l}\,\gamma^{l \bar{a}})(z),
\nonu \\
Q_{h+\frac{1}{2}}^{\lambda,\hat{A}}(z)
& = & \sqrt{2}\,(-4q)^{h-1}\sum_{i=0}^{h-1} \beta^{i}(h+1,\lambda)\,
\partial_z^{h-i-1}
((\partial_z^{i}b^{\bar{l} b})\, (t^{\hat{A}})_{b\bar{a}}\delta_{\bar{l} l}\,\gamma^{l \bar{a}})(z).
\label{QQA}
\eea
Note that the $\la$ dependent coefficient terms
in (\ref{WFBA}) appear nontrivially and
can be expressed as the known coefficients.
The various coefficients $a^{i}(h,\la), \al^i(h,\la)$
and $\beta^i(h,\la)$ appearing in (\ref{WFBA}) and (\ref{QQA})
are given in \cite{BdVnpb,BdVplb,AK2309}:
\bea 
 a^i(h, \la) \equiv \left(\begin{array}{c}
h-1 \\  i \\
 \end{array}\right) \, \frac{(-2\la-h+2)_{h-1-i}}{(h+i)_{h-1-i}}\, ,
 \qquad 0 \leq i \leq h-1\, ,
 \nonu \\
 \al^i(h, \la) \equiv \left(\begin{array}{c}
h-1 \\  i \\
 \end{array}\right) \, \frac{(-2\la-h+2)_{h-1-i}}{(h+i-1)_{h-1-i}}\, ,
 \qquad 0 \leq i \leq h-1\, ,
 \nonu \\
  \beta^i(h, \la) \equiv \left(\begin{array}{c}
h-2 \\  i \\
  \end{array}\right) \, \frac{(-2\la-h+2)_{h-2-i}}{(h+i)_{h-2-i}}\, ,
  \qquad 0 \leq i \leq h-2 \, ,
  \label{coeff}
  \eea
  where the rising Pochhammer symbols
  are used in $(a)_n \equiv a(a+1) \cdots (a+n-1)$
  with a nonnegative integer $n$.
  The parentheses in (\ref{coeff}) are the binomial coefficients.
  The newly obtained bosonic currents are
  given by $\tilde{W}^{\la}_h$ with 
(\ref{deformedexp}) and $\tilde{W}^{\la, \hat{A}}_h$
with (\ref{WFBA}) where $(t^{\hat{A}})_{b\bar{a}}$ is an adjoint
representation for $SU(K)$ generator. The fermionic currents
$Q_{h+\frac{1}{2}}^{\lambda}$ and $Q_{h+\frac{1}{2}}^{\lambda,\hat{A}}$
remain the same as the ones in \cite{AK2309}.

We would like to construct the new algebra
by  the bosonic currents $\tilde{W}^{\la}_h$ and
 $\tilde{W}^{\la, \hat{A}}_h$ and the fermionic currents 
$Q_{h+\frac{1}{2}}^{\lambda}$
and $Q_{h+\frac{1}{2}}^{\lambda,\hat{A}}$ together with (\ref{WFBA})
and (\ref{QQA}).
Because the relative coefficients in (\ref{deformedexp})
contain the $\la$ explicitly, the structure constants
in this new algebra will be different from those in \cite{AK2309}.

\subsection{The commutator between the bosonic currents}

First of all, we should calculate the OPE between
the bosonic currents obtained in previous subsection.
One way to determine that OPE between the singlets of
$SU(K)$ is as follows.
By calculating $(7)$ of \cite{PRSS} or (\ref{bosonic}),
 \bea
 4\comm{(\tilde{W}^{\la}_{\,h_1+1})_{m}}{(\tilde{W}^{\la}_{\,h_2+1})_{n}}
 & = &
 \comm{
   \acomm{(\bar{Q}^{\la}_{\frac{3}{2}})_{-\frac{1}{2}}}{
     (Q^{\la}_{h_1+\frac{1}{2}})_{m+\frac{1}{2}}}}
   {\acomm{(\bar{Q}^{\la}_{\frac{3}{2}})_{-\frac{1}{2}}}{
     (Q^{\la}_{h_2+\frac{1}{2}})_{n+\frac{1}{2}}}}
 \nonu \\
 & = &
 \acomm{(\bar{Q}^{\la}_{\frac{3}{2}})_{-\frac{1}{2}}}{\comm{(Q^{\la}_{h_2+\frac{1}{2}})_{n+\frac{1}{2}}}{\acomm{(\bar{Q}^{\la}_{\frac{3}{2}})_{-\frac{1}{2}}}{(Q^{\la}_{h_1+\frac{1}{2}})_{m+\frac{1}{2}}}}}
 \nonu \\
  &=&
  2 \acomm{(\bar{Q}^{\la}_{\frac{3}{2}})_{-\frac{1}{2}}}{\comm{(Q^{\la}_{h_2+\frac{1}{2}})_{n+\frac{1}{2}}}{(\tilde{W}^{\la}_{\,h_1+1})_m}},
\label{inter}
\eea
where the Jacobi identities are used
several times and the previous relation (\ref{bosonic}) is used.

Now the commutator inside of the curly bracket in (\ref{inter})
can be determined
from (\ref{WFBA}) and (\ref{QQA}) according to (\ref{fundOPE})
and it turns out that
\bea 
& & 
  -2\sum_{h_3=1}^{h_1+h_2\,\, }\sum_{k=0}^{h_1+h_2-h_3}
\,
(-1)^{h_3}\,(4q)^{h_1+h_2-h_3-1}\,(2h_3)!
\bigg(\,\,\tilde{T}_R^{\,\,h_1+1,h_2,h_3,k}\,\,[m+h_1]_{h_1+h_2-h_3-k}\,[n+h_2]_k
\nonu \\
& &-
\tilde{T}_L^{\,\,h_1+1,h_2,h_3,k}\,\,[m+h_1]_{k}\,[n+h_2]_{h_1+h_2-h_3-k}
\bigg)\,
\acomm{(\bar{Q}^{\la}_{\frac{3}{2}})_{-\frac{1}{2}}}{(Q_{h_3+\frac{1}{2}}^{\lambda})_{m+n+\frac{1}{2}}}.
\label{inter1}
\eea 
The structure constants are given in Appendix $A$ explicitly.
Furthermore, the above anticommutator can be replaced by
the bosonic current from (\ref{bosonic})
and it is given by
\bea
& &
  -4\sum_{h_3=1}^{h_1+h_2\,\, }\sum_{k=0}^{h_1+h_2-h_3}
\,
(-1)^{h_3}\,(4q)^{h_1+h_2-h_3-1}\,(2h_3)!
\bigg(\,\,\tilde{T}_R^{\,\,h_1+1,h_2,h_3,k}\,\,[m+h_1]_{h_1+h_2-h_3-k}\,[n+h_2]_k
\nonu \\
& & - 
\tilde{T}_L^{\,\,h_1+1,h_2,h_3,k}\,\,[m+h_1]_{k}\,[n+h_2]_{h_1+h_2-h_3-k}
\bigg)\,
v(\tilde{W}^{\la}_{\,h_3+1})_{m+n}.
\label{ttexp}
 \eea
 This implies that the structure constants in the OPE
 between the
 bosonic currents are written in terms of
 those in the OPE between the bosonic current and the fermionic
 current.
 
On the other hand, 
we obtain the commutator between the bosonic currents
directly from (\ref{WFBA}) with the help of (\ref{fundOPE})
\bea   
4\comm{(\tilde{W}^{\la}_{\,h_1+1})_{m}}{(\tilde{W}^{\la}_{\,h_2+1})_{n}}
   & = & 4\sum_{h_3=2}^{h_1+h_2+1\,\, }\sum_{k=0}^{h_1+h_2-h_3+1}
\,
(-1)^{h_3}\,(4q)^{h_1+h_2-h_3}\,(2h_3-2)! \nonu \\
&\times& 
\bigg(\,\,\tilde{S}_R^{\,\,h_1+1,h_2+1,h_3,k}\,\,[m+h_1]_{h_1+h_2-h_3-k+1}\,[n+h_2]_k
\nonu \\
&-&
\tilde{S}_L^{h_1+1,h_2+1,h_3,k}[m+h_1]_{k}[n+h_2]_{h_1+h_2-h_3-k+1}
\bigg)
(\tilde{W}^{\la}_{h_3})_{m+n}.
\label{ssexp}
\eea
Again, the structure constants are given in Appendix $A$.

By comparing (\ref{ttexp}) with (\ref{ssexp}),
there are nontrivial relations between the structure constants
as follows
 \bea
 \tilde{S}_R^{\,\,h_1+1,h_2+1,h_3,k}(\la)=
-\tilde{T}_R^{\,\,h_1+1,h_2,h_3-1,k}(\la)\,,
\quad
\tilde{S}_L^{\,\,h_1+1,h_2+1,h_3,k}(\la)=
-\tilde{T}_L^{\,\,h_1+1,h_2,h_3-1,k}(\la).
\label{Iden}
\eea
This can be checked by considering these structure constants
from Appendix $A$ directly. Then we are left with
two independent structure constants in our new algebra.

Therefore, the commutator relation between the bosonic currents
is summarized by (\ref{ssexp}) or (\ref{ttexp}).
The other two commutators between the bosonic currents
having $SU(K)$ indices can be obtained similarly and they are
given in Appendix $A$ (the second and the third
relations in Appendix (\ref{SEVEN})).

\subsection{The three commutators
  between the bosonic currents and the fermionic
currents}

We have seen the commutators between the bosonic singlet current
and the fermionic singlet current in (\ref{inter1}).
Then the remaining three commutators we are left with
can be determined
by using the proper delta symbols, $f$ symbols and $d$ symbols for the
$SU(K)$ (the last three relations of (\ref{SEVEN})).
The complete seven commutator relations are summarized in Appendix
$A$.


\subsection{The $q \rightarrow 0$ limit}

By taking the following transformations for the currents,
\bea
\tilde{W}^{\lambda}_{h} & \longrightarrow &
q^{h-2}\,\tilde{W}^{\lambda}_{h}\,,\qquad \tilde{W}^{\lambda,\hat{A}}_{h}\longrightarrow q^{h}\,\tilde{W}^{\lambda,\hat{A}}_{h}\,,\qquad 
  \nonu  \\
    Q^{\lambda}_{h+\frac{1}{2}}
   & \longrightarrow & 
    q^{h-\frac{3}{2}}Q^{\lambda}_{h+\frac{1}{2}}\,,\qquad 
    Q^{\lambda,\hat{A}}_{h+\frac{1}{2}}
   \longrightarrow 
    q^{h+\frac{1}{2}}Q^{\lambda,\hat{A}}_{h+\frac{1}{2}}, 
    \label{rescalingone}
\eea    
we obtain the $\la$ deformed
${\cal N}=1$ supersymmetric $w^K_{\infty}$ algebra
with $U(K)$ symmetry after $q \rightarrow 0$ limit
in Appendix (\ref{SEVEN})
\bea
\comm{(\tilde{W}^{\lambda}_{h_1})_{m}}{(\tilde{W}^{\lambda}_{h_2})_{n}}
& = & \Big((h_2-1)m-(h_1-1)n \Big)   (\tilde{W}^{\lambda}_{h_1+h_2-2})_{m+n},
\nonu \\
\comm{(\tilde{W}^{\lambda}_{h_1})_{m}}{(\tilde{W}^{\lambda,\hat{A}}_{h_2})_{n}}
& = &
\Big((h_2-1)m-(h_1-1)n \Big)   (\tilde{W}^{\lambda,\hat{A}}_{h_1+h_2-2})_{m+n},
\nonu \\
\comm{(\tilde{W}^{\lambda,\hat{A}}_{h_1})_{m}}{(\tilde{W}^{\lambda,\hat{B}}_{h_2})_{n}}
& = &
-\frac{i}{4}f^{\hat{A}\hat{B}\hat{C}}
(\tilde{W}^{\lambda,\hat{C}}_{h_1+h_2-1})_{m+n},
\nonu \\
\comm{(\tilde{W}^{\lambda}_{h_1})_{m}}{(Q^{\lambda}_{h_2+\frac{1}{2}})_{r}}
& = &
\Big(h_2\,m-(h_1-1)r+\frac{h_1-1}{2} \Big)(Q^{\lambda}_{h_1+h_2-\frac{3}{2}})_{m+r},
\nonu \\
\comm{(\tilde{W}^{\lambda}_{h_1})_{m}}{(Q^{\lambda,\hat{A}}_{h_2+\frac{1}{2}})_{r}}
& = &
\Big(h_2\,m-(h_1-1)r+\frac{h_1-1}{2} \Big)(Q^{\lambda,\hat{A}}_{h_1+h_2-\frac{3}{2}})_{m+r},
\nonu \\
\comm{(\tilde{W}^{\lambda,\hat{A}}_{h_1})_{m}}{(Q^{\lambda}_{h_2+\frac{1}{2}})_{r}}
& = &
\Big(h_2\,m-(h_1-1)r+\frac{h_1-1}{2} \Big)(Q^{\lambda,\hat{A}}_{h_1+h_2-\frac{3}{2}})_{m+r},
\nonu \\
\comm{(\tilde{W}^{\lambda,\hat{A}}_{h_1})_{m}}{(Q^{\lambda,\hat{B}}_{h_2+\frac{1}{2}})_{r}}
& = &
-\frac{i}{4}f^{\hat{A}\hat{B}\hat{C}}   (Q^{\lambda,\hat{C}}_{h_1+h_2-\frac{1}{2}})_{m+r}, 
\label{sevencomm}
\eea
where
$
h_1,h_2  =  2,3, \cdots $ (the weights for the
fermionic currents hold for $h_1,h_2=1$),
the adjoint indices $ 
\hat{A}, \hat{B} = 1, 2, \cdots, (K^2-1)$,
and the modes are $
m  =  0, \pm 1, \pm 2, \cdots $, 
 and $r = \pm \frac{1}{2}, \pm \frac{3}{2}, \cdots $.
This implies that  the $\la$ deformation of the
${\cal N}=1$ topological $W_{\infty}^{\text{top}}$ algebra
at $q \rightarrow 0$
limit shares with common behavior for the case without any
deformation in \cite{Ahn2202} because the mode dependent terms
or constant coefficients on the
right hand sides of (\ref{sevencomm}) do not depend on the $\la$.
Of course, the currents themselves do depend on the $\la$
explicitly.
During the twisting procedure, the original weights
for the fermionic currents is increased by $\frac{1}{2}$
($h_2 +\frac{1}{2} \rightarrow h_2 + 1$)
from the fourth and fifth relations of (\ref{sevencomm}). 
The first three relations of (\ref{sevencomm})
are 
the $w^K_{\infty}$ algebra
with $U(K)$ symmetry \footnote{As described in \cite{Ahn2202},
  the weight $1$ current can be added and the corresponding $w_{1+\infty}^K$
algebra can be obtained as in \cite{PRS1}.}
and the last four relations   
are the additional commutator relations in
its ${\cal N}=1$ supersymmetric extension
\footnote{
  We can calculate the commutators
  from the corresponding OPEs. For example,
  from the following OPE,
  \begin{align*}
    \tilde{W}^{\la}_{h_1}(z)\,Q^{\la}_{h_2+\frac{1}{2}}(w)
&=\frac{1}{(z-w)^2}\,(h_1+h_2-1)\,Q^{\la}_{h_1+h_2-\frac{3}{2}}(w)
+\frac{1}{(z-w)}\,(h_1-1)\,\partial Q^{\la}_{h_1+h_2-\frac{3}{2}}(w)+\cdots,
  \end{align*}
  we can calculate the fourth commutator of (\ref{sevencomm}).
  For $h_1=2$, the weight is given by $(h_2+1)$, not
  $(h_2+\frac{1}{2})$.}.

\section{ A truncation of ${\cal N}=2$ supersymmetric 
  $W_{1+\infty}[\la]$ algebra
}

We present the various supersymmetric  algebras
and their particular limits.

\subsection{ At $\la =0$}

In this case, the conformal weights for $(b,c)$ are equal to
$(\frac{1}{2},\frac{1}{2})$ and they
for $(\beta,\gamma)$ are
equal to
$(0,1)$ from the previous section.
Then we identify $b$ with $c$ and denote it by
$\psi$ while we denote the $(\beta,\gamma)$ by $(\phi, \pa \phi)$
respectively \cite{BdVnpb,BdVplb}.
In this subsection, we use the following OPEs
\bea
\psi^{i,a}(z)\,\psi^{j,b}(w)=\frac{1}{(z-w)}\delta^{ij}\delta^{ab}+\cdots\,,\quad
\partial \phi^{i,a}(z)\,
\partial\phi^{j,b}(w)=\frac{1}{(z-w)^2}\delta^{ij}\delta^{ab}+\cdots\,.
\label{redOPE}
\eea
There exist two independent free fields compared to the ones in
previous section.
There are vector indices $a, b $
of $SO(K)$ and 
there are  also vector indices $i, j $ of
$SO(N)$ \cite{EGR} (See also \cite{DMS} for the construction of
real independent free fermions.) \footnote{The corresponding
  bosonic subalgebra is denoted by $W_{\frac{1+\infty}{2}} \times
  W_{\frac{\infty}{2}} $ algebra
  where the first factor is realized by the fermions
  while the second factor is  realized by
  the bosons
  in \cite{BPRSS,BdVnpb,BdVplb}. In the abstract,
  we denote its ${\cal N}=1$ version
  by ${\cal N}=1$ supersymmetric $W_{\frac{\infty}{2}}$
  algebra.
  See also related work in \cite{BDZplb}
  and \cite{Ahn1106,Ahn1202} for the field contents and
  free field realization.}.

\subsubsection{The bosonic and fermionic currents}

Based on the construction of \cite{BdVnpb,BdVplb}, we
determine the 
following currents
\bea
 V^{(h)+}_{ab}
 & =& \sum_{r=0}^{h-1}\tilde{a}^r(h,0)\,(\partial^r \phi^{i,b})\delta_{ij}
 (\partial^{h-r}\phi^{j,a})
+\sum_{r=0}^{h-1}\tilde{a}^r(h,\tfrac{1}{2})\,(\partial^{r}\psi^{i,b})\delta_{ij}(\partial^{h-1-r} \psi^{j,a})
\nonu \\
&=&
\sum_{r=0}^{h-1}a^r(h,0)\,\partial^{h-1-r}\Big((\partial^r \phi^{i,b})\delta_{ij}\partial
 \phi^{j,a}\Big)
 +\sum_{r=0}^{h-1}a^r(h,\tfrac{1}{2})\,\partial^{h-1-r}\Big((\partial^{r}\psi^{i,b})\delta_{ij} \psi^{j,a}\Big),
 \nonu \\
V^{(h)-}_{ab}&=& -\frac{(h-1)}{(2h-1)}\sum_{r=0}^{h-1}\tilde{a}^r(h,0)\,
(\partial^r \phi^{i,b})\delta_{ij}(\partial^{h-r}\phi^{j,a})
\nonu \\
& + & \frac{h}{(2h-1)}\sum_{r=0}^{h-1}\tilde{a}^r(h,\tfrac{1}{2})\,
(\partial^{r}\psi^{i,b})\delta_{ij}(\partial^{h-1-r} \psi^{j,a})
\nonu \\
&=& -\frac{h-1}{2h-1}\sum_{r=0}^{h-1}a^r(h,0)\,\partial^{h-1-r}\Big(
(\partial^r \phi^{i,b})\delta_{ij}\partial\phi^{j,a}\Big)
\nonu \\
& + & \frac{h}{2h-1}\sum_{r=0}^{h-1}a^r(h,\tfrac{1}{2})\,
\partial^{h-1-r}\Big(
(\partial^{r}\psi^{i,b})\delta_{ij} \psi^{j,a}\Big),
\nonu \\
Q^{(h)\pm}_{ab}&= &
\sum_{r=0}^{h-1}\tilde{\alpha}^r(h,0)\,(\partial^r \phi^{i,b})\delta_{ij}(\partial^{h-1-r}\psi^{j,a})
\mp\sum_{r=0}^{h-2}\tilde{\beta}^r(h,0)\,(\partial^{r}\psi^{i,b})\delta_{ij}(\partial^{h-1-r} \phi^{j,a})
\nonu \\
&=& \sum_{r=0}^{h-1}\alpha^r(h,0)\,\partial^{h-1-r}
\Big((\partial^r \phi^{i,b})\delta_{ij}\psi^{j,a}\Big)
\mp\sum_{r=0}^{h-2}\beta^r(h,0)\,\partial^{h-2-r}
\Big((\partial^{r}\psi^{i,b})\delta_{ij} \partial \phi^{j,a}\Big),
\label{VVQ}
\eea
where the structure constants appearing in (\ref{VVQ})
are given by
\bea
\tilde{a}^{r}(h,\lambda)&=&
\binom{h-1}{r}\,\frac{(-1)^r}{(h)_{h-1}}\,
                               (2\lambda-h)_{r}\,(2-h-2\lambda)_{h-1-r}\,,
\nonu    \\
\tilde{\alpha}^{r}(h,\lambda)&=&
2\,\binom{h-1}{r}\,\frac{(-1)^r}{(h)_{h-1}}\,
                               (2\lambda-h+1)_{r}\,(2-h-2\lambda)_{h-1-r}\,,
\nonu    \\
\tilde{\beta}^{r}(h,\lambda)&=&
\binom{h-2}{r}\,\frac{(-1)^r}{(h)_{h-2}}\,
                               (2\lambda-h+1)_{r}\,(2-h-2\lambda)_{h-2-r}\,.
\label{tildeexpression}
\eea
which appear in \cite{BdVnpb,BdVplb}.
The $h$ in the fermionic current is greater than $1$ in (\ref{VVQ}).

By considering the linear combinations,
we obtain the following currents
\bea
W_{F,h}^{ab}&= &
n_{W_{F,h}}\,\frac{(-1)^h}{\sum_{i=0}^{h-1}a^{i}(h,\frac{1}{2})}\Bigg[\frac{(h-1)}{(2h-1)}V^{(h)+}_{ab}+V^{(h)-}_{ab}\Bigg]\,, \nonu
\\
W_{B,h}^{ab}&= &
n_{W_{B,h}}\,\frac{(-1)^h}{\sum_{i=0}^{h-1}a^{i}(h,0)}\Bigg[\frac{h}{2h-1}V^{(h)+}_{ab}-V^{(h)-}_{ab}\Bigg]\,,\nonu \\
Q_{h+\frac{1}{2}}^{ab}&=&
n_{Q_h}\,\frac{(-1)^{h+1}h}{2\sum_{i=0}^{h-1}\beta^{i}(h+1,0)}\Bigg[Q^{(h+1)-}_{ab}-Q^{(h+1)+}_{ab}\Bigg]\,, \nonu \\
\bar{Q}_{h+\frac{1}{2}}^{ba}&=&
n_{Q_h}\,\frac{(-1)^{h+1}}{2\sum_{i=0}^{h}\alpha^{i}(h+1,0)}\Bigg[Q^{(h+1)-}_{ab}+Q^{(h+1)+}_{ab}\Bigg]\,.
\label{WFWBQ}
\eea
Note that we can express these currents in terms of
previous parameters in (\ref{tildeexpression}).

By realizing the identity between the coefficients
in the fermionic currents, there exists only one independent
fermionic current
\footnote{
  There exists a relation
  \bea
   Q_{h+\frac{1}{2}}^{\,\,ab}&=(-1)^{h+1}\,\bar{Q}_{h+\frac{1}{2}}^{\,\,ab}
\nonu
\eea
for $h > 1$.
We have the normalization factors
\bea
 n_{W_F,h}=q^{h-2}\,\frac{2^{h-3}(h-1)!}{(2h-3)!!}\,,\quad
n_{W_B,h}=q^{h-2}\,\frac{2^{h-3}\,h!}{(2h-3)!!}\,,\quad
n_{Q,h}=q^{h-1}\,\frac{2^{h-\frac{1}{2}}\,h!}{(2h-1)!!}\,.
\nonu
\eea
}.

\subsubsection{The five OPEs (and (anti)commutators)
}

It is straightforward to
calculate the following OPEs from (\ref{redOPE})
\bea
W^{ab}_{F,h_1}(z)\,W^{cd}_{F,h_2}(w)
& = & N\,\delta^{h_1 h_2}\Big((-1)^{h_1+1}\delta^{ac}\delta^{bd}-\delta^{ad}\delta^{bc}\Big)\,q^{2(h_1-2)}c_{W_{F,h_1}}\,
\partial_z^{2h_1-1}\bigg[\frac{1}{(z-w)}\bigg]
\nonu \\
&-&
\frac{1}{2}\sum_{h=-1}^{h_1+h_2-3}q^h\,p_F^{h_1,h_2,h}(\partial_z^m,
\partial_w^n,0)\,
\bigg[
(-1)^{h_1+h}\,\frac{\delta^{ac}\,W^{bd}_{F,h_1+h_2-h-2}(w)}{(z-w)}
\nonu \\
&+ & \frac{\delta^{ad}\,W^{cb}_{F,h_1+h_2-h-2}(w)}{(z-w)}
+(-1)^{h}\,\frac{\delta^{bc}\,W^{ad}_{F,h_1+h_2-h-2}(w)}{(z-w)}
\nonu \\
& +& (-1)^{h_2+h}\,\frac{\delta^{bd}\,W^{ac}_{F,h_1+h_2-h-2}(w)}{(z-w)}
\bigg]+ \cdots,
\nonu \\
W^{ab}_{B,h_1}(z)\,W^{cd}_{B,h_2}(w)
& = &
N\,\delta^{h_1 h_2}\Big((-1)^{h_1+1}\delta^{ac}\delta^{bd}-\delta^{ad}\delta^{bc}\Big)\,q^{2(h_1-2)}c_{W_{B,h_1}}\,\partial_z^{2h_1-1}\bigg[\frac{1}{(z-w)}\bigg] \nonu
\\
&-&
\frac{1}{2}\sum_{h=-1}^{h_1+h_2-4}q^h\,p_B^{h_1,h_2,h}(\partial_z^m,
\partial_w^n,0)
\bigg[
(-1)^{h_1+h}\,\frac{\delta^{ac}\,W^{bd}_{B,h_1+h_2-h-2}(w)}{(z-w)}
\nonu \\
&+ & \frac{\delta^{ad}\,W^{cb}_{B,h_1+h_2-h-2}(w)}{(z-w)}
+(-1)^{h}\,\frac{\delta^{bc}\,W^{ad}_{B,h_1+h_2-h-2}(w)}{(z-w)}
\nonu \\
& + & (-1)^{h_2+h}\,\frac{\delta^{bd}\,W^{ac}_{B,h_1+h_2-h-2}(w)}{(z-w)}
\bigg]+ \cdots,
\nonu \\
W^{ab}_{F,h_1}(z)\,Q^{cd}_{h_2+\frac{1}{2}}(w)
& = &
-\sum_{h=-1}^{h_1+h_2-3}q^h\,q_F^{h_1,h_2+\frac{1}{2},h}(\partial_z^m,
\partial_w^n,0)\, \nonu
\\
&\times & \bigg[
(-1)^{h}\,\frac{\delta^{ad}\,Q^{cb}_{h_1+h_2-h-\frac{3}{2}}(w)}{(z-w)}
+(-1)^{h_1+h}\,\frac{\delta^{bd}\,Q^{ca}_{h_1+h_2-h-\frac{3}{2}}(w)}{(z-w)}
\bigg]+\cdots, \nonu
\\
W^{ab}_{B,h_1}(z)\,Q^{cd}_{h_2+\frac{1}{2}}(w)
& = &
-\sum_{h=-1}^{h_1+h_2-3}q^h\,q_B^{h_1,h_2+\frac{1}{2},h}(\partial_z^m,
\partial_w^n,0)\,
\nonu \\
&\times &
\bigg[
(-1)^{h_1+h}\,\frac{\delta^{ac}\,Q^{bd}_{h_1+h_2-h-\frac{3}{2}}(w)}{(z-w)}
+(-1)^{h}\,\frac{\delta^{bc}\,Q^{ad}_{h_1+h_2-h-\frac{3}{2}}(w)}{(z-w)}
\bigg]+\cdots,
\nonu \\
Q^{ab}_{F,h_1+\frac{1}{2}}(z)\,Q^{cd}_{h_2+\frac{1}{2}}(w)
& = &
N\,\delta^{h_1 h_2}\,\delta^{ac}\delta^{bd}\,\,q^{2(h_1-1)}\,(-1)^{h_1+1}\,
c_{Q_{h_1}}\,\partial_z^{2h_1}\bigg[\frac{1}{(z-w)}\bigg]
\nonu \\
&-&
\sum_{h=0}^{h_1+h_2-1}q^h\,(-1)^{h_1}\,o_F^{h_1+\frac{1}{2},h_2+\frac{1}{2},h}
(\partial_z^m,\partial_w^n,0)\,
\frac{\delta^{ac}\,W^{bd}_{F,h_1+h_2-h}(w)}{(z-w)}
\label{fiveOPEs}
\\
&
- & \sum_{h=0}^{h_1+h_2-2}q^h\,(-1)^{h_2+h}\,o_B^{h_1+\frac{1}{2},h_2+\frac{1}{2},h}
(\partial_z^m,\partial_w^n,0)\,
\frac{\delta^{bd}\,W^{ac}_{B,h_1+h_2-h}(w)}{(z-w)} + \cdots .
\nonu
\eea
The various structure constants are given in Appendix (\ref{structla}) and
the corresponding differential operators can be obtained
by replacing the arguments
$(m, n)$ with the $m$ derivatives of $z$ and
the $n$ derivatives of $w$ respectively at $\la =0$.
The central terms in (\ref{fiveOPEs}) contain 
\bea
&c_{W_{F,h}}=\frac{2^{2(h-3)}\big((h-1)!\big)^2}{(2h-3)!!(2h-1)!!}\,,
c_{W_{B,h}}=\frac{2^{2(h-3)}(h-2)!h!}{(2h-3)!!(2h-1)!!}
\,,
c_{Q_{h}}=\frac{2^{2(h-1)}(h-1)!h!}{\big((2h-1)!!\big)^2}\,.
\label{severalc}
\eea

Therefore,  the OPEs between the first three
currents in (\ref{WFWBQ}) are given by (\ref{fiveOPEs}).
The corresponding (anti)commutators can be obtained
by using the procedures described in \cite{AK2309}. 
See Appendix (\ref{fiverelation}).




\subsubsection{The $q \rightarrow 0$ limit}

By adding the bosonic currents and making the contractions with
Kronecker delta (and for fermionic currents we can do similarly)
and rescaling them by $q$ factors, 
\bea
W_h \equiv \frac{1}{2}\,
(W_{F,h}^{a b} + W_{B,h}^{a b})\, \delta_{b a} &
\rightarrow & q^{h-2}\, W_{h}, \qquad
Q_{h+\frac{1}{2}} \equiv  \frac{1}{ \sqrt{2}} \, Q_{h+\frac{1}{2}}^{a b}\,
\delta_{b a} \rightarrow  q^{h+\frac{1}{2}-\frac{3}{2}}\,
Q_{h+\frac{1}{2}},
\label{rescalingtwo}
\eea
we obtain the following (anti)commutators
\bea
\comm\Big{(W_{h_1})_m}{(W_{h_2})_n}&=& \Big((h_2-1)m-(h_1-1)n \Big)
(W_{h_1+h_2-2})_{m+n}, 
\nonu \\
    \comm\Big{(W_{h_1})_m}{(Q_{h_2+\frac{1}{2}})_r} &=&
\Big((h_2-\frac{1}{2})m-(h_1-1)r\Big)(Q_{h_1+h_2-\frac{3}{2}})_{m+r},
\nonu \\
\acomm\Big{(Q_{h_1+\frac{1}{2}})_r}{(Q_{h_2+\frac{1}{2}})_s} &= & 2
\,(-1)^{h_2+1}
\,(W_{h_1+h_2})_{r+s}.
\label{threecomm}
\eea
The central terms for $h_1=h_2=2$ in the first commutator
and for $h_1=h_2=1$ in the third anticommutator
are present but we do not write them here.
Note that by adding the two bosonic currents,
the $\frac{1}{q}$ terms are vanishing in Appendix (\ref{qzerolimit}).
In the first commutator, the conformal weights
for $h_1$ and $h_2$ are even  \footnote{ We can denote
  this bosonic subalgebra as $w^K_{ \frac{ \infty}{2}}$ algebra.}.
The anticommutator satisfies that
the conformal weights for $h_1$ and $h_2$ are both even or both odd.
Then for the former, after we introduce the
imaginary number $i$ in the mode for fermionic current,
we can absorb the minus sign on the right hand side of
last relation in (\ref{threecomm}).
We will describe the more general (anti)commutators in (\ref{nine})
where there are no restrictions for the weights
on the bosonic and fermionic currents.
As done for $SU(K)$ adjoints, we can multiply
the generators of $SO(K)$ properly into the above (anti)commutators
and obtain the corresponding (anti)commutators with $SO(K)$
group indices.

\subsection{ At $\la =\frac{1}{4}$}

As noticed in \cite{BdVnpb,BdVplb},
there exists a subalgebra generated by
\bea
&& V^{(h),+}_{\la}, h=2,4,6, \cdots, \qquad
V^{(h),-}_{\la}, h=1,3,5, \cdots, 
\nonu \\
&& Q^{(h),+}_{\la}, h=2,4,6, \cdots, \qquad
Q^{(h),-}_{\la}, h=2,4,6, \cdots,
\label{constraints}
\eea
where the weights for the bosonic currents
$V^{(h),\pm}_{\la}$ are given by $h$ and
those for the fermionic currents
$ Q^{(h),\pm}_{\la}$ are given by $(h-\frac{1}{2})$.

For $h_1,h_2$ even and  $h_3$ odd, there exists an identity
on the structure constants
for the particular value $\la =\frac{1}{4}$
\bea
&        &
    \sum_{k=0}^{h_1+h_2-h_3-1}\bigg[
\Big(
S_{F,R}^{\,\,h_1,h_2,h_3,k}(\tfrac{1}{4})
\pm S_{B,R}^{\,\,h_1,h_2,h_3,k}(\tfrac{1}{4})
\Big)
\partial_z^{h_1+h_2-h_3-k-1}\partial_w^k
\nonu \\
& &
\mp
\Big(
S_{F,L}^{\,\,h_1,h_2,h_3,k}(\tfrac{1}{4})
\pm S_{B,L}^{\,\,h_1,h_2,h_3,k}(\tfrac{1}{4})
\Big)
\partial_z^k\partial_w^{h_1+h_2-h_3-k-1}
\bigg]=0\,,
\label{cond1}
    \eea
and  
for $h_1,h_2$ even and $ h_3$ even, 
the similar identity on the structure constants
for the particular value $\la =\frac{1}{4}$
is given by
\bea
& &
    \sum_{k=0}^{h_1+h_2-h_3-1}\bigg[
\Big(
S_{F,R}^{\,\,h_1,h_2,h_3,k}(\tfrac{1}{4})
\pm S_{B,R}^{\,\,h_1,h_2,h_3,k}(\tfrac{1}{4})
\Big)
\partial_z^{h_1+h_2-h_3-k-1}\partial_w^k
\nonu \\
&
&
\pm
\Big(
S_{F,L}^{\,\,h_1,h_2,h_3,k}(\tfrac{1}{4})
\pm S_{B,L}^{\,\,h_1,h_2,h_3,k}(\tfrac{1}{4})
\Big)
\partial_z^k\partial_w^{h_1+h_2-h_3-k-1}
\bigg]=0.
    \label{cond2}
\eea

For example,
in the commutator of
$
\comm{\Big(V^{(h_1),+}_{\frac{1}{4}} \Big)_m}{\Big( V^{(h_2),+}_{\frac{1}{4}} \Big)_n}
$,
there are $\Big(V^{(h_3),+}_{\frac{1}{4}} \Big)_{m+n}$
and
$\Big(V^{(h_3),-}_{\frac{1}{4}} \Big)_{m+n}$ on the right hand side,
for any positive integers $h_1,h_2$ and $h_3$.
For $h_3$ odd with two even $h_1$ and $h_2$,
due to the identity (\ref{cond1}),
the summation over $h_3$ with
 $\Big(V^{(h_3),+}_{\frac{1}{4}} \Big)_{m+n}$
for this commutator runs over only even number.
There is no contribution from the $h_3$ odd case.
Similarly,
for $h_3$ even with  two even $h_1$ and $h_2$,
due to the identity (\ref{cond2}),
the summation over $h_3$ with
 $\Big(V^{(h_3),-}_{\frac{1}{4}} \Big)_{m+n}$
for this commutator runs over only odd number.
There is no contribution from the $h_3$ even case.
Therefore, we observe that the right hand side of this
particular commutator contains the bosonic fields
described by (\ref{constraints}).

We summarize the whole (anti)commutators in Appendix $C$
described by Appendix (\ref{appcone}) and (\ref{appctwo}).

\subsection{ At $\la =\frac{1}{2}$}

In this case, the conformal weights for $(\beta,\gamma)$ are equal to
$(\frac{1}{2},\frac{1}{2})$ and they
for $(b,c)$ are
equal to
$(1,0)$. Then we identify $\beta$ with $\gamma$ and denote it by
$\psi$ while we denote the $(b,c)$ by $(\pa \phi,  \phi)$
respectively \cite{BdVnpb,BdVplb}.
In this subsection, we use the previous OPEs (\ref{redOPE}).

By identifying the following relations
\bea
    W^{\la =\frac{1}{2}}_{F,h} &=&  -W_{B,h}^{\la =0}\,,\qquad
    W^{\la =\frac{1}{2}}_{B,h} =  W^{\la =0}_{F,h}\,,
    \qquad
    Q_{h+\frac{1}{2}}^{\la =\frac{1}{2}}  = 
    (-1)^{h+1}\,Q^{\la =0}_{h+\frac{1}{2}}\,, \qquad
    \label{laonehalf}
    \eea
    where the currents at $\la =0$ are given in (\ref{VVQ}),
    (\ref{WFWBQ}),
    we obtain the similar (anti)commutator relations
    by noting the relations (\ref{laonehalf}).
    
\subsection{ At a general $\la$}

We can also analyze the previous description
for the general $\la$ by considering
the original $(\beta, \gamma)$, $(b, c)$ system (\ref{fundOPE})
\footnote{In this case we denote the algebra as
  ${ \cal N}=2$ $W_{1+ \infty}[ \la]$ algbra by following the footnote
  $2$ of \cite{AK2309}, compared to the one in
  \cite{AK2309}. This notation holds for the previous subsection
with $\la = \frac{1}{4}$.}.
Ler us introduce the following redefined bosonic
and fermionic currents
\bea
W_{h}& \equiv &
\Big( W^{\la}_{F,\,h}+W^{\la}_{B,\,h}\Big)\,,\qquad 
    W^{\hat{A}}_{h} \equiv
    \Big( W^{\la,\hat{A}}_{F,\,h}+W^{\la,\hat{A}}_{B,\,h}\Big)\,,
    \nonu \\
    Q_{h+\frac{1}{2}}& \equiv
    & \frac{1}{\sqrt{2}}\, \Big( Q^{\la}_{h+\frac{1}{2}}
    + \bar{Q}^{\la}_{h+\frac{1}{2}} \Big)\,,
    \qquad 
    Q_{h+\frac{1}{2}}^{\hat{A}} \equiv  \frac{1}{\sqrt{2}}\, \Big(
    Q^{\la,\hat{A}}_{h+\frac{1}{2}}
    + \bar{Q}^{\la,\hat{A}}_{h+\frac{1}{2}}\Big) \,.
\label{redefined}
\eea
The reason for the combination of
bosonic currents in (\ref{redefined}) is that
when we consider each OPE between the
bosonic and fermionic currents
separately, the lowest order term
contains $\frac{1}{q^2}$ which diverges as $q \rightarrow 0$ limit.
Therefore, in order to eliminate these divergent terms,
we should consider the above addition for the bosonic currents.
Furthermore, for the fermionic currents,
we should consider the above combination because
in the anticommutator, the left hand side
should contain the same fermionic current.

\subsubsection{The $q \rightarrow 0$ limit}

By taking the following transformations for the currents,
\bea
W_{h} & \longrightarrow &
q^{h-2}\,W_{h}\,,\qquad W^{\hat{A}}_{h}
\longrightarrow q^{h}\,W^{\hat{A}}_{h}\,,\qquad 
  \nonu  \\
    Q_{h+\frac{1}{2}}
   & \longrightarrow & 
    q^{h-1}Q_{h+\frac{1}{2}}\,,\qquad 
    Q^{\hat{A}}_{h+\frac{1}{2}}
   \longrightarrow 
    q^{h+1}Q^{\hat{A}}_{h+\frac{1}{2}}, 
    \label{rescalingthree}
\eea    
we summarize the $q \rightarrow 0$ limit for the
nontrivial nine (anti)commutators from Appendix (\ref{appd}) as follows:
\bea
\comm\Big{(W_{h_1})_m}{(W_{h_2})_n}&=&\Big((h_2-1)m-(h_1-1)n \Big)
(W_{h_1+h_2-2})_{m+n}, 
\nonu \\
  \comm{(W_{\,h_1})_m}{(W_{\,h_2}^{\hat{A}})_n}&=&
  \Big((h_2-1)m-(h_1-1)n\Big)\,(W_{\,h_1+h_2-2}^{\hat{A}})_{m+n},
    \nonu \\
\comm{(W_{\,h_1}^{\hat{A}})_m}{(W_{\,h_2}^{\hat{B}})_n}&= &
-\frac{1}{4}\,i\,f^{\hat{A}\hat{B}\hat{C}}
    \,(W^{\hat{C}}_{h_1+h_2-1})_{m+n},
    \nonu \\
    \comm\Big{(W_{h_1})_m}{(Q_{h_2+\frac{1}{2}})_r} &=&
\Big((h_2-\frac{1}{2})m-(h_1-1)r\Big)(Q_{h_1+h_2-\frac{3}{2}})_{m+r},
\nonu \\
\comm{(W_{h_1})_m}{(Q_{h_2+\frac{1}{2}}^{\hat{A}})_r}&=&
\Big((h_2-\frac{1}{2})m-(h_1-1)r\Big)\,
(Q_{h_1+h_2-\frac{3}{2}}^{\hat{A}})_{m+r},
\nonu \\
\comm{(W_{h_1}^{\hat{A}})_m}{(Q_{h_2+\frac{1}{2}})_r}
&=&
\Big((h_2-\frac{1}{2})m-(h_1-1)r\Big)\,
(Q_{h_1+h_2-\frac{3}{2}}^{\hat{A}})_{m+r},
\nonu \\
\comm{(W_{h_1}^{\hat{A}})_m}{(Q^{\hat{B}}_{h_2+\frac{1}{2}})_r}&=&
-\frac{1}{4}i\,f^{\hat{A}\hat{B}\hat{C}}\,(Q^{
  \hat{C}}_{h_1+h_2-\frac{1}{2}})_{m+r},
\nonu \\
\acomm\Big{(Q_{h_1+\frac{1}{2}})_r}{(Q_{h_2+\frac{1}{2}})_s} &=& 2
\,(W_{h_1+h_2})_{r+s},
\nonu \\
\acomm{(Q_{h_1+\frac{1}{2}})_r}{(Q^{\hat{A}}_{h_2+\frac{1}{2}})_s}&=&
2
\,
(W^{\hat{A}}_{h_1+h_2})_{r+s},
\label{nine}
\eea
where the weight $h =1,2,3, \cdots$ in
$Q_{h+\frac{1}{2}}$ while
the weight $h =1,2,3, \cdots$ in
$W_{\,h}$.
Note that
the weight $h_1 =1$ in
$W_{\,h_1}$ cannot appear in the fourth, fifth and sixth relations
in (\ref{nine}) from Appendix (\ref{appd}).
The central terms are ignored.
This is other type of $\la$ deformed
${\cal N}=1$ supersymmetric $w^K_{1+\infty}$ algebra
with $U(K)$ symmetry after $q \rightarrow 0$ limit
\footnote{In the construction of \cite{BPRSS}, there is no
  bosonic current from the bosonic fields
  of weight $1$, we cannot combine it
  with the bosonic current from the fermionic fields as in
  (\ref{redefined}). This implies the bosonic subalgebra
is given by $w_{\infty}^K$ after $q \rightarrow 0$ limit.}.
Note that the first with $h_1=h_2=2$,
fourth with $h_1=2, h_2=1$ and eighth
with $h_1=h_2=1$ relations
of (\ref{nine}) consist of the standard ${\cal N}=1$
superconformal algebra with vanishing central charges
\cite{BKT,FQS} \footnote{See also previous works
related to the extensions of this algebra in \cite{Sezgin89,PS,BDZ,CKZ}.}.

\section{ An extension of Lie superalgebra $PSU(2,2|4)$   }

The weight-$\frac{1}{2}$ conjugate pairs of symplectic
boson fields \cite{GOW} are denoted by $(\la^{\al}, \mu_{\al}^{\dagger})$
and $(\mu^{\dot{\al}},\la_{\dot{\al}}^{\dagger})$ where
$\al, \dot{\al} =1, 2$
while four weight-$\frac{1}{2}$ complex fermions
are denoted by 
$(\psi^a, \psi_a^{\dagger})$ where $a=1,2,3,4$ \cite{GG2104,GG2105}.
The $\al$ and $\dot{\al}$ are spinor indices
with respect to two different $SU(2)$'s and $\psi^a(\psi_a^{\dagger})$
transforms in the fundamental(antifundamental) representation of $SU(4)$.
The conformal dimension-$\frac{1}{2}$ fields,
$(\la^{\al}, \mu_{\al}^{\dagger})$
and $(\mu^{\dot{\al}},\la_{\dot{\al}}^{\dagger})$, are bosonic
and they satisfy quasi statistics. 
See also the work from the worldsheet description in \cite{Jiang2110}.

Their nontrivial operator product expansions
in the left-moving sector of the worldsheet theory
are given by
\bea
\la^{\alpha}(z) \, \mu_{\beta}^{\dagger}(w) & = & \frac{1}{(z-w)} \,
\de^{\alpha}_{\beta}+ \cdots,
\nonu \\
\mu^{\dot{\al}}(z) \la_{\dot{\beta}}^{\dagger}(w) & = &
\frac{1}{(z-w)} \,
\de^{\dot{\alpha}}_{\dot{\beta}} + \cdots,
\nonu \\
\psi^{a}(z) \, \psi_b^{\dagger}(w) &=&
\frac{1}{(z-w)} \,
\de^{a}_{b}+ \cdots.
\label{ope}
\eea

The components of ambitwistor fields \cite{Berkovits}
are given by \bea
Z^I \equiv (\la^{\al}, \mu^{\dot{\al}},\psi^a),
\qquad
Y_J \equiv (\mu_{\al}^{\dagger}, \la_{\dot{\al}}^{\dagger},
\psi_a^{\dagger}).
\label{ZY}
\eea
Then the above OPE (\ref{ope}) with (\ref{ZY}) can be written as
\bea
Z^I(z) \, Y_J(w) = \frac{1}{(z-w)}\, \de^{I}_J + \cdots.
\label{ZYope}
\eea

By using the quadratic terms
\bea
J^I_{\,\,\, J} \equiv Y_J \, Z^I,
\label{Jij}
\eea
the generators of Lorentz symmetry $  \mathcal{L}^{\alpha}_{\,\,\,\beta}$,
$  \mathcal{\dot{L}}^{\dot{\alpha}}_{\,\,\,\dot{\beta}}$,
the generators of $R$ symmetry $  \mathcal{R}^{a}_{\,\,\,b}$,
the generators of super translations $  \mathcal{Q}^{a}_{\,\,\,\alpha}$,
$  \mathcal{\dot{Q}}^{\dot{\alpha}}_{\,\,\,a}$ and $
\mathcal{P}^{\dot{\alpha}}_{\,\,\,\beta}$,
the generators of super conformal boosts $  \mathcal{S}^{\alpha}_{\,\,\,a}$,
$  \mathcal{\dot{S}}^{a}_{\,\,\,\dot{\alpha}}$ and $  \mathcal{K}^{\alpha}_{
  \,\,\,\dot{\beta}}$
are described explicitly.
There are also the $U(1)$ hyper charge $  \mathcal{B}$,
the central charge $  \mathcal{C}$ and the dilatation generator
$  \mathcal{D}$.

We can construct the stress energy tensor which is 
\bea
    T
    &  = &
    \frac{1}{2}\Big(\,
\la^{\alpha}\partial\mu^{\dagger}_{\alpha}+\mu^{\dot{\alpha}}\partial \la^{\dagger}_{\dot{\alpha}}
-\psi^{a}\partial \psi^{\dagger}_{a}-\partial\la^{\alpha}\mu^{\dagger}_{\alpha}
-\partial \mu^{\dot{\alpha}}\la^{\dagger}_{\dot{\alpha}}
+\partial \psi^{a}\psi^{\dagger}_{a}
\,\Big)
\nonu \\
&\equiv & L^{\alpha}_{\,\,\,\alpha,\,2}
+\dot{L}^{\dot{\alpha}}_{\,\,\,\dot{\alpha},\,2}
+R^{a}_{\,\,\,a,\,2},
\label{stress}
\eea
by considering the derivatives.
It is obvious that the quadratic terms (\ref{Jij})
are weight $1$ primary operators under (\ref{stress}).
The precise definitions of the generators are given by
the next subsection.

\subsection{The extension of Lie superalgebra $PSU(2,2|4)$ }

We can construct
the following generalization of the generators in
the Lie superalgebra $PSU(2,2|4)$ together with (\ref{ope}),
(\ref{ZY}) and (\ref{ZYope})
\bea
     L^{\alpha}_{\,\,\,\beta,\,h} & = & \sum_{k=0}^{h-1}\,a_{h,k} \,
    \partial^{h-k-1}\mu^{\dagger}_{\beta}  \,  \partial^k \la^{\alpha}\,,
    \qquad
  \dot{L}^{\dot{\alpha}}_{\,\,\,\dot{\beta},\,h}=\sum_{k=0}^{h-1}\,a_{h,k} \,
    \partial^{h-k-1}   \la^{\dagger}_{\dot{\beta}} \, \partial^k \mu^{\dot{\alpha}}\,,
       \nonu \\
R^{a}_{\,\,\,b,\,h} & = & \sum_{k=0}^{h-1}\,a_{h,k} \,
    \partial^{h-k-1} \psi^{\dagger}_{b} \,  \partial^k \psi^{a}  \,,
    \qquad
 Q^{a}_{\,\,\,\alpha,\,h}=\sum_{k=0}^{h-1}\,a_{h,k} \,
 \partial^{h-k-1} \mu^{\dagger}_{\alpha} \,   \partial^k \psi^{a}
 \,,
        \nonu \\
\dot{Q}^{\dot{\alpha}}_{\,\,\,a,\,h} & = & \sum_{k=0}^{h-1}\,a_{h,k} \,
        \partial^{h-k-1} \psi^{\dagger}_{a} \,  \partial^k  \mu^{\dot{\alpha}}
        \,,
            \qquad
P^{\dot{\alpha}}_{\,\,\,\beta,\,h}=\sum_{k=0}^{h-1}\,a_{h,k} \,
    \partial^{h-k-1} \mu^{\dagger}_{\beta} \,  \partial^k  \mu^{\dot{\alpha}} \,,
    \nonu \\
S^{\alpha}_{\,\,\,a,\,h} & = & \sum_{k=0}^{h-1}\,a_{h,k} \,
    \partial^{h-k-1} \psi^{\dagger}_{a} \,  \partial^k  \la^{\alpha}
    \,,
       \qquad
 \dot{S}^{a}_{\,\,\,\dot{\alpha},\,h}=\sum_{k=0}^{h-1}\,a_{h,k} \,
 \partial^{h-k-1} \la^{\dagger}_{\dot{\alpha}} \,  \partial^k  \psi^{a}
 \,,
    \nonu \\    
K^{\alpha}_{\,\,\,\dot{\beta},\,h} & = & \sum_{k=0}^{h-1}\,a_{h,k} \,
    \partial^{h-k-1} \la^{\dagger}_{\dot{\beta}} \,  \partial^k
    \la^{\alpha}  \,,
\label{fullexp}
    \eea
    by allowing the various derivatives to (\ref{Jij}).
The relative coefficients in (\ref{fullexp})
are determined by the conditions for the
quasiprimary operators.
They are given by
\bea
a_{h,k}\equiv (-1)^k \binom{h-1}{k}^2\,a_{h,0}, \qquad
 a_{h,0} \equiv \frac{2^{h-3}(h-1)!}{(2h-3)!!}\,q^{h-2}. 
\label{NOR}
\eea
In this normalization of (\ref{NOR}), $a_{1,0}=\frac{1}{4 q}$
\footnote{
The bosonic currents can be rewritten as, by rearranging the terms,
\bea
L^{\alpha}_{\,\,\,\beta,\,h}(z)&=& (4 q)^{h-2}\sum_{i=0}^{h-1} a^{i}
(h,\lambda=\tfrac{1}{2}) \,
\partial_z^{h-i-1}
\big((\partial^{i}_z \mu^{\dagger}_{\beta})\, \lambda^{\alpha} \big)(z),
\nonu
\\
\dot{L}^{\dot{\alpha}}_{\,\,\,\dot{\beta},\,h}(z) & = & (4 q)^{h-2}\sum_{i=0}^{h-1}
 a^{i}
(h,\la=\tfrac{1}{2}) \,
\partial_z^{h-i-1}
\big((\partial^{i}_z\lambda^{\dagger}_{\dot{\beta}})\, \mu^{\dot{\alpha}}\big)(z),
\nonu       \\
R^{a}_{\,\,\,b,\,h}(z)&=& (4 q)^{h-2}\sum_{i=0}^{h-1} a^{i}
(h,\la+\tfrac{1}{2}=\tfrac{1}{2}) \,
\partial_z^{h-i-1}
\big((\partial^{i}_z\psi^{\dagger}_{b})\psi^{a}\big)(z),
\nonu    \\
P^{\dot{\alpha}}_{\,\,\,\beta,\,h}(z)&=& (4 q)^{h-2}\sum_{i=0}^{h-1} a^{i}
(h,\la=\tfrac{1}{2}) \,
\partial_z^{h-i-1}
\big((\partial^{i}_z \mu^{\dagger}_{\beta} )\,    \mu^{\dot{\alpha}}\big)(z) \,,
\nonu
\\
K^{\alpha}_{\,\,\,\dot{\alpha},\,h}(z)&=& (4 q)^{h-2}\sum_{i=0}^{h-1} a^{i}
(h,\la=\tfrac{1}{2}) \,
\partial_z^{h-i-1}
\big((\partial^{i} \la^{\dagger}_{\dot{\alpha}}) \, \la^{\alpha}\big)(z)  \,.
\nonu
\eea
Then the current $R^{a}_{\,\,\,b,\,h}$ corresponds to  $ W_{F,h}^{\la=0}$
while others do to
$W_{B,h}^{\la=\frac{1}{2}}$,
up to the group indices. For the fermionic currents,
there exist some degrees of freedom in the overall normalization
factors although the same normalization factors compared to
the bosonic currents are taken in this
paper.}.

Then we obtain the following generators
corresponding to the extension of the Lie superalgebra $PSU(2,2|4)$
from (\ref{fullexp})
\bea
    \mathcal{L}^{\alpha}_{\,\,\,\beta,\,h}
 & = &
   L^{\alpha}_{\,\,\,\beta,\,h}
-\frac{1}{2}\delta^{\alpha}_{\beta}\,L^{\gamma}_{\,\,\,\gamma,\,h}\,,
\qquad
    \dot{\mathcal{L}}^{\dot{\alpha}}_{\,\,\dot{\beta},\,h}
  =
   \dot{L}^{\dot{\alpha}}_{\,\,\,\dot{\beta},\,h}
-\frac{1}{2}\delta^{\dot{\alpha}}_{\dot{\beta}}\,\dot{L}^{\dot{\gamma}}_{\,\,\,\dot{\gamma}, \,h}\,,
\qquad
\mathcal{R}^{a}_{\,\,\,b,\,h}
=R^a_{\,\,\,b,\,h}-\frac{1}{4}\delta^{a}_b\,R^{c}_{\,\,\,c,\,h}\,,
\nonu \\
\mathcal{Q}^a_{\,\,\,\alpha,\,h} & = & Q^a_{\,\,\,\alpha,\,h}
\,,
\qquad
\dot{\mathcal{Q}}^{\dot{\alpha}}_{\,\,\,a,\,h}
 =\dot{Q}^{\dot{\alpha}}_{\,\,\,a,\,h}
\,,\qquad
\mathcal{P}^{\dot{\alpha}}_{\,\,\,\beta,\,h}=P^{\dot{\alpha}}_{\,\,\,\beta,\,h}\,,
\nonu \\
\mathcal{S}^{\alpha}_{\,\,\,a,\,h}&=& S^{\alpha}_{\,\,\,a,\,h}
\,,\qquad
\dot{\mathcal{S}}^{a}_{\,\,\,\dot{\alpha},\,h}=\dot{S}^{a}_{\,\,\,\dot{\alpha},\,h}
\,,\qquad
\mathcal{K}^{\alpha}_{\,\,\,\dot{\beta},\,h}=K^{\alpha}_{\,\,\,\dot{\beta},\,h},
\nonu \\
\mathcal{B}_h & = &
\frac{1}{2} \big(L^\alpha_{\,\,\,\alpha,\,h}+
\dot{L}^{\dot{\alpha}}_{\,\,\,\dot{\alpha},\,h}\big)\,,
\mathcal{C}_h=\frac{1}{2}\big(L^\alpha_{\,\,\,\alpha,\,h}+\dot{L}^{\dot{\alpha}}_{\,\,\,\dot{\alpha},\,h}
+R^{a}_{\,\,\,a,\,h}\big)\,,
\mathcal{D}_h=\frac{1}{2}\big(L^\alpha_{\,\,\,\alpha,\,h}-
\dot{L}^{\dot{\alpha}}_{\,\,\,\dot{\alpha},\,h}\big)\,.
\label{12objects}
\eea
Note that we have $\mathcal{R}^{a}_{\,\,\,a,\,h}=0$ from the third relation
of (\ref{12objects}).
Furthermore, there are following currents
by contractions in each $SU(2)$, $SU(2)$ and $SU(4)$ index respectively
\bea
  \mathcal{U}_h&= & L^{\gamma}_{\,\,\,\gamma,\,h}
  \,,\qquad
  \dot{\mathcal{U}}_h=\dot{L}^{\dot{\gamma}}_{\,\,\,\dot{\gamma},\,h}\,,\qquad
  \mathcal{V}_h=R^{c}_{\,\,\,c,\,h}\,,
  \label{threeu}
\eea  
which can be written in terms of the last three generators in
(\ref{12objects}).

\subsection{ The construction of bosonic current}

Because the stress energy tensor is given by
(\ref{stress}), we consider the following extension
of stress energy tensor, 
by adding the currents in (\ref{threeu}),
\bea
  W_{h} \equiv
  L^{\alpha}_{\,\,\,\alpha,\,h}+\dot{L}^{\dot{\alpha}}_{\,\,\,\dot{\alpha},\,h}+R^{a}_{\,\,\,a,\,h}\, = 2 \, \mathcal{C}_h .
\label{Wexpression}
  \eea
Then
its commutator relation satisfies
\footnote{
\label{pvalues}
  According to Appendix $B$,
  there exists a relation 
  \bea
    p_B^{h_1,h_2,h}(m,n,\tfrac{1}{2})&=p_F^{h_1,h_2,h}(m,n,0)\,
    =\frac{1}{2(h+1)!}  \,  \phi_{h+2}^{h_1 ,h_2}(0,1)\,N^{h_1,h_2}_{h}(m,n).
    \nonu
    \eea 
In particular, the mode dependent function leads to
\bea
N^{h_1,h_2}_{-1}(m,n)&=1\,,\qquad
   N^{h_1,h_2}_{0}(m,n)&=2\,\Big((h_2-1)m-(h_1-1)n\Big)\,.
   \nonu
   \eea
   Moreover,
   \bea
   \phi_{-1+2}^{h_1 ,h_2}(0,1)&=1\,,\qquad
   \phi_{0+2}^{h_1 ,h_2}(0,1)&=1\,.
   \nonu
   \eea
   This implies
   \bea
   p_B^{h_1,h_2,-1}(m,n,\tfrac{1}{2})=\frac{1}{2}, \qquad
  p_B^{h_1,h_2,0}(m,n,\tfrac{1}{2})=\Big((h_2-1)m-(h_1-1)n\Big).
   \nonu
\eea}
\bea
    \comm{(W_{h_1})_m}{(W_{h_2})_n}&=
    \sum_{h=0}^{(h_1+h_2-3)/2}
q^{2h}\,p_B^{h_1,h_2,2h}(m,n,\tfrac{1}{2})\,
(W_{h_1+h_2-2h-2})_{m+n}\,.
\label{WWcomm}
\eea

Let us write down some of the OPEs as follows:
\bea
W_1(z)\,W_2(w)&= & \frac{1}{(z-w)^2}\,W_1(w)+\cdots\,,
\nonu \\
W_1(z)\,W_3(w)&=& \frac{1}{(z-w)^2}\,2\,W_2(w)+\cdots\,,
\nonu \\
W_2(z)\,W_2(w)&=&
\frac{1}{(z-w)^2}\,2\,W_2(w)+\frac{1}{(z-w)}\,\partial W_2(w)+\cdots\,,
\nonu \\
W_2(z)\,W_3(w)&= &
\frac{1}{(z-w)^4}\,16q^2\,W_1(w)+\frac{1}{(z-w)^2}\,3\, W_3(w)+\frac{1}{(z-w)}\,\partial W_3(w)+\cdots\,,\nonu
\\
W_{3}(z)\,W_{3}(w)
&=& \frac{1}{(z-w)^4}\,64q^2\,W_{2}(w)
+\frac{1}{(z-w)^3}\,32q^2\,\partial W_{2}(w)
\nonu \\
&+&
\frac{1}{(z-w)^2}\,\bigg[\,4\,W_{4}+\frac{48\,q^2}{5}\,\partial^2 W_{2}\,\bigg](w)
\nonu \\
&+&
\frac{1}{(z-w)}\,\bigg[\,2\,\partial W_{4}+\frac{32\,q^2}{15}\,\partial^3 W_{2}\,\bigg](w)+\cdots.
\label{W123}
\eea
We can check these OPEs (\ref{W123})
are the same as the equation $(3.5)$
of \cite{AK2309}
with $\la=0$ and vanishing central terms: $W_{1+\infty}[\la=0]$ algebra.
In general, when we take $\la \rightarrow 0$ in the first equation
of Appendix $(D.1)$ of \cite{AK2309}, the corresponding commutator
without a central term becomes the one in (\ref{WWcomm}).

\subsection{ The OPEs between the bosonic current and other generators}

It is straightforward to calculate the following
commutator relations with (\ref{Wexpression})
\bea
 \comm{(W_{h_1})_m}{(\mathcal{L}^{\alpha}_{\,\,\,\beta,\,h_2})_n} &= 
& \sum_{h=0}^{(h_1+h_2-3)/2}
q^{2h}\,p_B^{h_1,h_2,2h}(m,n,\tfrac{1}{2})\,(\mathcal{L}^{\alpha}_{\,\,\,\beta,\,h_1+h_2-2h-2})_{m+n}
\,,
\nonu \\
\comm{(W_{h_1})_m}{(\dot{\mathcal{L}}^{\dot{\alpha}}_{\,\,\,\dot{\beta},\,h_2})_n}
&= &
\sum_{h=0}^{(h_1+h_2-3)/2}
q^{2h}\,p_B^{h_1,h_2,2h}(m,n,\tfrac{1}{2})\,(\dot{\mathcal{L}}^{\dot{\alpha}}_{\,\,\,\dot{\beta},\,h_1+h_2-2h-2})_{m+n}
\,,    
\nonu \\
\comm{(W_{h_1})_m}{(\mathcal{R}^{a}_{\,\,\,b,\,h_2})_n}&=
& \sum_{h=0}^{(h_1+h_2-3)/2}
q^{2h}\,p_B^{h_1,h_2,2h}(m,n,\tfrac{1}{2})\,(\mathcal{R}^{a}_{\,\,\,b,\,h_1+h_2-2h-2})_{m+n}
\,, \nonu
\\
\comm{(W_{h_1})_m}{(\mathcal{Q}^{a}_{\,\,\,\alpha,\,h_2})_n}
& =& \sum_{h=0}^{(h_1+h_2-3)/2}
q^{2h}\,p_B^{h_1,h_2,2h}(m,n,\tfrac{1}{2})\,
(\mathcal{Q}^{a}_{\,\,\,\alpha,\,h_1+h_2-2h-2})_{m+n}\,, 
\nonu \\
  \comm{(W_{h_1})_m}{(\dot{\mathcal{Q}}^{\dot{\alpha}}_{\,\,\,a,\,h_2})_n}& =
& \sum_{h=0}^{(h_1+h_2-3)/2}
q^{2h}\,p_B^{h_1,h_2,2h}(m,n,\tfrac{1}{2})\,
(\dot{\mathcal{Q}}^{\dot{\alpha}}_{\,\,\,a,\,h_1+h_2-2h-2})_{m+n}\,, 
\nonu \\
\comm{(W_{h_1})_m}{(\mathcal{P}^{\dot{\alpha}}_{\,\,\,\beta,\,h_2})_n}&=
& \sum_{h=0}^{(h_1+h_2-3)/2}
q^{2h}\,p_B^{h_1,h_2,2h}(m,n,\tfrac{1}{2})\,
(\mathcal{P}^{\dot{\alpha}}_{\,\,\,\beta,\,h_1+h_2-2h-2})_{m+n}\,, 
\nonu \\
  \comm{(W_{\,h_1})_m}{(\mathcal{S}^{\alpha}_{\,\,\,a,\,h_2})_n} &=
& \sum_{h=0}^{(h_1+h_2-3)/2}
q^{2h}\,p_B^{h_1,h_2,2h}(m,n,\tfrac{1}{2})\,
(\mathcal{S}^{\alpha}_{\,\,\,a,\,h_1+h_2-2h-2})_{m+n}\,,  
\nonu \\
 \comm{(W_{\,h_1})_m}{(\dot{\mathcal{S}}^{a}_{\,\,\,\dot{\alpha},\,h_2})_n}& =
& \sum_{h=0}^{(h_1+h_2-3)/2}
q^{2h}\,p_B^{h_1,h_2,2h}(m,n,\tfrac{1}{2})\,
(\dot{\mathcal{S}}^{a}_{\,\,\,\dot{\alpha},\,h_1+h_2-2h-2})_{m+n}\,,
\nonu \\
\comm{(W_{\,h_1})_m}{(\mathcal{K}^{\alpha}_{\,\,\,\dot{\beta},\,h_2})_n} &=&
\sum_{h=0}^{(h_1+h_2-3)/2}
q^{2h}\,p_B^{h_1,h_2,2h}(m,n,\tfrac{1}{2})\,
(\mathcal{K}^{\alpha}_{\,\,\,\dot{\beta},\,h_1+h_2-2h-2})_{m+n}\,,
\nonu \\
\comm{(W_{h_1})_m}{(\mathcal{B}_{\,h_2})_n} &=&
-q^{2(h_1-2)}\,2\,\delta_{h_1,h_2}\,c_{W_{F,h_1}}(m)\,\delta_{m+n}
\nonu \\
&+&
 \sum_{h=0}^{(h_1+h_2-3)/2}
q^{2h}\,p_B^{h_1,h_2,2h}(m,n,\tfrac{1}{2})\,
(\mathcal{B}_{h_1+h_2-2h-2})_{m+n}\,,
\nonu \\
\comm{(W_{h_1})_m}{(\mathcal{D}_{\,h_2})_n} &=
& \sum_{h=0}^{(h_1+h_2-3)/2}
q^{2h}\,p_B^{h_1,h_2,2h}(m,n,\tfrac{1}{2})\,
(\mathcal{D}_{h_1+h_2-2h-2})_{m+n}\,,
\label{commfinal}
\eea
where the central term appearing in (\ref{commfinal})
together with (\ref{severalc}) is given by
\bea
c_{W_{F,h}}(m)&=\frac{2^{2(h-3)}((h-1)!)^2}{
  (2h-3)!!(2h-1)!!}[m+h-1]_{2h-1}\,.
\label{cexpression}
\eea
Note that the structure constants (mode dependent terms)
in (\ref{commfinal})
are given by the $N$ function according to the footnote \ref{pvalues}.
We can obtain the corresponding OPEs of (\ref{commfinal})
by using the procedure in \cite{AK2309}.

\subsection{ The $q \rightarrow 0 $ limit}
\label{qlimit}

We obtain the $q \rightarrow 0$ limit for the commutators
in (\ref{commfinal}) by rescaling of $q^{h-2}$ for the bosonic currents
and $q^{h-1}$ for the fermionic currents with the help of the
footnote \ref{pvalues}
\bea
\comm{(W_{h_1})_m}{(\mathcal{L}^{\alpha}_{\,\,\,\beta,\,h_2})_n}
&=& \Big((h_2-1)m-(h_1-1)n\Big)\,(\mathcal{L}^{\alpha}_{\,\,\,\beta,\,h_1+h_2-2})_{m+n}\,,
\nonu \\
\comm{(W_{h_1})_m}{(\mathcal{\dot{L}}^{\dot{\alpha}}_{\,\,\,\dot{\beta},\,h_2})_n}
&=& \Big((h_2-1)m-(h_1-1)n\Big)\,(\mathcal{\dot{L}}^{\dot{\alpha}}_{\,\,\,\dot{\beta},\,h_1+h_2-2})_{m+n}\,,
\nonu \\
\comm{(W_{h_1})_m}{(\mathcal{R}^{a}_{\,\,\,b,\,h_2})_n}
&=& \Big((h_2-1)m-(h_1-1)n\Big)\,(\mathcal{R}^{a}_{\,\,\,b,\,h_1+h_2-2})_{m+n}\,,
\nonu \\
 \comm{(W_{h_1})_m}{(\mathcal{Q}^{a}_{\,\,\,\alpha,\,h_2})_n}
 & =&
 \Big((h_2-1)m-(h_1-1)n\Big)\,(\mathcal{Q}^{a}_{\,\,\,\alpha,\,h_1+h_2-2})_{m+n}\,, 
\nonu \\
 \comm{(W_{h_1})_m}{(\dot{\mathcal{Q}}^{\dot{\alpha}}_{\,\,\,a,\,h_2})_n} &=&
\Big((h_2-1)m-(h_1-1)n\Big)\,(\dot{\mathcal{Q}}^{\dot{\alpha}}_{\,\,\,a,\,h_1+h_2-2})_{m+n}\,,   
\nonu \\
 \comm{(W_{h_1})_m}{(\mathcal{P}^{\dot{\alpha}}_{\,\,\,\beta,\,h_2})_n} &=&
\Big((h_2-1)m-(h_1-1)n\Big)\,(\mathcal{P}^{\dot{\alpha}}_{\,\,\,\beta,\,h_1+h_2-2})_{m+n}\,, \nonu \\
 \comm{(W_{h_1})_m}{(\mathcal{S}^{\alpha}_{\,\,\,a,\,h_2})_n} &=&
\Big((h_2-1)m-(h_1-1)n\Big)\,(\mathcal{S}^{\alpha}_{\,\,\,a,\,h_1+h_2-2})_{m+n}\,, 
\nonu \\
 \comm{(W_{h_1})_m}{(\mathcal{\dot{S}}^{a}_{\,\,\,\dot{\alpha},\,h_2})_n} &=&
\Big((h_2-1)m-(h_1-1)n\Big)\,(\mathcal{\dot{S}}^{a}_{\,\,\,\dot{\alpha},\,h_1+h_2-2})_{m+n}\,, 
\nonu \\
 \comm{(W_{h_1})_m}{(\mathcal{K}^{\alpha}_{\,\,\,\dot{\beta},\,h_2})_n} &=&
\Big((h_2-1)m-(h_1-1)n\Big)\,(\mathcal{K}^{\alpha}_{\,\,\,\dot{\beta},\,h_1+h_2-2})_{m+n}\,,
\nonu \\
 \comm{(W_{h_1})_m}{(\mathcal{B}_{h_2})_n} &=&
 \Big((h_2-1)m-(h_1-1)n\Big)\,
 (\mathcal{B}_{h_1+h_2-2})_{m+n} \, ,
\nonu \\
 \comm{(W_{h_1})_m}{(W_{h_2})_n} &=&
\Big((h_2-1)m-(h_1-1)n\Big)\,(W_{h_1+h_2-2})_{m+n}\,,
\nonu \\
 \comm{(W_{h_1})_m}{(\mathcal{D}_{h_2})_n} &=&
\Big((h_2-1)m-(h_1-1)n\Big)\,(\mathcal{D}_{h_1+h_2-2})_{m+n}\,,
\label{qlimitwithW}
\eea
where the central term appearing in the third relation from the below
in (\ref{qlimitwithW})
is ignored  \footnote{The bosonic subalgebra
  $w_{ 1+ \infty}$ algebra appears in the second relation from the below
in (\ref{qlimitwithW}).}.

Furthermore,
the anticommutators are given by
\bea
\acomm{(\mathcal{Q}^{a}_{\,\,\,\alpha,\,h_1})_m}{(\dot{\mathcal{Q}}^{
    \dot{\beta}}_{\,\,\,b,\,h_2})_n} &=&
\frac{1}{4} \,
\delta^{a}_{b}\,(\mathcal{P}^{\dot{\beta}}_{\,\,\,\alpha,\,h_1+h_2-1})_{m+n},
\nonu \\
  \acomm{(\mathcal{Q}^{a}_{\,\,\,\alpha,\,h_1})_m}{(\mathcal{S}^{\beta}_{\,\,\,b,\,h_2})_n} &=&
\frac{1}{4} 
\Bigg(
\delta^{\beta}_{\alpha}\,(\mathcal{R}^{a}_{\,\,\,b,\,h_1+h_2-1})_{m+n}
+\delta^{a}_{b}\,(\mathcal{L}^{\beta}_{\,\,\,\alpha,\,h_1+h_2-1})_{m+n}
\nonu \\
& 
+& \delta^{a}_{b} \delta^{\beta}_{\alpha}\,\bigg[
\frac{1}{2}\,
(\mathcal{U}_{h_1+h_2-1})_{m+n}
+\frac{1}{4}\,
(\mathcal{V}_{h_1+h_2-1})_{m+n}\bigg]
\Bigg),    
\nonu \\
\acomm{(\dot{\mathcal{Q}}^{\dot{\alpha}}_{\,\,\,a,\,h_1})_m}{(\dot{\mathcal{S}}^{b}_{\,\,\,\dot{\beta},\,h_2})_n} &=&
\frac{1}{4} 
\Bigg(\,
\delta^{\dot{\alpha}}_{\dot{\beta}}\,(\mathcal{R}^{b}_{\,\,\,a,\,h_1+h_2-1})_{m+n}+\delta^{b}_{a}\,(\dot{\mathcal{L}}^{\dot{\alpha}}_{\,\,\,\dot{\beta},\,h_1+h_2-1})_{m+n}
\nonu \\
&
+& \delta^{b}_{a} \delta^{\dot{\alpha}}_{\dot{\beta}}\,\bigg[\frac{1}{2}\,
(\dot{\mathcal{U}}_{h_1+h_2-1})_{m+n}
+\frac{1}{4}\,
(\mathcal{V}_{h_1+h_2-1})_{m+n}\bigg]
\Bigg),
\nonu \\
\acomm{(\mathcal{S}^{\alpha}_{\,\,\,a,\,h_1})_m}{(\dot{\mathcal{S}}^{
    b}_{\,\,\,\dot{\beta},\,h_2})_n} &= &
\frac{1}{4} \,
\delta^{b}_{a}\,(\mathcal{K}^{\alpha}_{\,\,\,\dot{\beta},\,h_1+h_2-1})_{m+n},
\label{fouranticomm}
\eea
where the central terms appearing in
(\ref{fouranticomm}), which can be seen from
Appendix $E$ and contain the term (\ref{cexpression}),
are ignored.

Other commutators after $q \rightarrow 0$ limit
with appropriate rescalings for the modes of
the currents can be determined from the final results in
Appendix $E$.

\section{ The application in the celestial conformal field theory  }

According to the celestial holography,
we present the soft current algebra by combining
the algebras obtained in previous sections with the
insertion of the helicity and
propose the corresponding bulk theory
based on the ${\cal N}=1$ supergravity theory.

\subsection{  The soft current algebra in section $2$}

The algebra presented in (\ref{sevencomm}) appears also in \cite{Ahn2111}
where there is no $\la$ deformation in the two dimensional conformal field
theory.
For a given ${\cal N}=2$ supersymmetric $W^K_{1+\infty}[\la]$ algebra
\cite{AK2309},
via topological twisting \cite{PRSS}, the corresponding ${\cal N}=1$
supersymmetric topological $W^K_{\infty}[\la]$ \footnote{
  In an abstract, we use a simplified notation for this as
 $W_{\infty}$.}
is obtained in Appendix $A$.
By using the rescalings for the modes
(\ref{rescalingone}), the previous algebra (\ref{sevencomm})
is determined explicitly.
As discussed in \cite{Ahn2111,Ahn2202}, according to
the celestial holography, the corresponding OPEs (or commutators)
in the ${\cal N}=1$ supersymmetric Einstein Yang-Mills theory
are obtained in \cite{FSTZ}, even for
the presence of the deformation parameter
$\la$. The soft gravitons correspond to the bosonic
$SU(K)$  singlet current while the soft gluons do the bosonic
$SU(K)$ adjoint current.
Similarly, the soft gravitinos correspond to the fermionic
$SU(K)$  singlet current while the soft gluinos do the fermionic
$SU(K)$ adjoint current. 

By writing down the helicities explicitly,
the soft current algebra is summarized by
\bea
\comm{(\tilde{W}^{\lambda,+2}_{h_1})_{m}}{(\tilde{W}^{\lambda,+2}_{h_2})_{n}}
& = & \Big((h_2-1)m-(h_1-1)n \Big)   (\tilde{W}^{\lambda,+2}_{h_1+h_2-2})_{m+n},
\nonu \\
\comm{(\tilde{W}^{\lambda,+2}_{h_1})_{m}}{(\tilde{W}^{\lambda,\hat{A},+1}_{h_2})_{n}}
& = &
\Big((h_2-1)m-(h_1-1)n \Big)   (\tilde{W}^{\lambda,\hat{A},+1}_{h_1+h_2-2})_{m+n},
\nonu \\
\comm{(\tilde{W}^{\lambda,\hat{A},+1}_{h_1})_{m}}{(\tilde{W}^{\lambda,\hat{B},+1}_{h_2})_{n}}
& = &
-\frac{i}{4}f^{\hat{A}\hat{B}\hat{C}}
(\tilde{W}^{\lambda,\hat{C},+1}_{h_1+h_2-1})_{m+n},
\nonu \\
\comm{(\tilde{W}^{\lambda,+2}_{h_1})_{m}}{(Q^{\lambda,+\frac{3}{2}}_{h_2+\frac{1}{2}})_{r}}
& = &
\Big(h_2\,m-(h_1-1)r+\frac{h_1-1}{2} \Big)(Q^{\lambda,+\frac{3}{2}}_{h_1+h_2-\frac{3}{2}})_{m+r},
\nonu \\
\comm{(\tilde{W}^{\lambda,+2}_{h_1})_{m}}{(Q^{\lambda,\hat{A},+\frac{1}{2}}_{h_2+\frac{1}{2}})_{r}}
& = &
\Big(h_2\,m-(h_1-1)r+\frac{h_1-1}{2} \Big)(Q^{\lambda,\hat{A},+\frac{1}{2}}_{h_1+h_2-\frac{3}{2}})_{m+r},
\nonu \\
\comm{(\tilde{W}^{\lambda,\hat{A},+1}_{h_1})_{m}}{(Q^{\lambda,+\frac{3}{2}}_{h_2+\frac{1}{2}})_{r}}
& = &
\Big(h_2\,m-(h_1-1)r+\frac{h_1-1}{2} \Big)(Q^{\lambda,\hat{A},+\frac{1}{2}}_{h_1+h_2-\frac{3}{2}})_{m+r},
\nonu \\
\comm{(\tilde{W}^{\lambda,\hat{A},+1}_{h_1})_{m}}{(Q^{\lambda,\hat{B},+\frac{1}{2}}_{h_2+\frac{1}{2}})_{r}}
& = &
-\frac{i}{4}f^{\hat{A}\hat{B}\hat{C}}   (Q^{\lambda,\hat{C},+\frac{1}{2}}_{h_1+h_2-\frac{1}{2}})_{m+r}. 
\label{sevencommsoft}
\eea
The analysis for the first three commutators
is done in \cite{Strominger2105}.
See also the soft symmetry analysis from the Carrollian
amplitudes (which comes from
other flat space holography) studied in \cite{MRY}.

\subsection{The soft current algebra in subsection $3.1$}

For a given ${\cal N}=2$ supersymmetric $W^K_{1+\infty}[\la]$ algebra
\cite{AK2309},
by introducing the real free fields at fixed $\la=0$,
the corresponding ${\cal N}=1$
supersymmetric $W_{1+\infty}[\la=0]$ (or $W_{\frac{\infty}{2}}$)
is obtained in Appendix $B$.
By using the rescalings for the modes
(\ref{rescalingtwo}), the previous algebra (\ref{threecomm})
is obtained explicitly.
Compared to the one in (\ref{sevencomm}), there exists an anticommutator
between the modes for the fermionic current.
The analysis for the soft current algebra
we will not present them in this paper
can be done similarly by following the procedure in next subsection. 

\subsection{ Other soft current algebra in subsection $3.4$}

The ${\cal N}=2$ supersymmetric $W^K_{1+\infty}[\la]$ algebra
\cite{AK2309} provides 
the corresponding ${\cal N}=1$
supersymmetric $W^K_{1+\infty}[\la]$ by using the relations
(\ref{redefined}) \footnote{ We use a simplified notation for this
as  $W_{1+\infty}[\la]$ in an abstract. }.
By using the rescalings for the modes
(\ref{rescalingthree}), the algebra (\ref{nine})
is determined explicitly. In this case,
there are no restrictions on the weights for the modes of bosonic currents
compared to the one (\ref{threecomm}) where
the weights for those are even. Similar behavior for the fermionic currents
occurs.
We expect that
in the celestial conformal field theory, the
anticommutator should be described
by the soft gravitinos
with the helicity $\pm \frac{3}{2}$, and
the soft photinos (or the soft spin $\frac{1}{2}$ field)
with the helicity $\pm \frac{1}{2}$
\footnote{The analysis on this subsection
  is based on the intensive discussion for the various
  aspects on the celestial holography with M. Pate during CA's visit
  at NYU in Sept., 2022-Feb., 2023.}.

\subsubsection{ The ${\cal N}=1$
supergravity theory}

Let us consider the (simple) ${\cal N}=1$ supergravity
found in \cite{FvF}. See also \cite{DZ}. The corresponding soft algebra,
by putting the helicities in the superscript (and recalling the
construction of \cite{MRSV,BHS} in the context of the symplecton algebra),
can be described as
\bea
\comm\Big{(W^{+2}_{h_1})_m}{(W^{\pm 2}_{h_2})_n}&=&\Big((h_2-1)m-(h_1-1)n \Big)
(W^{ \pm 2}_{h_1+h_2-2})_{m+n}, 
\nonu \\
 \comm\Big{(W^{+2}_{h_1})_m}{(Q^{ \pm \frac{3}{2}}_{h_2+\frac{1}{2}})_r} &=&
    \Big((h_2-\frac{1}{2})m-(h_1-1)r\Big)(Q^{ \pm \frac{3}{2}}_{
      h_1+h_2-\frac{3}{2}})_{m+r},
    \nonu \\
\acomm\Big{(Q^{+\frac{3}{2}}_{h_1+\frac{1}{2}})_r}{(Q^{-\frac{3}{2}}_{h_2+\frac{1}{2}})_s} &=& 2
\,(W^{-2}_{h_1+h_2})_{r+s}.
\label{5exp}
\eea
From the one of the gravitational action(or Lagrangian)
$\frac{1}{\kappa^2} \, e \, R$
where
$e
\equiv
\mbox{det} \, e^a_{\mu}
$ for vierbein $e^a_{\mu}$
and $R$
is a scalar curvature
\cite{FvF}, the linear gravitational coupling
$\kappa$ term contains three gravitons with
two derivatives \cite{CSS}.
Because there exist two derivatives, the dimension of the three point
vertex is given by  $d_V = 3 +2 =5$ where $3$ comes from each dimension
of graviton \cite{PRSY}. Then the sum of helicities
becomes $(d_V-3)$ \cite{HPS}.
This implies that the sum of helicities should be
equal to $2$.
Therefore, the first commutator
in (\ref{5exp}) leads to the helicities $(+2, \pm 2, \mp 2)$
for three gravitons when $d_V=5$.
Let us emphasize that
the helicity on the right hand side in (anti)commutators
appears negatively.
All
the possible helicities from $(\pm 2, \pm \frac{3}{2})$
satisfying $d_V=5$ are listed in (\ref{5exp})
\footnote{
   \label{mathcal}
  We can check this by using the following mathematica
  calculation:
  \bea
&& {\tt Clear[SoT]}\nonu \\
&& {\tt SoT[x_,y_,z_]:=True/;x+y+z==2} \nonu \\
&& {\tt SoT[x_,y_,z_]:=False/;x+y+z!=2} \nonu \\
&& {\tt Do[If[SoT[x,y,z],Print[{x,y,z}]],
  \{x,\{-2,2,-\frac{3}{2},\frac{3}{2}\}\},\{y,\{-2,2,-\frac{3}{2},
      \frac{3}{2}\}\},\{z,\{-2,2,-\frac{3}{2},\frac{3}{2}\}\}]}
  \nonu
  \eea
  Here, $x$ $y$ stand for the helicities appearing on the left hand side
  while $z$ stands for the helicity appearing 
  on the right hand side  of (anti)commutator.
  Among nine outputs, the independent five outputs occur as in
  (\ref{5exp}).}.

From the next
gravitational action
$\ep^{\mu \nu \rho \sigma}\, \bar{\psi}_{\mu}\, \gamma_5 \, \gamma_{\nu}\,
D_{\rho} \, \psi_{\sigma}$ where $\psi_{\mu}$ is a spin $\frac{3}{2}$
Majorana spinor and $D_{\mu}$ is the gravitationally covariant
derivative \cite{FvF},
it is known that $\gamma_5$ is a constant while $\gamma_{\nu}$
contains the above vierbein \cite{VanNieuwenhuizen}
where the vierbein $e^a_{\mu}$ is expanded
around the flat Minkowski spacetimes
\cite{Woodard}. The linear $\kappa$ term contains a graviton.
Because there exists a single derivative, the dimension of the three point
vertex is given by  $d_V = \frac{3}{2} + \frac{3}{2}+1 +1 =5$
where $\frac{3}{2}$ comes from each dimension
of gravitino and $1$ does from the dimension of graviton.
Then we can write down the second commutator relation in (\ref{5exp})
and  the helicities are given by $(+2, \pm \frac{3}{2}, \mp \frac{3}{2})$
for graviton and two gravitinos. Moreover,
the helicities in the last relation in
(\ref{5exp}) are given by $(+\frac{3}{2},-\frac{3}{2}, +2)$
for two gravitinos and graviton.
See also \cite{FSTZ} for the presence of  this kind of
interaction \footnote{
Recently, in \cite{CDPR}, the three point interactions
corresponding to the first two commutators of (\ref{5exp})
are calculated. The first one comes from the product of two Yang-Mills
vertices involving the gluons while
the second one comes from the product of a super Yang-Mills vertex
involving two gluinos and one gluon and a Yang-Mills vertex involving
gluons together with various polarization tensors \cite{BE}.}.


\subsubsection{ The ${\cal N}=1$
supersymmetric Maxwell Einstein theory}

It is known that the spins $(2, \frac{3}{2})$ gauge multiplet
for supergravity \cite{FvF} (See also \cite{DZ}) is coupled to
the spins $(1, \frac{1}{2})$
vector multiplet in the Maxwell Einstein theory in \cite{FSv}.
This can be also obtained from the simplest
${\cal N}=3$ supergravity theory \cite{FSZ}
by consistent truncation.

We summarize the soft current algebra with the explicit helicities
appearing in the superscript of the modes for the currents
\bea
\comm\Big{(W^{+2}_{h_1})_m}{(W^{\pm 1}_{h_2})_n}&=&\Big((h_2-1)m-(h_1-1)n \Big)
(W^{\pm 1}_{h_1+h_2-2})_{m+n}, 
\nonu \\
\comm\Big{(W^{+2}_{h_1})_m}{(W^{\pm 2}_{h_2})_n}&=&\Big((h_2-1)m-(h_1-1)n \Big)
(W^{ \pm 2}_{h_1+h_2-2})_{m+n}, 
\nonu \\
\comm\Big{(W^{+1}_{h_1})_m}{(W^{-1}_{h_2})_n}&=&\Big((h_2-1)m-(h_1-1)n \Big)
(W^{- 2}_{h_1+h_2-2})_{m+n}, 
\nonu \\
 \comm\Big{(W^{+2}_{h_1})_m}{(Q^{ \pm \frac{3}{2}}_{h_2+\frac{1}{2}})_r} &=&
    \Big((h_2-\frac{1}{2})m-(h_1-1)r\Big)(Q^{ \pm \frac{3}{2}}_{
      h_1+h_2-\frac{3}{2}})_{m+r},
    \nonu \\
     \comm\Big{(W^{+2}_{h_1})_m}{(Q^{\pm \frac{1}{2}}_{h_2+\frac{1}{2}})_r} &=&
    \Big((h_2-\frac{1}{2})m-(h_1-1)r\Big)(Q^{ \pm \frac{1}{2}}_{
      h_1+h_2-\frac{3}{2}})_{m+r},
\nonu \\
    \comm\Big{(W^{+1}_{h_1})_m}{(Q^{+\frac{3}{2}}_{h_2+\frac{1}{2}})_r} &=&
    \Big((h_2-\frac{1}{2})m-(h_1-1)r\Big)(Q^{+\frac{1}{2}}_{
      h_1+h_2-\frac{3}{2}})_{m+r},
    \nonu \\
     \comm\Big{(W^{+1}_{h_1})_m}{(Q^{-\frac{1}{2}}_{h_2+\frac{1}{2}})_r} &=&
    \Big((h_2-\frac{1}{2})m-(h_1-1)r\Big)(Q^{-\frac{3}{2}}_{
      h_1+h_2-\frac{3}{2}})_{m+r},
\nonu \\
\acomm\Big{(Q^{+\frac{3}{2}}_{h_1+\frac{1}{2}})_r}{(Q^{-\frac{3}{2}}_{h_2+\frac{1}{2}})_s} &=& 2
\,(W^{-2}_{h_1+h_2})_{r+s},
\nonu \\
\acomm\Big{(Q^{+\frac{1}{2}}_{h_1+\frac{1}{2}})_r}{(Q^{-\frac{1}{2}}_{h_2+\frac{1}{2}})_s} &=& 2
\,(W^{-2}_{h_1+h_2})_{r+s},
\nonu \\
\acomm\Big{(Q^{+\frac{3}{2}}_{h_1+\frac{1}{2}})_r}{(Q^{-\frac{1}{2}}_{h_2+\frac{1}{2}})_s} &=& 2
\,(W^{-1}_{h_1+h_2})_{r+s}.
\label{14exp}
\eea
The second, fourth, and eighth relations
of (\ref{14exp}) can be analyzed similarly as done before.

From the one of the matter action
$e \, g^{\mu \rho}\, g^{\nu \sigma}\, F_{\mu \nu} \, F_{\rho \sigma}$
\cite{FSv}
with $F_{\mu \nu} \equiv \pa_{\mu}\, A_{\nu}- \pa_{\nu}\, A_{\mu}$ and $e\equiv
\mbox{det} \, e^a_{\mu}
$,
we observe the interaction between the graviton and two photons
by expanding the vierbein $e^a_{\mu}$ around the flat Minkowski spacetimes
\cite{Woodard} as before
and noticing that the linear term in the gravitational coupling
$\kappa$ contains the graviton.
Because there exist two derivatives, the dimension of the three point
vertex is given by  $d_V = 3 +2 =5$ where $3$ comes from each dimension
of graviton and photon \cite{PRSY}. Then sum of helicities
becomes $(d_V-3)$ \cite{HPS}.
This implies that the sum of helicities should be
equal to $2$.
Therefore, we conclude that the first commutator
in (\ref{14exp}) shows the helicities $(+2, \pm 1, \mp 1)$
for graviton, photon and photon.
Note that the helicity on the right hand side in (anti)commutators
appears negatively.
The third commutator
in (\ref{14exp}) can be analyzed similarly
and  the helicities are given by $(+1, - 1, +2)$
for two photons and graviton.
See also \cite{EJN} for the presence of  this kind of
interaction.

%

From the next matter action
$e\, \bar{\la}\, \gamma^{\mu}\, D_{\mu} \, \la$ where
$\la$ is a Majorana spin $\frac{1}{2}$ field  \cite{FSv},
there exists a single derivative, and
$d_V = \frac{3}{2} + \frac{3}{2}+1 +1 =5$
where $\frac{3}{2}$ comes from each dimension
of photino \footnote{In \cite{VanNieuwenhuizen}, the terminology
'neutrino' is used.}
and $1$ does from the dimension of graviton as before.
The fifth commutator relation in (\ref{14exp})
contains the helicities by $(+2, \pm \frac{1}{2}, \mp \frac{1}{2})$
for graviton and two photinos. Moreover,
the helicities in the ninth relation in
(\ref{14exp}) are given by $(+\frac{1}{2},-\frac{1}{2}, +2)$
for two photinos and graviton.
See also \cite{FSTZ} for the presence of  this kind of
interaction.

From the last
matter action
$e \, \kappa \, \bar{\psi}_{\mu} \, \gamma^{\alpha}\, \gamma^{\beta}\,
\gamma^{\mu} \, \la \, F_{\alpha \beta}$ at the level of
linear $\kappa$ \cite{FSv},
there exists a single derivative, and
$d_V = \frac{3}{2} + \frac{3}{2}+1 +1 =5$
where $\frac{3}{2}$ comes from each dimension
of gravitino, and photino and $1$ does
from the dimension of photon.
In this case, the vierbein plays the role of
Kronecker delta appearing in the lowest order of
$\kappa$($\kappa$ independent term) because this action has a linear
$\kappa$ term.
The sixth commutator relation in (\ref{14exp})
contains the helicities by $(+1, + \frac{3}{2}, - \frac{1}{2})$
for photon, gravitino and  photino. Moreover,
the helicities in the seventh relation in
(\ref{14exp}) are given by $(+1,-\frac{1}{2}, +\frac{3}{2})$
for photon, photino and gravitino.
Finally,
the last anticommutator relation in (\ref{14exp})
contains the helicities by $(+\frac{3}{2}, - \frac{1}{2}, +1)$
for gravitino, photino and the photon.
See also \cite{EJN,FSTZ} for the presence of  this kind of
interaction.
Note that some
possible helicities from $(\pm 2, \pm \frac{3}{2}, \pm 1, \pm
\frac{1}{2})$ satisfying $d_V=5$ are listed in (\ref{14exp})
and the other possible
helicities $(+\frac{1}{2}, +\frac{1}{2}, +1)$
and $(+\frac{3}{2}, +\frac{3}{2}, -1)$ are not present
\footnote{This can be checked by using the similar calculation
  done in the footnote \ref{mathcal} by further considering
  the helicities $\pm 1$ and $\pm \frac{1}{2}$.}
in this ${\cal N}=1$ supersymmetric Maxwell Einstein theory 
\footnote{In \cite{FGSv}, the Yang-Mills theory is
  coupled to the simple ${\cal N}=1$ supergravity.
  It is straightforward to describe the corresponding soft algebra
  from the second, third, fifth, sixth, seventh and ninth equations
  of (\ref{nine}) as done in the ${\cal N}=1$ Maxwell Einstein
  theory.}.

\subsubsection{ The ${\cal N}=1$
supergravity theory coupled to the scalar multiplet }

The globally supersymmetric multiplet
contains a scalar field $A$ (denoted by the helicity $+0$),
a pseudoscalar field $B$ (denoted by the helicity $-0$) and
a real spin $\frac{1}{2}$ field $\chi$.
The full locally supersymmetric Lagrangian is described in \cite{FFvBGS}.
We describe the corresponding soft algebra as follows:
\bea
\comm\Big{(W^{ \pm 0}_{h_1})_m}{(W^{\pm 0}_{h_2})_n}&=&
\Big((h_2-1)m-(h_1-1)n \Big)
(W^{ -2}_{h_1+h_2-2})_{m+n}, 
\nonu \\
\comm\Big{(W^{ \pm 0}_{h_1})_m}{(W^{+ 2}_{h_2})_n}&=&
\Big((h_2-1)m-(h_1-1)n \Big)
(W^{ \pm 0}_{h_1+h_2-2})_{m+n}, 
\nonu \\
\comm\Big{(W^{+2}_{h_1})_m}{(W^{\pm 2}_{h_2})_n}&=&\Big((h_2-1)m-(h_1-1)n \Big)
(W^{ \pm 2}_{h_1+h_2-2})_{m+n}, 
\nonu \\
 \comm\Big{(W^{+2}_{h_1})_m}{(Q^{ \pm \frac{3}{2}}_{h_2+\frac{1}{2}})_r} &=&
    \Big((h_2-\frac{1}{2})m-(h_1-1)r\Big)(Q^{ \pm \frac{3}{2}}_{
      h_1+h_2-\frac{3}{2}})_{m+r},
    \nonu \\
     \comm\Big{(W^{+2}_{h_1})_m}{(Q^{\pm \frac{1}{2}}_{h_2+\frac{1}{2}})_r} &=&
    \Big((h_2-\frac{1}{2})m-(h_1-1)r\Big)(Q^{ \pm \frac{1}{2}}_{
      h_1+h_2-\frac{3}{2}})_{m+r},
    \nonu \\
      \comm\Big{(W^{\pm 0}_{h_1})_m}{(Q^{+ \frac{1}{2}}_{h_2+\frac{1}{2}})_r} &=&
    \Big((h_2-\frac{1}{2})m-(h_1-1)r\Big)(Q^{ - \frac{3}{2}}_{
      h_1+h_2-\frac{3}{2}})_{m+r},
    \nonu \\
     \comm\Big{(W^{ \pm 0}_{h_1})_m}{(Q^{+ \frac{3}{2}}_{h_2+\frac{1}{2}})_r} &=&
    \Big((h_2-\frac{1}{2})m-(h_1-1)r\Big)(Q^{ - \frac{1}{2}}_{
      h_1+h_2-\frac{3}{2}})_{m+r},
\nonu \\
\acomm\Big{(Q^{+\frac{3}{2}}_{h_1+\frac{1}{2}})_r}{(Q^{-\frac{3}{2}}_{h_2+\frac{1}{2}})_s} &=& 2
\,(W^{-2}_{h_1+h_2})_{r+s},
\nonu \\
\acomm\Big{(Q^{+\frac{1}{2}}_{h_1+\frac{1}{2}})_r}{(Q^{-\frac{1}{2}}_{h_2+\frac{1}{2}})_s} &=& 2
\,(W^{-2}_{h_1+h_2})_{r+s},
\nonu \\
\acomm\Big{(Q^{+\frac{1}{2}}_{h_1+\frac{1}{2}})_r}{(Q^{+\frac{3}{2}}_{h_2+\frac{1}{2}})_s} &=& 2
\,(W^{ \pm 0}_{h_1+h_2})_{r+s}.
\label{16exp}
\eea
The third, fourth, fifth, eighth and ninth relations
of (\ref{16exp}) can be analyzed similarly as done before.
Of course, the role of spin $\frac{1}{2}$ fields
appearing in (\ref{14exp}) and (\ref{16exp}) is different from
each other.

From
$e \, g^{\mu \nu} \, \pa_{\mu} \, A \, \pa_{\nu} \, A$
and
$e \, g^{\mu \nu} \, \pa_{\mu} \, B \, \pa_{\nu} \, B$ \cite{FFvBGS},
we observe the interaction between the graviton and two scalars
by expanding the vierbein $e^a_{\mu}$ around the flat Minkowski spacetimes
and noticing that the linear term in the gravitational coupling
$\kappa$ contains the graviton as before.
Because there exist two derivatives, the dimension of the three point
vertex is given by  $d_V = 5$. Then sum of helicities
becomes $(d_V-3)$.
This implies that the sum of helicities should be
equal to $2$.
Therefore, we conclude that the first commutator
in (\ref{16exp}) shows the helicities $(\pm 0, \pm 0, +2)$
for scalar, scalar and graviton. The second commutator can be analyzed
similarly.

From
$e \, \kappa \, \bar{\psi}_{\mu}\, \slashed{\pa}
\, A \, \gamma^{\mu} \, \chi$
and
$e \, \kappa \, \bar{\psi}_{\mu}\, \gamma^5 \,
\slashed{\pa} \, B \, \gamma^{\mu} \, \chi$,
there exists a single derivative, and
$d_V = \frac{3}{2} + \frac{3}{2}+1 +1 =5$
where $\frac{3}{2}$ comes from each dimension
of gravitino, and spin $\frac{1}{2}$  field $\chi$ and $1$ does
from the dimension of scalar.
In this case, the vierbein plays the role of Kronecker delta.
The sixth commutator relation in (\ref{16exp})
contains the helicities by $(\pm 0, + \frac{1}{2}, + \frac{3}{2})$
for scalar, spin $\frac{1}{2}$ field and gravitino. Moreover,
the helicities in the seventh relation in
(\ref{16exp}) are given by $(\pm 0, +\frac{3}{2}, +\frac{1}{2})$
for scalar, gravitino and  spin $\frac{1}{2}$ field.
The last anticommutator can be described similarly.
Note that all
the possible helicities from $(\pm 2, \pm \frac{3}{2}, \pm
\frac{1}{2}, 0)$  satisfying $d_V=5$ are listed in (\ref{16exp})
with the help of the footnote \ref{mathcal}
\footnote{
  \label{finalfootnote}
  We can check that
  the soft current algebras in (\ref{5exp}), (\ref{14exp}) and (\ref{16exp})
can arise   
in the soft operators \cite{BRS1,BRS2} after ignoring the $R$
symmetry $SU(8)_R$ indices, along the line of \cite{BEF}.
That is, we do not have
their equation $(2.23)$ of \cite{BRS2} which does not have
$\frac{1}{(z-w)}$ term, the first equation of
Appendix $(A.1)$
having the helicity $(+\frac{3}{2},+\frac{3}{2},-1)$,
the first equation of Appendix $(A.2)$ with the helicity $(+1,+1,0)$,
the first three equations of Appendix $(A.3)$ where the helicity is
given by $(+\frac{1}{2},+\frac{1}{2},+1)$,
the second equation of Appendix $(A.9)$ having the helicity
$(+\frac{3}{2},-1,+\frac{3}{2})$,
the first equation of Appendix $(A.12)$ where the helicity
is $(+1,+\frac{1}{2},+\frac{1}{2})$ and
the first equations of Appendix $(A.13)$ with the helicity
$(+1,0,+1)$ in our case, but we do have
their remaining equations. We can find the similar interaction
with the helicity $(+1,+1,0)$ in \cite{MNS,MN}.

All
the possible helicities from $(\pm 2, \pm \frac{3}{2}, \pm 1, \pm
\frac{1}{2}, 0)$  satisfying $d_V=5$ can be obtained
by using the procedure presented in the footnote \ref{mathcal}.
The above helicities $(+\frac{3}{2}, +\frac{3}{2}, -1)$
corresponding to the interaction which is a
generalization of $ e\, \kappa \,  \bar{\psi}_{\mu}\,
F^{\mu \, \nu} \, \psi_{\nu}$,
$(+\frac{1}{2},+\frac{1}{2},+1)$
corresponding to the interaction which is a generalization of
$e\, \kappa \, \bar{\chi} \, \sigma^{\mu \, \nu} \, \chi \, F_{\mu \, \nu}
$
and $(+1,+1,0)$
corresponding to the interaction which is a generalization of
$ e\, \kappa \, A \, g^{\mu \, \rho} \, g^{\nu \, \sigma}\, F_{\mu \, \nu
}\,
F_{\rho \, \sigma}$
satisfy the
$d_V=5$ condition but do not appear in the simple supergravity
theory coupled to the matter of this paper. We expect that these
interactions will appear in more supersymmetric (or extended) supergravity
theories by putting the appropriate $R$ symmetry indices on the fields.
The various OPEs between the operators having the
Euler Beta function can be determined explicitly.}.

\subsection{ Towards the soft current algebra in section $4$}

The Lie superalgebra $PSU(2,2|4)$
appears in \cite{HLS,FKTv} when we construct the conformal supergravity
by gauging the superconformal symmetry of flat space theories \cite{FT}.
See also \cite{FV}.
The physical fields are the graviton corresponding to
the generator $\mathcal{P}^{\dot{\alpha}}_{\,\,\,\beta}$,
four gravitinos corresponding to the generators
$\mathcal{Q}^{a}_{\,\,\,\alpha}$
and $SU(4)$
vectors corresponding to the $R$ symmetry
generators $\mathcal{R}^{a}_{\,\,\,b}$.
Note that in the lower supersymmetric case
with Lie superalgebra $SU(2,2|{\cal N}=1,2,3)$,
  there exists an additional $U(1)$ vector
  corresponding to the generator $\mathcal{R}^{a}_{\,\,\,a}$.
See also \cite{Eberhardt,BL,DO}.

From the second relation from the last in (\ref{qlimitwithW}),
we observe the commutator relation between the soft gravitons
with helicities $(+2,\pm 2 )$
producing the soft gravitons with helicity $\mp 2$.
From the remaining relations in (\ref{qlimitwithW}),
the corresponding soft algebra can be read off directly.
For the anticommutators, we can take
a summation over the spinor indices in the second relation of
(\ref{fouranticomm}) by multiplying $\delta^{\al}_{\beta}$.
Then on the right hand side, there exist
the mode for $\mathcal{R}^{a}_{\,\,\,b, h_1+h_2-1}$, the mode
for $W_{h_1+h_2-1}$ and the mode for
the dilatation generator ${\cal D}_{h_1+h_2-1}$. 

Moreover, when we consider the commutator
between 
$(\mathcal{Q}^{a}_{\,\,\,\alpha, h_1})_m$
and $({\cal D}_{h_2})_n$, the $\frac{1}{q^2}$
term arises on the right hand side of this commutator.
At the moment, we cannot remove this divergent term.
One way to obtain the closed algebra under the particular limit
for the parameter is to consider $\la$ deformed version around
(\ref{ope}) by introducing $\la$ dependent weights of
symplectic bosons and complex fermions.
It is an open problem to observe whether the presence of $\la$
affects the above anticommutator and leads to the well defined
closed algebra.
There is other possibility for the overall normalizations for the
fermionic currents to remove the divergent terms. 





\section{ Conclusions and outlook}

The new supersymmetric algebras obtained in this paper are given by
the ones in Appendices $A,B,C$ and $E$. 
Note that there exists a nonsupersymmetric algebra
in Appendix $C$.
The corresponding $q \rightarrow 0$ limits
for Appendices $A, B, D$ and $E$  are considered.
The algebra corresponding to Appendix $D$ appears in \cite{AK2309}.
The dual descriptions for these algebras in Appendices
$A$ and $D$ (or $B$ or $C$)
under the $q \rightarrow 0 $ limit  are obtained in section $5$:
See the equations
(\ref{sevencommsoft}), (\ref{5exp}), (\ref{14exp}) and (\ref{16exp}).

In the subsection $5.3$,
some ${\cal N}=1$ supergravity theories
 are interpreted as the duals for the soft current algebras
found in this paper in the celestial conformal field theory.
We expect that as we increase the number of supersymmetry,
it is more possible to cover the interactions we do not consider
in this paper. See the footnote \ref{finalfootnote}. It is known that
there exist ${\cal N}=2$ supergravity in \cite{Fv},
${\cal N}=3$ supergravity in \cite{Freedman,FSZ}, and
${\cal N}=4$ supergravity in \cite{Das,CS1,CS2,CSF}.
It is natural to study the ${\cal N}=4$ supersymmetric
$W_{1+\infty}[\la]$
algebra \cite{Ahn2205,Ahn2208,AK2309} by considering the appropriate
$R$ symmetries depending on the number of supersymmetry.
It is known in \cite{AK2309} that
the small ${\cal N}=4$ superconformal algebra
\cite{Ademolloetal1,Ademolloetal2,AGK} is reproduced in \cite{CPS}.
See also the relevant works of \cite{BLLPRv,BHS-1}.
It is an open problem to observe whether
the ${\cal N}=3$ supersymmetric
$W_{1+\infty}[\la]$ exists or not from the above
${\cal N}=4$ supersymmetric
$W_{1+\infty}[\la]$
algebra by truncation or from the direct construction by using the free
field realization (\ref{fundOPE}).

So far, we have considered the case of the dimension of three point vertex
$d_V=5$ but we can think of the higher derivative cases
\cite{HPS,MRSV,RSYV1} where $d_V$ can be either $d_V=7$ or
$d_V=9$.
It would be interesting to
consider the massive case \cite{DFR} in the context of \cite{HP}.
We are not sure whether the ${\cal N}=1$ supersymmetric
Maxwell Einstein theory with scalars exists or not which combines
the previous subsection $5.3.2$ with the subsection $5.3.3$.
It is an open problem to obtain the ${\cal N}=1$ supergravity
theory coupled to the most general matter.
As mentioned in the introduction, as a next step,
it is an open problem to construct the soft current algebra
for the ${\cal N}=8$ supergravity theory studied in \cite{BRS1,BRS2}.
It is not obvious to observe how we can construct the
${\cal N}=8$ supersymmetric $W_{1+\infty}[\la]$ algbra
in two dimensions explicitly.

\vspace{.7cm}

\centerline{\bf Acknowledgments}

CA thanks M. Pate for  the discussion on the possibility of
nontrivial anticommutators between the fermions in the context of
\cite{HPS} and on the several aspects of three point vertices in
\cite{PRSY}, M. Rocek  for the explanation on the birth of
supergravity \cite{FvF},
T. Rahnuma and R.K. Singh for an email
correspondence on \cite{BRS1,BRS2}
and J.-H. Park for the discussion on the presence of supertranslation
in the double field theory. 
This work was supported by the National
Research Foundation of Korea(NRF) grant funded by the
Korea government(MSIT) 
(No. 2023R1A2C1003750).

\newpage

\appendix

\renewcommand{\theequation}{\Alph{section}\mbox{.}\arabic{equation}}

\section{ The seven commutators (and OPEs) of section $2$ }

The seven commutator relations in section $2$ can be summarized as
\bea
\comm{(\tilde{W}_{h_1}^{\lambda})_m}{(\tilde{W}_{h_2}^{\lambda})_n}
&=&
 \sum_{h_3=2}^{h_1+h_2-1\,\, }\sum_{k=0}^{h_1+h_2-h_3-1}
\,
(-1)^{h_3}\,(4q)^{h_1+h_2-h_3-2}\,(2h_3-2)!\nonu \\
&\times&
\bigg(\,\,\tilde{S}^{\,\,h_1,h_2,h_3,k}_{R}\, [m+h_1-1]_{h_1+h_2-h_3-k-1}\,[n+h_2-1]_{k}
\nonu \\
&-&
\tilde{S}^{\,\,h_1,h_2,h_3,k}_{L}\,  [m+h_1-1]_{k}\,[n+h_2-1]_{h_1+h_2-h_3-k-1}\,\,\bigg)\,(\tilde{W}_{h_3}^{\lambda})_{m+n},    
\nonu \\
\comm{(\tilde{W}_{h_1}^{\lambda})_m}{(\tilde{W}_{h_2}^{\lambda,\hat{A}})_n}
&= &  \sum_{h_3=2}^{h_1+h_2-1\,\, }\sum_{k=0}^{h_1+h_2-h_3-1}
\,
(-1)^{h_3}\,(4q)^{h_1+h_2-h_3-2}\,(2h_3-2)!\nonu \\
&\times&
\bigg(\,\,\tilde{S}^{\,\,h_1,h_2,h_3,k}_{R}\, [m+h_1-1]_{h_1+h_2-h_3-k-1}\,[n+h_2-1]_{k}
\nonu \\
&-&
\tilde{S}^{\,\,h_1,h_2,h_3,k}_{L}\,  [m+h_1-1]_{k}\,[n+h_2-1]_{h_1+h_2-h_3-k-1}\,\,\bigg)\,(\tilde{W}_{h_3}^{\lambda,\hat{A}})_{m+n},    
\nonu \\
\comm{(\tilde{W}_{h_1}^{\lambda,\hat{A}})_m}{(\tilde{W}_{h_2}^{\lambda,\hat{B}})_n}
&=&
\frac{1}{2}\sum_{h_3=2}^{h_1+h_2-1\,\, }\sum_{k=0}^{h_1+h_2-h_3-1}
\,
(-1)^{h_3}\,(4q)^{h_1+h_2-h_3-2}\,(2h_3-2)! \nonu \\
&\times&
\Bigg[ \bigg(\,\,\tilde{S}^{\,\,h_1,h_2,h_3,k}_{R}\, [m+h_1-1]_{h_1+h_2-h_3-k-1}\,[n+h_2-1]_{k}
\nonu \\
&-&
\tilde{S}^{\,\,h_1,h_2,h_3,k}_{L}\,  [m+h_1-1]_{k}\,[n+h_2-1]_{h_1+h_2-h_3-k-1}\,\,\bigg)
\nonu \\
& \times & \bigg(
\frac{2}{K}\delta^{\hat{A}\hat{B}}\,\,(\tilde{W}_{h_3}^{\lambda})_{m+n}
+d^{\hat{A}\hat{B}\hat{C}}\,(\tilde{W}_{h_3}^{\lambda,\hat{C}})_{m+n}
\bigg)
\nonu \\
&+&
\bigg(\,\,\tilde{S}^{\,\,h_1,h_2,h_3,k}_{R}\, [m+h_1-1]_{h_1+h_2-h_3-k-1}\,[n+h_2-1]_{k}
\nonu \\
&+ &
\tilde{S}^{\,\,h_1,h_2,h_3,k}_{L}\,  [m+h_1-1]_{k}\,[n+h_2-1]_{h_1+h_2-h_3-k-1}\,\,\bigg)\nonu \\
& \times &
i\,f^{\hat{A}\hat{B}\hat{C}}\,(\tilde{W}_{h_3}^{\lambda,\hat{C}})_{m+n} \,\Bigg],
\nonu \\
\comm{(\tilde{W}_{h_1}^{\lambda})_m}{(Q_{h_2+\frac{1}{2}}^{\lambda})_r}
&= & \sum_{h_3=1}^{h_1+h_2-1\,\, }\sum_{k=0}^{h_1+h_2-h_3-1}
\,
(-1)^{h_3}\,(4q)^{h_1+h_2-h_3-2}\,(2h_3)! \nonu \\
&\times & 
\bigg(\,\,\tilde{T}_R^{\,\,h_1,h_2,h_3,k}\,\,[m+h_1-1]_{h_1+h_2-h_3-k-1}\,[r+h_2-\tfrac{1}{2}]_k
\nonu \\
&-&
\tilde{T}_L^{\,\,h_1,h_2,h_3,k}\,\,[m+h_1-1]_{k}\,[r+h_2-\tfrac{1}{2}]_{h_1+h_2-h_3-k-1}
\bigg)\,
(Q_{h_3+\frac{1}{2}}^{\lambda})_{m+r},    
\nonu \\
\comm{(\tilde{W}_{h_1}^{\lambda})_m}{(Q_{h_2+\frac{1}{2}}^{\lambda,\hat{A}})_r}
&=& \sum_{h_3=1}^{h_1+h_2-1\,\, }\sum_{k=0}^{h_1+h_2-h_3-1}
\,
(-1)^{h_3}\,(4q)^{h_1+h_2-h_3-2}\,(2h_3)! \nonu \\
&\times& 
\bigg(\,\,\tilde{T}_R^{\,\,h_1,h_2,h_3,k}\,\,[m+h_1-1]_{h_1+h_2-h_3-k-1}\,[r+h_2-\tfrac{1}{2}]_k
\nonu \\
&-&
\tilde{T}_L^{\,\,h_1,h_2,h_3,k}\,\,[m+h_1-1]_{k}\,[r+h_2-\tfrac{1}{2}]_{h_1+h_2-h_3-k-1}
\bigg)\,
(Q_{h_3+\frac{1}{2}}^{\lambda,\hat{A}})_{m+r},  
\nonu \\
\comm{(W_{h_1}^{\lambda,\hat{A}})_m}{(Q_{h_2+\frac{1}{2}}^{\lambda})_r} & = &
\comm{(\tilde{W}_{h_1}^{\lambda})_m}{(Q_{h_2+\frac{1}{2}}^{\lambda,\hat{A}})_r},
\nonu \\
\comm{(\tilde{W}_{h_1}^{\lambda,\hat{A}})_m}{(Q_{h_2+\frac{1}{2}}^{\lambda,\hat{B}})_r}
&=&
\frac{1}{2}\sum_{h_3=1}^{h_1+h_2-1\,\, }\sum_{k=0}^{h_1+h_2-h_3-1}
\,
(-1)^{h_3}\,(4q)^{h_1+h_2-h_3-2}\,(2h_3)!
\nonu \\
&\times&
\Bigg[ \bigg(\,\,\tilde{T}^{\,\,h_1,h_2,h_3,k}_{R}\, [m+h_1-1]_{h_1+h_2-h_3-k-1}\,[n+h_2-\tfrac{1}{2}]_{k}
\nonu \\
&-&
\tilde{T}^{\,\,h_1,h_2,h_3,k}_{L}\,  [m+h_1-1]_{k}\,[n+h_2-\tfrac{1}{2}]_{h_1+h_2-h_3-k-1}\,\,\bigg)
\nonu \\
& \times & \bigg(
\frac{2}{K}\delta^{\hat{A}\hat{B}}\,\,(Q_{h_3+\frac{1}{2}}^{\lambda})_{m+r}
+d^{\hat{A}\hat{B}\hat{C}}\,(Q_{h_3+\frac{1}{2}}^{\lambda, \hat{C}})_{m+r}
\bigg)
\nonu \\
&+&
\bigg(\,\,\tilde{T}^{\,\,h_1,h_2,h_3,k}_{R}\, [m+h_1-1]_{h_1+h_2-h_3-k-1}\,[n+h_2-\tfrac{1}{2}]_{k}
\nonu \\
&+ &
\tilde{T}^{\,\,h_1,h_2,h_3,k}_{L}\,  [m+h_1-1]_{k}\,[n+h_2-\tfrac{1}{2}]_{h_1+h_2-h_3-k-1}\,\,\bigg)\nonu \\
& \times &
i\,f^{\hat{A}\hat{B}\hat{C}}\,(Q_{h_3+\frac{1}{2}}^{\lambda, \hat{C}})_{m+r}  \,\Bigg],
\label{SEVEN} 
\eea
where the structure constants are given by
\bea
\tilde{S}^{\,\,h_1,h_2,h_3,k}_{\,R}(\lambda)&=&
\sum_{i_1=1}^{h_1-1}\sum_{i_2=1}^{h_2-1}\,\sum_{r=0}^{i_1+i_2}
\Bigg[\frac{(-1)^{i_1+i_2}}{(h_3+r-1)!}\,
\beta^{i_1-1}(h_1,\lambda)
\beta^{i_2-1}(h_2,\lambda)\nonu \\
&\times&
\binom{r-1}{h_3-2}\binom{i_2}{i_1+i_2-r}\binom{1-h_3+r}{1-
h_2 +i_2+k}
\prod_{j=0}^{r-h_3}(1-r-2\lambda+j)\Bigg],
\nonu \\
\tilde{S}^{\,\,h_1,h_2,h_3,k}_{\,L}(\lambda)&=&
\sum_{i_1=1}^{h_1-1}\sum_{i_2=1}^{h_2-1}\,\sum_{r=0}^{i_1+i_2}
\Bigg[\frac{(-1)^{i_1+i_2}}{(h_3+r-1)!}\,
\beta^{i_1-1}(h_1,\lambda)
\beta^{i_2-1}(h_2,\lambda)\nonu \\
&\times&
\binom{r-1}{h_3-2}\binom{i_1}{i_1+i_2-r}\binom{1-h_3+r}{1-
h_1 +i_1+k}
\prod_{j=0}^{r-h_3}(1-r-2\lambda+j)\Bigg],
\nonu \\
\tilde{T}_{\,R}^{\,\,h_1,h_2,h_3,k}(\lambda)&=&
\sum_{i_1=0}^{h_1-1}\sum_{i_2=0}^{h_2-1}\,\sum_{r=0}^{i_1+i_2}
\Bigg[\frac{(-1)^{i_1+i_2}}{(h_3+r+1)!}\,
\Big(2a^{i_1}(h_1,\lambda+\tfrac{1}{2})-\alpha^{i_1}(h_1,\lambda)\Big)\beta^{i_2}(h_2+1,\lambda)
\nonu \\
&\times&
\binom{r}{h_3-1}\binom{i_2}{i_1+i_2-r}\binom{1-h_3+r}{1-h_2+
i_2+k}
\prod_{j=0}^{r-h_3}(-r-2\lambda+j)
\Bigg],
\nonu \\
\tilde{T}_{\,L}^{\,\,h_1,h_2,h_3,k}(\lambda)&=&
\sum_{i_1=0}^{h_1-1}\sum_{i_2=0}^{h_2-1}\,\sum_{r=0}^{i_1+i_2}
\Bigg[\frac{(-1)^{i_1+i_2}}{(h_3+r+1)!}\,
\Big(2a^{i_1}(h_1,\lambda)-\alpha^{i_1}(h_1,\lambda)\Big)\beta^{i_2}(h_2+1,\lambda)
\nonu \\
&\times&
\binom{r}{h_3-1}\binom{i_1}{i_1+i_2-r}\binom{1-h_3+r}{1-h_1+
i_1+k}
\prod_{j=0}^{r-h_3}(-r-2\lambda+j)
\Bigg].
\label{finalappa}
\eea
Note that there are some identities between the structure constants
in (\ref{Iden}).
The first and the fourth of Appendix
(\ref{SEVEN}) appeared in (\ref{ssexp}) and
(\ref{inter1})
respectively. Of course, the above structure constants
become the ones in \cite{Ahn2202} for the vanishing $\la$.

The corresponding seven OPEs can be obtained by
the description in \cite{AK2309} from the above
commutator relations Appendix (\ref{SEVEN}) together with Appendix
(\ref{finalappa}).

\section{ The details of the subsection $3.1$ }

\subsection{ The five (anti)commutators in the subsection $3.1$ }

For convenience, we present the five
(anti)commutators corresponding to (\ref{fiveOPEs})
\bea
\comm{(W_{F,h_1}^{ab})_m}{(W_{F,h_2}^{cd})_n}
&=&
N\,\delta^{h_1 h_2}\Big((-1)^{h_1}\delta^{ac}\delta^{bd}+\delta^{ad}\delta^{bc}\Big)q^{2(h_1-2)}\,c_{W_{F,h_1}}\,
[m+h_1-1]_{2h_1-1} \nonu \\
&
+&
\frac{1}{2}\sum_{h=-1}^{h_1+h_2-3}
q^h \,p_F^{h_1,h_2,h}(m,n,0)
\nonu \\
&\times & \Bigg[
\delta^{a c}(-1)^{h_1}\,(W_{F,h_1+h_2-h-2}^{bd})_{m+n}
+\delta^{a d}(-1)^{h}\,(W_{F,h_1+h_2-h-2}^{cb})_{m+n}
\nonu \\
&
+& \delta^{b c}\,(W_{F,h_1+h_2-h-2}^{ad})_{m+n}
+\delta^{b d}(-1)^{h_2}\,(W_{F,h_1+h_2-h-2}^{ac})_{m+n}
\Bigg],
\nonu \\
\comm{(W_{B,h_1}^{ab})_m}{(W_{B,h_2}^{cd})_n}
&=&
N\,\delta^{h_1 h_2}\Big((-1)^{h_1}\delta^{ac}\delta^{bd}+\delta^{ad}\delta^{bc}\Big)q^{2(h_1-2)}\,c_{W_{B,h_1}}\,
[m+h_1-1]_{2h_1-1} \nonu \\
&
+ & \frac{1}{2}\sum_{h=-1}^{h_1+h_2-4}
q^h \,p_B^{h_1,h_2,h}(m,n,0)
\nonu \\
&\times & \Bigg[
\delta^{a c}(-1)^{h_1}\,(W_{B,h_1+h_2-h-2}^{bd})_{m+n}
+\delta^{a d}(-1)^{h}\,(W_{B,h_1+h_2-h-2}^{cb})_{m+n}
\nonu \\
&
+& \delta^{b c}\,(W_{B,h_1+h_2-h-2}^{ad})_{m+n}
+\delta^{b d}(-1)^{h_2}\,(W_{B,h_1+h_2-h-2}^{ac})_{m+n}
\Bigg],
\nonu \\
\comm{(W_{F,h_1}^{ab})_m}{(Q_{h_2+\frac{1}{2}}^{cd})_r}
&=& \sum_{h=-1}^{h_1+h_2-3}
q^h \,q_F^{h_1,h_2+\frac{1}{2},h}(m,r,0)
\nonu \\
&\times & \Bigg[
\delta^{a d}\,(Q_{h_1+h_2-h-\frac{3}{2}}^{cb})_{m+r}
+\delta^{b d}(-1)^{h_1}\,(Q_{F,h_1+h_2-h-\frac{3}{2}}^{ca})_{m+r}
\Bigg],
\nonu \\
\comm{(W_{B,h_1}^{ab})_m}{(Q_{h_2+\frac{1}{2}}^{cd})_r}
&=& \sum_{h=-1}^{h_1+h_2-3}
q^h \,q_B^{h_1,h_2+\frac{1}{2},h}(m,r,0)
\nonu \\
&\times & \Bigg[
\delta^{a c}\,(-1)^{h_1}\,(Q_{h_1+h_2-h-\frac{3}{2}}^{bd})_{m+r}
+\delta^{b c}\,(W_{F,h_1+h_2-h-\frac{3}{2}}^{ad})_{m+r}
\Bigg],
\nonu \\
\acomm{(Q_{h_1+\frac{1}{2}}^{ab})_r}{(Q_{h_1+\frac{1}{2}}^{cd})_s}
&=&
N\,\delta^{h_1 h_2}\Big((-1)^{h_1+1}\delta^{ac}\delta^{bd}\Big)q^{2(h_1-1)}\,
c_{Q_{h_1}}\,
[r+h_1-\tfrac{1}{2}]_{2h_1} \nonu \\
&
-& \sum_{h=0}^{h_1+h_2-1}
q^h \,o_F^{h_1+\frac{1}{2},h_2+\frac{1}{2},h}(r,s,0)
\,\delta^{a c}\,(-1)^{h_1+h}\,(W_{F,h_1+h_2-h}^{bd})_{r+s}
\nonu \\
&
-& \sum_{h=0}^{h_1+h_2-2}
q^h \,o_B^{h_1+\frac{1}{2},h_2+\frac{1}{2},h}(r,s,0)
\,\delta^{b d}\,(-1)^{h_2}\,(W_{B,h_1+h_2-h}^{ac})_{r+s}.
\label{fiverelation}
\eea


The generalized hypergeometric function
is denoted by
\bea
\phi_{r}^{h_1 ,h_2}(\Lambda,a)  \equiv \ _4F_3\left[
\begin{array}{c|}
\frac{1}{2} + \Lambda \ ,  \frac{1}{2} - \Lambda  \ , \frac{1+a-r}{ 2}\ , \frac{a-r}{2}\\
\frac{3}{2}-h_1 \ , \frac{3}{2} -h_2\ , \frac{1}{2}+ h_1+h_2-r
\end{array}  \ 1\right],
\label{phi}
\eea
and mode dependent function
is given by
\bea
N^{h_1,h_2}_{h}(m,n)
\!&
\equiv \!&
\sum_{l=0 }^{h+1}(-1)^l
\left(\begin{array}{c}
h+1 \\  l \\
\end{array}\right)
[h_1-1+m]_{h+1-l}[h_1-1-m]_l
\nonu \\
\!& \times \!& [h_2-1+n]_l [h_2-1-n]_{h+1-l}.
\label{Ndef}
\eea
Let us introduce
the three kinds of structure constants in \cite{AK2009}
together with Appendix (\ref{phi}) and Appendix (\ref{Ndef})
\bea
\mathrm{BB}^{h_1,h_2}_{r,\,\pm}(m,n; \mu )
&\equiv&
 -\frac{1  }{ (r-1)!} N_{r-2}^{h_1, h_2}(m,n) \Bigg[
\phi_{r}^{h_1 ,h_2}(\mu,1)  \pm \phi_{r}^{h_1 ,h_2}(1-\mu,1)  
\Bigg],
\nonu\\
\mathrm{BF}^{h_1,h_2+\frac{1}{2}}_{r,\,\pm}(m,\rho; \mu )
&\equiv&
 -\frac{1  }{ (r-1)!} N_{r-2}^{h_1, h_2+\frac{1}{2}}(m,\rho) \Bigg[
 \phi_{r+1}^{h_1 ,h_2+1}(\mu,\frac{3\pm1}{2})
 \nonu \\
 & \pm & \phi_{r+1}^{h_1 ,h_2+1}(1-\mu,\frac{3\pm1}{2})  
\Bigg],
\nonu\\
\mathrm{FF}^{h_1+\frac{1}{2},h_2+\frac{1}{2}}_{r,\,\pm}(\rho,\omega; \mu )
&\equiv&
-\frac{1  }{ (r-1)!}N_{r-2}^{h_1+\frac{1}{2}, h_2+\frac{1}{2}}(\rho,\omega)
\Bigg[
 \phi_{r+1}^{h_1+1 ,h_2+1}(\mu,\frac{3\pm1}{2})  \nonu \\
 & \pm & \phi_{r+1}^{h_1+1 ,h_2+1}(1-\mu,\frac{3\pm1}{2})  
 \Bigg].
\label{3struct}
\eea

Then the structure constants in section $3$, from Appendix (\ref{3struct}),
are given by
\bea
p_{F,h}^{h_1,h_2}(m,n,\la) & = & -\frac{1}{4} \Bigg[
  \mathrm{BB}^{h_1,h_2}_{h+2,\,+} +
   \mathrm{BB}^{h_1,h_2}_{h+2,\,-} \Bigg]_{\mu = 2 \la},
\nonu \\
p_{B,h}^{h_1,h_2}(m,n,\la) & = & -\frac{1}{4} \Bigg[
  \mathrm{BB}^{h_1,h_2}_{h+2,\,+} -
   \mathrm{BB}^{h_1,h_2}_{h+2,\,-} \Bigg]_{\mu = 2 \la},
\nonu \\
q_{F,2h}^{h_1,h_2+\frac{1}{2}}(m,n,\la) & = &  \Bigg[
 -\frac{1}{8} \mathrm{BF}^{h_1,h_2+\frac{1}{2}}_{2h+2,\,+} +
 \frac{(2h_1-2h-3)}{16(h+1)}\,
 \mathrm{BF}^{h_1,h_2+\frac{1}{2}}_{2h+2,\,-} \Bigg]_{\mu = 2 \la},
\nonu \\
q_{F,2h+1}^{h_1,h_2+\frac{1}{2}}(m,n,\la) & = &  \Bigg[
 \frac{1}{8} \mathrm{BF}^{h_1,h_2+\frac{1}{2}}_{2h+3,\,+} -
 \frac{(h_1-h-2)}{4(2h+3)}\,
 \mathrm{BF}^{h_1,h_2+\frac{1}{2}}_{2h+3,\,-} \Bigg]_{\mu = 2 \la},
\nonu \\
q_{B,2h}^{h_1,h_2+\frac{1}{2}}(m,n,\la) & = &  \Bigg[
 -\frac{1}{8} \mathrm{BF}^{h_1,h_2+\frac{1}{2}}_{2h+2,\,+} -
 \frac{(2h_1-2h-3)}{16(h+1)}\,
 \mathrm{BF}^{h_1,h_2+\frac{1}{2}}_{2h+2,\,-} \Bigg]_{\mu = 2 \la},
\nonu \\
q_{B,2h+1}^{h_1,h_2+\frac{1}{2}}(m,n,\la) & = &  \Bigg[
 -\frac{1}{8} \mathrm{BF}^{h_1,h_2+\frac{1}{2}}_{2h+3,\,+} -
 \frac{(h_1-h-2)}{4(2h+3)}\,
 \mathrm{BF}^{h_1,h_2+\frac{1}{2}}_{2h+3,\,-} \Bigg]_{\mu = 2 \la},
\nonu \\
o_{F,2h}^{h_1+\frac{1}{2},h_2+\frac{1}{2}}(m,n,\la) & = &  \Bigg[
 -\mathrm{FF}^{h_1+\frac{1}{2},h_2+\frac{1}{2}}_{2h+1,\,+} -
 \frac{2(h_1+h_2-h)}{(2h+1)}\,
 \mathrm{FF}^{h_1+\frac{1}{2},h_2+\frac{1}{2}}_{2h+1,\,-} \Bigg]_{\mu = 2 \la},
\nonu \\
o_{F,2h+1}^{h_1+\frac{1}{2},h_2+\frac{1}{2}}(m,n,\la) & = &  \Bigg[
 \mathrm{FF}^{h_1+\frac{1}{2},h_2+\frac{1}{2}}_{2h+2,\,+} +
 \frac{2(h_1+h_2-h)-1}{2(h+1)}\,
 \mathrm{FF}^{h_1+\frac{1}{2},h_2+\frac{1}{2}}_{2h+2,\,-} \Bigg]_{\mu = 2 \la},
\nonu \\
o_{B,2h}^{h_1+\frac{1}{2},h_2+\frac{1}{2}}(m,n,\la) & = &  \Bigg[
 -\mathrm{FF}^{h_1+\frac{1}{2},h_2+\frac{1}{2}}_{2h+1,\,+} +
 \frac{2(h_1+h_2-h)}{(2h+1)}\,
 \mathrm{FF}^{h_1+\frac{1}{2},h_2+\frac{1}{2}}_{2h+1,\,-} \Bigg]_{\mu = 2 \la},
\nonu \\
o_{B,2h+1}^{h_1+\frac{1}{2},h_2+\frac{1}{2}}(m,n,\la) & = &  \Bigg[
- \mathrm{FF}^{h_1+\frac{1}{2},h_2+\frac{1}{2}}_{2h+2,\,+} +
 \frac{2(h_1+h_2-h)-1}{2(h+1)}\,
 \mathrm{FF}^{h_1+\frac{1}{2},h_2+\frac{1}{2}}_{2h+2,\,-} \Bigg]_{\mu = 2 \la}.
\label{structla}
\eea
At $\la=0$, the above structure constants reduce to
the ones in \cite{Odake,AKK1910}.

\subsection{ The $q \rightarrow 0$ limit at $\la=0$ in the
  subsection $3.1$ }

We present some $q$ terms for Appendix (\ref{fiverelation}) 
after taking the rescalings
in (\ref{rescalingtwo})
\bea
     \comm\Big{(W_{F,h_1}^{ab})_m}{(W_{F,h_2}^{cd})_n}&
     =&
     q^{4(h_1-2)}\,N\,\delta^{h_1 h_2}\Big((-1)^{h_1}\delta^{ac}\delta^{bd}+\delta^{ad}\delta^{bc}\Big)\,c_{W_{F,h_1}}\,
[m+h_1-1]_{2h_1-1}
\nonu \\
&+& \frac{1}{4q^2}\,
\Bigg[
\delta^{a c}(-1)^{h_1}\,(W_{F,h_1+h_2-1}^{bd})_{m+n}
-\delta^{a d}\,(W_{F,h_1+h_2-1}^{cb})_{m+n}
\nonu \\
& + & \delta^{b c}\,(W_{F,h_1+h_2-1}^{ad})_{m+n}
+\delta^{b d}(-1)^{h_2}\,(W_{F,h_1+h_2-1}^{ac})_{m+n}
\Bigg]
\nonu \\
&
+ & q^0\, \frac{1}{2}\,\Big((h_2-1)m-(h_1-1)n \Big)\,
\nonu \\
&\times & \Bigg[
\delta^{a c}(-1)^{h_1}\,(W_{F,h_1+h_2-2}^{bd})_{m+n}
+\delta^{a d}\,(W_{F,h_1+h_2-2}^{cb})_{m+n}
\nonu \\
& + & \delta^{b c}\,(W_{F,h_1+h_2-2}^{ad})_{m+n}
+\delta^{b d}(-1)^{h_2}\,(W_{F,h_1+h_2-2}^{ac})_{m+n}
\Bigg]
+\cdots, 
\nonu \\
\comm\Big{(W_{B,h_1}^{ab})_m}{(W_{B,h_2}^{cd})_n} & = & 
q^{4(h_1-2)}\,N\,\delta^{h_1 h_2}\Big((-1)^{h_1}\delta^{ac}\delta^{bd}+\delta^{ad}\delta^{bc}\Big)\,c_{W_{B,h_1}}\,
[m+h_1-1]_{2h_1-1}
\nonu \\
&+ & \frac{1}{4q^2} \,
\Bigg[
\delta^{a c}(-1)^{h_1}\,(W_{B,h_1+h_2-1}^{bd})_{m+n}
-\delta^{a d}\,(W_{B,h_1+h_2-1}^{cb})_{m+n}
\nonu \\
&+ & \delta^{b c}\,(W_{B,h_1+h_2-1}^{ad})_{m+n}
+\delta^{b d}(-1)^{h_2}\,(W_{B,h_1+h_2-1}^{ac})_{m+n}
\Bigg]
\nonu \\
&
+& q^0\,\frac{1}{2}\Big((h_2-1)m-(h_1-1)n \Big)\,
\nonu \\
&\times & \Bigg[
\delta^{a c}(-1)^{h_1}\,(W_{B,h_1+h_2-2}^{bd})_{m+n}
+\delta^{a d}\,(W_{B,h_1+h_2-2}^{cb})_{m+n}
\nonu \\
& + & \delta^{b c}\,(W_{B,h_1+h_2-2}^{ad})_{m+n}
+\delta^{b d}(-1)^{h_2}\,(W_{B,h_1+h_2-2}^{ac})_{m+n}
\Bigg]
+ \cdots,
\nonu \\
\comm{(W_{F,h_1}^{ab})_m}{(Q_{h_2+\frac{1}{2}}^{cd})_r}
& = & -\frac{1}{4q^2}\,\Bigg[
\delta^{a d}\,(Q_{h_1+h_2-\frac{1}{2}}^{cb})_{m+r}
+\delta^{b d}(-1)^{h_1}\,(Q_{F,h_1+h_2-\frac{1}{2}}^{ca})_{m+r}
\Bigg]
\nonu \\
&
+&
q^0\frac{(2h_2^2-h_1-4h_2+2h_1 h_2+1)\Big((2h_2-1)m-2(h_1-1)n\Big)}{2(2h_2-1)(2h_1+2h_2-3)}
\nonu \\
&\times & \Bigg[
\delta^{a d}\,(Q_{h_1+h_2-\frac{3}{2}}^{cb})_{m+r}
+\delta^{b d}(-1)^{h_1}\,(Q_{F,h_1+h_2-\frac{3}{2}}^{ca})_{m+r}
\Bigg]
+ \cdots,  
\nonu \\
\comm{(W_{B,h_1}^{ab})_m}{(Q_{h_2+\frac{1}{2}}^{cd})_r}
&=& \frac{1}{4q^2}\,
\Bigg[
\delta^{a c}\,(-1)^{h_1}\,(Q_{h_1+h_2-h-\frac{1}{2}}^{bd})_{m+r}
+\delta^{b c}\,(Q_{h_1+h_2-h-\frac{1}{2}}^{ad})_{m+r}
\Bigg]
\nonu \\
&
+&
q^0\frac{(2h_2^2-h_1-4h_2+2h_1 h_2+2)\Big((2h_2-1)m-2(h_1-1)n\Big)}{2(2h_2-1)(2h_1+2h_2-3)}
\nonu \\
&\times &
\Bigg[
\delta^{a c}\,(-1)^{h_1}\,(Q_{h_1+h_2-h-\frac{3}{2}}^{bd})_{m+r}
+\delta^{b c}\,(Q_{h_1+h_2-h-\frac{3}{2}}^{ad})_{m+r}
\Bigg]
+ \cdots,  
\nonu \\
\acomm{(Q_{h_1+\frac{1}{2}}^{ab})_r}{(Q_{h_2+\frac{1}{2}}^{cd})_s}
&=&
q^{4(h_1-1)}\,N\,\delta^{h_1 h_2}\Big((-1)^{h_1+1}\delta^{ac}\delta^{bd}\Big)\,c_{Q_{h_1}}\,
[r+h_1-\tfrac{1}{2}]_{2h_1} \nonu \\
&
- &
q^0\,2 \,\delta^{a c}\,(-1)^{h_1}\,(W_{F,h_1+h_2}^{bd})_{r+s}
\nonu \\
&
-& q^0\,2 \,\delta^{b d}\,(-1)^{h_2}\,(W_{B,h_1+h_2}^{ac})_{r+s}
\quad
\nonu \\
&
-&
q^2\,\frac{4(2h_1h_2-h_1-h_2+1)\Big((2h_2-1)r-(2h_1-1)s\Big))}{(2h_1-1)(2h_2-1)}
\nonu \\
& \times & \delta^{a c}\,(-1)^{h_1}\,(W_{F,h_1+h_2-1}^{bd})_{r+s}
\nonu \\
&
-& q^2\,\frac{4(2h_1h_2-h_1-h_2)\Big((2h_2-1)r-(2h_1-1)s\Big))}{(2h_1-1)(2h_2-1)}
\nonu \\
& \times & \delta^{b d}\,(-1)^{h_2}\,(W_{B,h_1+h_2-1}^{ac})_{r+s}
\nonu \\
&+& \cdots,
\label{qzerolimit}
\eea
where the abbreviated parts in Appendix (\ref{qzerolimit})
are the higher orders of $q$.
We can easily remove the divergent $\frac{1}{q^2}$ terms
by linear combinations between the currents.

\section{ The (anti)commutators at $\la =\frac{1}{4}$
in the subsection $3.2$}

We present the (anti)commutators as follows:
\bea
    \comm{\Big(V^{(h_1),+}_{\frac{1}{4}} \Big)_m}{\Big( V^{(h_2),+}_{\frac{1}{4}} \Big)_n}
  &  = &
 \sum_{h_3=2,\,\text{even}}^{h_1+h_2-2}\sum_{k=0}^{\,\,h_1+h_2-h_3-1}
   (2h_3-1)!\, \nonu \\
&\times & \Big(
S_{F,R}^{\,\,h_1,h_2,h_3,k}(\tfrac{1}{4})
+S_{B,R}^{\,\,h_1,h_2,h_3,k}(\tfrac{1}{4})
\Big)
 \nonu \\
&\times &
[m+h_1-1]_{h_1+h_2-h_3-k-1}[n+h_2-1]_{k}
\,\Big(V^{(h_3),+}_{\frac{1}{4}} \Big)_{m+n}
\nonu \\
& + & \sum_{h_3=1,\,\text{odd}}^{h_1+h_2-1}\sum_{k=0}^{\,\,h_1+h_2-h_3-1}
  2 \,(2h_3-1)!\, \nonu \\
&\times & \Big(
S_{F,R}^{\,\,h_1,h_2,h_3,k}(\tfrac{1}{4})
-S_{B,R}^{\,\,h_1,h_2,h_3,k}(\tfrac{1}{4})
\Big)
\nonu \\
&\times &
[m+h_1-1]_{h_1+h_2-h_3-k-1}[n+h_2-1]_{k}
\,\Big(V^{(h_3),-}_{\frac{1}{4}} \Big)_{m+n},
\nonu \\
    \comm{\Big(V^{(h_1),-}_{\frac{1}{4}} \Big)_m}{\Big( V^{(h_2),-}_{\frac{1}{4}} \Big)_n}
  &  = &
 \sum_{h_3=2,\,\text{even}}^{h_1+h_2-2}\sum_{k=0}^{\,\,h_1+h_2-h_3-1}
 \frac{1}{4} (2h_3-1)!\,
\nonu
 \\
&\times & \Big(
S_{F,R}^{\,\,h_1,h_2,h_3,k}(\tfrac{1}{4})
+S_{B,R}^{\,\,h_1,h_2,h_3,k}(\tfrac{1}{4})
\Big) \nonu
\\
&\times &
[m+h_1-1]_{h_1+h_2-h_3-k-1}[n+h_2-1]_{k}
\,\Big(V^{(h_3),+}_{\frac{1}{4}} \Big)_{m+n}
\nonu \\
& + & \sum_{h_3=1,\,\text{odd}}^{h_1+h_2-1}\sum_{k=0}^{\,\,h_1+h_2-h_3-1}
\frac{1}{2} \,(2h_3-1)!\,
\nonu
\\
&\times & \Big(
S_{F,R}^{\,\,h_1,h_2,h_3,k}(\tfrac{1}{4})
-S_{B,R}^{\,\,h_1,h_2,h_3,k}(\tfrac{1}{4})
\Big) \nonu
\\
&\times &
[m+h_1-1]_{h_1+h_2-h_3-k-1}[n+h_2-1]_{k}
\,\Big(V^{(h_3),-}_{\frac{1}{4}} \Big)_{m+n},
\nonu \\
\comm{\Big(V^{(h_1),+}_{\frac{1}{4}} \Big)_m}{\Big( V^{(h_2),-}_{\frac{1}{4}} \Big)_n}
  &  = &
 \sum_{h_3=2,\,\text{even}}^{h_1+h_2-1}\sum_{k=0}^{\,\,h_1+h_2-h_3-1}
  \frac{1}{2} (2h_3-1)!\, \nonu \\
&\times & \Big(
-S_{F,R}^{\,\,h_1,h_2,h_3,k}(\tfrac{1}{4})
+S_{B,R}^{\,\,h_1,h_2,h_3,k}(\tfrac{1}{4})
\Big)
\nonu \\
&\times &
[m+h_1-1]_{h_1+h_2-h_3-k-1}[n+h_2-1]_{k}
\,\Big(V^{(h_3),+}_{\frac{1}{4}} \Big)_{m+n}
\nonu \\
& + & \sum_{h_3=1,\,\text{odd}}^{h_1+h_2-2}\sum_{k=0}^{\,\,h_1+h_2-h_3-1}
(2h_3-1)!\,\nonu \\
&\times & \Big(
-S_{F,R}^{\,\,h_1,h_2,h_3,k}(\tfrac{1}{4})
-S_{B,R}^{\,\,h_1,h_2,h_3,k}(\tfrac{1}{4})
\Big)
\nonu \\
&\times &
[m+h_1-1]_{h_1+h_2-h_3-k-1}[n+h_2-1]_{k}
\,\Big(V^{(h_3),-}_{\frac{1}{4}} \Big)_{m+n},
\nonu \\
    \comm{\Big(V^{(h_1),+}_{\frac{1}{4}} \Big)_m}{\Big( Q^{(h_2),+}_{\frac{1}{4}} \Big)_r}
  &  = &
   \sum_{h_3=2,\,\text{even}}^{h_1+h_2-2}\sum_{k=0}^{\,\,h_1+h_2-h_3-1}
   (2h_3-2)!\,
\nonu \\
&\times &
\Bigg[\,
\Big(
- T_{F}^{\,\,h_1,h_2-1,h_3-1,k}(\tfrac{1}{4})
+\bar{T}_{B}^{\,\,h_1,h_2-1,h_3-1,k}(\tfrac{1}{4})
\Big)\nonu \\
& \times & [m+h_1-1]_{h_1+h_2-h_3-k-1}[r+h_2-\tfrac{3}{2}]_k\,
\Big( Q^{(h_3),+}_{\frac{1}{4}} \Big)_{m+r}
\nonu \\
&
+ &
\Big(
 T_{F}^{\,\,h_1,h_2-1,h_3-1,k}(\tfrac{1}{4})
+\bar{T}_{B}^{\,\,h_1,h_2-1,h_3-1,k}(\tfrac{1}{4})
\Big)\nonu \\
&
\times & [m+h_1-1]_{h_1+h_2-h_3-k-1}[r+h_2-\tfrac{3}{2}]_k\,
\Big( Q^{(h_3),-}_{\frac{1}{4}} \Big)_{m+r}
\Bigg],
\nonu \\
    \comm{\Big(V^{(h_1),-}_{\frac{1}{4}} \Big)_m}{\Big( Q^{(h_2),+}_{\frac{1}{4}} \Big)_r}
  &  = &
   \sum_{h_3=2,\,\text{even}}^{h_1+h_2-1}\sum_{k=0}^{\,\,h_1+h_2-h_3-1}
  \frac{1}{2}  \,  (2h_3-2)!\,
\nonu \\
&\times &
\Bigg[\,
\Big(
T_{F}^{\,\,h_1,h_2-1,h_3-1,k}(\tfrac{1}{4})
+\bar{T}_{B}^{\,\,h_1,h_2-1,h_3-1,k}(\tfrac{1}{4})
\Big)
\nonu \\
& \times & [m+h_1-1]_{h_1+h_2-h_3-k-1}[r+h_2-\tfrac{3}{2}]_k\,
\Big( Q^{(h_3),+}_{\frac{1}{4}} \Big)_{m+r}
\nonu \\
&
+ &
\Big(
- T_{F}^{\,\,h_1,h_2-1,h_3-1,k}(\tfrac{1}{4})
+\bar{T}_{B}^{\,\,h_1,h_2-1,h_3-1,k}(\tfrac{1}{4})
\Big)
\nonu \\
&
\times & [m+h_1-1]_{h_1+h_2-h_3-k-1}[r+h_2-\tfrac{3}{2}]_k\,
\Big( Q^{(h_3),-}_{\frac{1}{4}} \Big)_{m+r}
\Bigg],
\nonu \\
    \comm{\Big(V^{(h_1),+}_{\frac{1}{4}} \Big)_m}{\Big( Q^{(h_2),-}_{\frac{1}{4}} \Big)_r}
  &  = &
   \sum_{h_3=2,\,\text{even}}^{h_1+h_2-2}\sum_{k=0}^{\,\,h_1+h_2-h_3-1}
   (2h_3-2)!\,
\nonu \\
&\times &
\Bigg[\,
\Big(
T_{F}^{\,\,h_1,h_2-1,h_3-1,k}(\tfrac{1}{4})
+\bar{T}_{B}^{\,\,h_1,h_2-1,h_3-1,k}(\tfrac{1}{4})
\Big)
\nonu \\
&\times & [m+h_1-1]_{h_1+h_2-h_3-k-1}[r+h_2-\tfrac{3}{2}]_k\,
\Big( Q^{(h_3),+}_{\frac{1}{4}} \Big)_{m+r}
\nonu \\
&
+ &
\Big(
- T_{F}^{\,\,h_1,h_2-1,h_3-1,k}(\tfrac{1}{4})
+\bar{T}_{B}^{\,\,h_1,h_2-1,h_3-1,k}(\tfrac{1}{4})
\Big) \nonu \\
&
\times & [m+h_1-1]_{h_1+h_2-h_3-k-1}[r+h_2-\tfrac{3}{2}]_k\,
\Big( Q^{(h_3),-}_{\frac{1}{4}} \Big)_{m+r}
\Bigg],
\nonu \\
    \comm{\Big(V^{(h_1),-}_{\frac{1}{4}} \Big)_m}{\Big( Q^{(h_2),-}_{\frac{1}{4}} \Big)_r}
  &  = &
   \sum_{h_3=2,\,\text{even}}^{h_1+h_2-1}\sum_{k=0}^{\,\,h_1+h_2-h_3-1}
   \frac{1}{2}(2h_3-2)!\,
\nonu \\
&\times &
\Bigg[\,
\Big(
-T_{F}^{\,\,h_1,h_2-1,h_3-1,k}(\tfrac{1}{4})
+\bar{T}_{B}^{\,\,h_1,h_2-1,h_3-1,k}(\tfrac{1}{4})
\Big)
\nonu \\
& \times & [m+h_1-1]_{h_1+h_2-h_3-k-1}[r+h_2-\tfrac{3}{2}]_k\,
\Big( Q^{(h_3),+}_{\frac{1}{4}} \Big)_{m+r}
\nonu \\
&
+ &
\Big( T_{F}^{\,\,h_1,h_2-1,h_3-1,k}(\tfrac{1}{4})
+\bar{T}_{B}^{\,\,h_1,h_2-1,h_3-1,k}(\tfrac{1}{4})
\Big)
\nonu \\
&
\times & [m+h_1-1]_{h_1+h_2-h_3-k-1}[r+h_2-\tfrac{3}{2}]_k\,
\Big( Q^{(h_3),-}_{\frac{1}{4}} \Big)_{m+r}
\Bigg],
\nonu \\
\acomm{\Big( Q^{(h_1),\pm}_{\frac{1}{4}} \Big)_r}{\Big( Q^{(h_2),\pm}_{\frac{1}{4}} \Big)_s}
  &  = &
   \sum_{h_3=2,\,\text{Even}}^{h_1+h_2-2}\sum_{k=0}^{\,\,h_1+h_2-h_3-2}
   (2h_3-1)!
 \nonu \\
&\times &
\Bigg[\,
\Big(
\pm U_{F}^{\,\,h_1-1,h_2-1,h_3,k}(\tfrac{1}{4})
\pm U_{B}^{\,\,h_2-1,h_1-1,h_3,k}(\tfrac{1}{4})
\Big)
\nonu
\\
& \times & [r+h_1-\tfrac{3}{2}]_{h_1+h_2-h_3-k-2}[s+h_2-\tfrac{3}{2}]_k\,
\Big( V^{(h_3),+}_{\frac{1}{4}} \Big)_{r+s}
\Bigg]   
\nonu \\
  & 
+&  \sum_{h_3=1,\,\text{odd}}^{h_1+h_2-1}\sum_{k=0}^{\,\,h_1+h_2-h_3-2}
  2\, (2h_3-1)!
 \nonu \\
&\times &
\Bigg[\,
\Big(
\pm U_{F}^{\,\,h_1-1,h_2-1,h_3,k}(\tfrac{1}{4})
\mp U_{B}^{\,\,h_2-1,h_1-1,h_3,k}(\tfrac{1}{4})
\Big)
\nonu \\
& \times & [r+h_1-\tfrac{3}{2}]_{h_1+h_2-h_3-k-2}[s+h_2-\tfrac{3}{2}]_k\,
\Big( V^{(h_3),-}_{\frac{1}{4}} \Big)_{r+s}
\Bigg],   
\nonu \\
\acomm{\Big( Q^{(h_1),\pm}_{\frac{1}{4}} \Big)_r}{\Big( Q^{(h_2),\mp}_{\frac{1}{4}} \Big)_s}
  &  = &
   \sum_{h_3=2,\,\text{Even}}^{h_1+h_2-2}\sum_{k=0}^{\,\,h_1+h_2-h_3-2}
   (2h_3-1)!
 \nonu \\
&\times &
\Bigg[\,
\Big(
\pm U_{F}^{\,\,h_1-1,h_2-1,h_3,k}(\tfrac{1}{4})
\mp U_{B}^{\,\,h_2-1,h_1-1,h_3,k}(\tfrac{1}{4})
\Big)
\nonu
\\
& \times & [r+h_1-\tfrac{3}{2}]_{h_1+h_2-h_3-k-2}[s+h_2-\tfrac{3}{2}]_k\,
\Big( V^{(h_3),+}_{\frac{1}{4}} \Big)_{r+s}
\Bigg]   
\nonu \\
  & 
+&  \sum_{h_3=1,\,\text{odd}}^{h_1+h_2-1}\sum_{k=0}^{\,\,h_1+h_2-h_3-2}
  2\, (2h_3-1)!
 \nonu \\
&\times &
\Bigg[\,
\Big(
\pm U_{F}^{\,\,h_1-1,h_2-1,h_3,k}(\tfrac{1}{4})
\pm U_{B}^{\,\,h_2-1,h_1-1,h_3,k}(\tfrac{1}{4})
\Big)
\nonu \\
& \times & [r+h_1-\tfrac{3}{2}]_{h_1+h_2-h_3-k-2}[s+h_2-\tfrac{3}{2}]_k\,
\Big( V^{(h_3),-}_{\frac{1}{4}} \Big)_{r+s}
\Bigg],
\label{appcone}
\eea
where the structure constants are, from \cite{AK2309},
\bea
S^{\,\,h_1,h_2,h_3,k}_{F,\,R}(\lambda)&=&
\sum_{i_1=0}^{h_1-1}\sum_{i_2=0}^{h_2-1}\,\sum_{r=0}^{i_1+i_2}
\Bigg[\frac{(-1)^{i_1+i_2}}{(h_3+r)!}\,
a^{i_1}(h_1,\lambda+\tfrac{1}{2})
a^{i_2}(h_2,\lambda+\tfrac{1}{2})\nonu \\
&\times&
\binom{r}{h_3-1}\binom{i_2}{i_1+i_2-r}\binom{1-h_3+r}{1-
h_2 +i_2+k}
\prod_{j=0}^{r-h_3}(-r-2\lambda+j)\Bigg],
\nonu \\
S^{\,\,h_1,h_2,h_3,k}_{F,\,L}(\lambda)&=&
\sum_{i_1=0}^{h_1-1}\sum_{i_2=0}^{h_2-1}\,\sum_{r=0}^{i_1+i_2}
\Bigg[\frac{(-1)^{i_1+i_2}}{(h_3+r)!}\,
a^{i_1}(h_1,\lambda+\tfrac{1}{2})
a^{i_2}(h_2,\lambda+\tfrac{1}{2})
\nonu \\
m&\times&
\binom{r}{h_3-1}\binom{i_1}{i_1+i_2-r}
\binom{1-h_3+r}{1-h_1+i_1+k}
\prod_{j=0}^{r-h_3}( -r-2\lambda+j)\Bigg],
\nonu \\
S^{\,\,h_1,h_2,h_3,k}_{B,\,R}(\lambda)&=&
\sum_{i_1=0}^{h_1-1}\sum_{i_2=0}^{h_2-1}\,\sum_{r=0}^{i_1+i_2}
\Bigg[\frac{(-1)^{i_1+i_2}}{(h_3+r)!}
\,a^{i_1}(h_1,\lambda\big)a^{i_2}(h_2,\lambda\big)
\nonu \\
&\times&
\binom{r}{h_3-1}
\binom{i_2}{i_1+i_2-r}\binom{1-h_3+r}{1-h_2+i_2+k}
\prod_{j=0}^{r-h_3}(1-r-2\lambda+j)\Bigg],
\nonu \\
S^{\,\,h_1,h_2,h_3,k}_{B,\,L}(\lambda)&=&
\sum_{i_1=0}^{h_1-1}\sum_{i_2=0}^{h_2-1}\,\sum_{r=0}^{i_1+i_2}
\Bigg[\frac{(-1)^{i_1+i_2}}{(h_3+r)!}
\,a^{i_1}(h_1,\lambda)a^{i_2}(h_2,\lambda)
\nonu \\
&\times&
\binom{r}{h_3-1}
\binom{i_1}{i_1+i_2-r}\binom{1-h_3+r}{1-h_1+i_1+k}
\prod_{j=0}^{r-h_3} (1-r-2\lambda+j)\Bigg],
\nonu \\
T_{F}^{\,\,h_1,h_2,h_3,k}(\lambda)&=&
\sum_{i_1=0}^{h_1-1}\sum_{i_2=0}^{h_2-1}\,\sum_{r=0}^{i_1+i_2}
\Bigg[\frac{(-1)^{i_1+i_2}}{(h_3+r+1)!}\,
a^{i_1}(h_1,\lambda+\tfrac{1}{2})\beta^{i_2}(h_2+1,\lambda)
\nonu \\
&\times&
\binom{r}{h_3-1}\binom{i_2}{i_1+i_2-r}\binom{1-h_3+r}{1-h_2+
i_2+k}
\prod_{j=0}^{r-h_3}(-r-2\lambda+j)
\Bigg],
\nonu \\
\bar{T}_{F}^{\,\,h_1,h_2,h_3,k}(h_1,\lambda)&=&
\sum_{i_1=0}^{h_1-1}\sum_{i_2=0}^{h_2}\,\sum_{r=0}^{i_1+i_2}
\Bigg[\frac{(-1)^{i_1+i_2}}{(h_3+r)!}\,
a^{i_1}(h_1,\lambda+\tfrac{1}{2})\alpha^{i_2}(h_2+1,\lambda)
\nonu \\
&\times&
\binom{r}{h_3}\binom{i_1}{i_1+i_2-r}\binom{-h_3+r}{1-h_1+
i_1+k}
\prod_{j=0}^{r-h_3-1}( 1-r-2\lambda+j)\Bigg],
\nonu \\
T_{B}^{\,\,h_1,h_2,h_3,k}(\lambda)&=&
\sum_{i_1=0}^{h_1-1}\sum_{i_2=0}^{h_2-1}\,\sum_{r=0}^{i_1+i_2}
\Bigg[\frac{(-1)^{i_1+i_2}}{(h_3+r+1)!}\,
a^{i_1}(h_1,\lambda)
\beta^{i_2}(h_2+1,\lambda)
\nonu \\
&\times&
\binom{r}{h_3-1}\binom{i_1}{i_1+i_2-r}\binom{1-h_3+r}{
  1-h_1+i_1+k}
\prod_{j=0}^{r-h_3}(-r-2\lambda+j)
\Bigg],
\nonu \\
\bar{T}_{B}^{\,\,h_1,h_2,h_3,k}(\lambda)&=&
\sum_{i_1=0}^{h_1-1}\sum_{i_2=0}^{h_2}\,\sum_{r=0}^{i_1+i_2}
\Bigg[\frac{(-1)^{i_1+i_2}}{(h_3+r)!}\,
a^{i_1}(h_1,\lambda)\alpha^{i_2}(h_2+1,\lambda)
\nonu \\
&\times&
\binom{r}{h_3}\binom{i_2}{i_1+i_2-r}\binom{-h_3+r}{-
h_2+i_2+k}
\prod_{j=0}^{r-h_3-1}(1-r-2\lambda+j)\Bigg],
\nonu \\
U_F^{\,\,h_1,h_2,h_3,k}(\lambda)&=&
\sum_{i_1=0}^{h_1-1}\sum_{i_2=0}^{h_2}\sum_{r=0}^{i_1+i_2}
\Bigg[\frac{(-1)^{i_1+i_2}}{(h_3+r)!}\,
\beta^{i_1}(h_1+1,\lambda)\alpha^{i_2}(h_2+1,\lambda)
\nonu \\
&\times&
\binom{r}{h_3-1}
\binom{i_2}{i_1+i_2-r}\binom{1-h_3+r}{-h_2+
i_2+k}
\prod_{j=0}^{r-h_3}(-r-2\lambda+j )\Bigg],
\nonu \\
U_B^{\,\,h_1,h_2,h_3,k}(\lambda)&=& 
\sum_{i_1=0}^{h_1-1}\sum_{i_2=0}^{h_2}\sum_{r=0}^{i_1+i_2}
\Bigg[\frac{(-1)^{i_1+i_2}}{(h_3+r)!}\,
\beta^{i_1}(h_1+1,\lambda)\alpha^{i_2}(h_2+1,\lambda)
\nonu \\
&\times&
\binom{r}{h_3-1}
\binom{i_1}{i_1+i_2-r}\binom{1-h_3+r}{1-h_1+i_1+k}
\prod_{j=0}^{r-h_3}( 1-r-2\lambda+j)
\Bigg].
\label{appctwo}
\eea
Note that on the right hand sides of Appendix (\ref{appcone}),
the dummy variables $h_3$ are restricted to be the even
for the modes of
bosonic currents having $+$ superscript and the odd
for the modes of
bosonic currents having $-$ superscript.
They are restricted to be the even
for the modes of
fermionic currents having $\pm$ superscript.

In \cite{BdVnpb,BdVplb},
there exists other subalgebra generated by
\bea
&& V^{(h),+}_{\la}, h=2,4,6, \cdots, \qquad
V^{(h),-}_{\la}, h=1,3,5, \cdots, 
\nonu \\
&& Q^{(h),+}_{\la}, h=1,3,5, \cdots, \qquad
Q^{(h),-}_{\la}, h=1,3,5, \cdots,
\label{constraints1}
\eea
which is the nonsupersymmetric case.
We can express the corresponding (anti)commutators similarly.
The first three commutators and the last
two anticommutators of Appendix (\ref{appcone}) remain the same.
Of course, the fermions are restricted to have odd number of weights
as in Appendix (\ref{constraints1}). 
For the next four commutators of Appendix (\ref{appcone}),
we need to write down for the odd number for the fermions
as in Appendix (\ref{constraints1}) with the extra minus signs on the
right hand sides. In other words, the structure constants
are opposite, compared to the supersymmetric case in
Appendix (\ref{appcone}).

\section{ The $q \rightarrow 0$ limit  at general $\la$ in the
  subsection $3.4$ }

We present some $q$ terms for Appendix $D$ of \cite{AK2309}
after a rescaling (\ref{rescalingthree}) as follows:
\bea
&&    \comm{(W^{\la}_{F,\,h_1})_m}{(W^{\la}_{F,\,h_2})_n}
    =
    q^{2(h_1+h_2-4)}\,K\,N\,c^{h_1,h_2}_{W_F}(\la)\,[m+h_1-1]_{h_1+h_2-1}
    \nonu \\
    & & +
    q^0\,    \Big((h_2-1)m-(h_1-1)n\Big)\,(W^{\la}_{F,\,h_1+h_2-2})_{m+n}
    + \cdots\,,
    \nonu \\
  &&  \comm{(W^{\la}_{F,\,h_1})_m}{(W^{\la,\hat{A}}_{F,\,h_2})_n}
    =
    q^0\,\Big((h_2-1)m-(h_1-1)n\Big)\,(W^{\la,\hat{A}}_{F,\,h_1+h_2-2})_{m+n}
 + \cdots
    \,,
    \nonu \\
    && \comm{(W^{\la,\hat{A}}_{F,\,h_1})_m}{(W^{\la,\hat{B}}_{F,\,h_2})_n}
    = q^{2(h_1+h_2-2)}
     \,N\,\delta^{\hat{A}\hat{B}}\, c^{h_1,h_2}_{W_F}(\la)\,[m+h_1-1]_{h_1+h_2-1}
     \nonu\\
    &&+
    q^4\,\delta^{\hat{A}\hat{B}}\,\frac{1}{K}\,\Big((h_2-1)m-(h_1-1)n\Big)\,(W^{\la}_{F,\,h_1+h_2-2})_{m+n}
    \nonu \\
    && +
    q^{2} \,\frac{1}{2}\,\Big((h_2-1)m-(h_1-1)n\Big)\,d^{\hat{A}\hat{B}\hat{C}}
\,(W^{\la,\hat{C}}_{F,\,h_1+h_2-2})_{m+n}
\nonu \\
&& - q^{0}\,\frac{i}{4}\,f^{\hat{A}\hat{B}\hat{C}}
\,(W^{\la,\hat{C}}_{F,\,h_1+h_2-1})_{m+n}
 + \cdots
\,,
    \nonu \\
   && \comm{(W^{\la}_{B,\,h_1})_m}{(W^{\la}_{B,\,h_2})_n}
    = q^{2(h_1+h_2-4)}\,K\,N\,c^{h_1,h_2}_{W_B}(\la)\,[m+h_1-1]_{h_1+h_2-1}
    \nonu \\
    &&+
    q^0\,\Big((h_2-1)m-(h_1-1)n\Big)\,(W^{\la}_{B,\,h_1+h_2-2})_{m+n}
 + \cdots
    \,,
     \nonu \\
    && \comm{(W^{\la}_{B,\,h_1})_m}{(W^{\la,\hat{A}}_{B,\,h_2})_n}
    =
     q^0\,\Big((h_2-1)m-(h_1-1)n\Big)\,(W^{\la,\hat{A}}_{B,\,h_1+h_2-2})_{m+n}
 + \cdots
     \,,
\nonu     \\
    && \comm{(W^{\la,\hat{A}}_{B,\,h_1})_m}{(W^{\la,\hat{B}}_{B,\,h_2})_n}
     =
     q^{2(h_1+h_2-2)}\,N\,\delta^{\hat{A}\hat{B}}\, c^{h_1,h_2}_{W_B}(\la)\,[m+h_1-1]_{h_1+h_2-1} \nonu \\
    &&+
    q^4\,\delta^{\hat{A}\hat{B}}\,\frac{1}{K}\,\Big((h_2-1)m-(h_1-1)n\Big)\,(W^{\la}_{B,\,h_1+h_2-2})_{m+n}
    \nonu \\
    &&+
    q^{2} \,\frac{1}{2}\,\Big((h_2-1)m-(h_1-1)n\Big)\,d^{\hat{A}\hat{B}\hat{C}}
\,(W^{\la,\hat{C}}_{B,\,h_1+h_2-2})_{m+n}
\nonu \\
&&- q^{0}\,\frac{i}{4}\,f^{\hat{A}\hat{B}\hat{C}}
\,(W^{\la,\hat{C}}_{B,\,h_1+h_2-1})_{m+n}
 + \cdots
\,,
\nonu \\
%
%
%
 &&   \comm{(W^{\la}_{F,\,h_1})_m}{(Q^{\la}_{h_2+\frac{1}{2}})_r}
    =-\frac{1}{4q^2}\,(Q^{\la}_{h_1+h_2-\frac{1}{2}})_{m+r}
+ \cdots     \nonu \\
    && +   q^0
      \begin{cases}
        \frac{(h_2-1)(h_2-2\la)}{(2h_2-1)}\,m,
          \,\, \text{If}\, h_1=1
       \nonu  \\
    \frac{(2h_2^2-h_1-4h_2+2h_1 h_2+1+2\la)((2h_2-1)m-2(h_1-1)r)}{2(2h_2-1)(2h_1+2h_2-3)},
      \,\, \text{If}\, h_1>1
    \end{cases}
    \Bigg\}
    (Q^{\la}_{h_1+h_2-\frac{3}{2}})_{m+r},
    \nonu \\
   && \comm{(W^{\la}_{F,\,h_1})_m}{(Q^{\la,\hat{A}}_{h_2+\frac{1}{2}})_r}
    =
    -\frac{1}{4q^2}\,(Q^{\la,\hat{A}}_{h_1+h_2-\frac{1}{2}})_{m+r}
+ \cdots \nonu   \\
  && +  q^0
    \begin{cases}
      \frac{(h_2-1)(h_2-2\la)}{(2h_2-1)}\, m,  \,\,
      \text{If}\,h_1=1
       \nonu  \\
    \frac{(2h_2^2-h_1-4h_2+2h_1 h_2+1+2\la)((2h_2-1)m-2(h_1-1)r)}{2(2h_2-1)(2h_1+2h_2-3)}, 
     \,\, \text{If}\,h_1>1
    \end{cases}
    \Bigg\}(Q^{\la,\hat{A}}_{h_1+h_2-\frac{3}{2}})_{m+r},
    \nonu \\
   && \comm{(W^{\la,\hat{A}}_{F,\,h_1})_m}{(Q^{\la}_{h_2+\frac{1}{2}})_r} =
    \comm{(W^{\la}_{F,\,h_1})_m}{(Q^{\la,\hat{A}}_{h_2+\frac{1}{2}})_r},
    \nonu \\
   && \comm{(W^{\la,\hat{A}}_{F,\,h_1})_m}{(Q^{\la,\hat{B}}_{h_2+\frac{1}{2}})_r}
    =
    -q^2\frac{1}{4}\,\delta^{\hat{A}\hat{B}}\,\frac{1}{K}\,(Q^{\la}_{h_1+h_2-\frac{1}{2}})_{m+r}
    \nonu       \\
    && -
    q^0\,\frac{1}{8}\,( d^{\hat{A}\hat{B}\hat{C}} +
          i\,f^{\hat{A}\hat{B}\hat{C}}  )\,(Q^{\la,\hat{C}}_{h_1+h_2-\frac{1}{2}})_{m+r}
          + \cdots,
\nonu \\
&& \comm{(W^{\la}_{B,\,h_1})_m}{(Q^{\la}_{h_2+\frac{1}{2}})_r}
    = \frac{1}{4q^2}\,(Q^{\la}_{h_1+h_2-\frac{1}{2}})_{m+r}
+ \cdots    \nonu \\
    && +  q^0
   \begin{cases}
     \frac{(h_2-1)(h_2-1+2\la)}{(2h_2-1)}\, m,  \,\,
     \text{If}\,h_1=1
       \nonu \\
    \frac{(2h_2^2-h_1-4h_2+2h_1 h_2+2-2\la)((2h_2-1)m-2(h_1-1)r)}{2(2h_2-1)(2h_1+2h_2-3)}, 
     \,\, \text{If}\,h_1>1
    \end{cases}
    \Bigg\}(Q^{\la}_{h_1+h_2-\frac{3}{2}})_{m+r},
      \nonu \\
   &&   \comm{(W^{\la}_{B,\,h_1})_m}{(Q^{\la,\hat{A}}_{h_2+\frac{1}{2}})_r}
    = \frac{1}{4q^2}\,(Q^{\la,\hat{A}}_{h_1+h_2-\frac{1}{2}})_{m+r}
 + \cdots   \nonu \\
  && +  q^0
    \begin{cases}
      \frac{(h_2-1)(h_2-1+2\la)}{(2h_2-1)}\, m,
      \,\, \text{If}\,h_1=1
        \nonu \\
    \frac{(2h_2^2-h_1-4h_2+2h_1 h_2+2-2\la)((2h_2-1)m-2(h_1-1)r)}{2(2h_2-1)(2h_1+2h_2-3)},
     \,\, \text{If}\,h_1>1
    \end{cases}
    \Bigg\}(Q^{\la,\hat{A}}_{h_1+h_2-\frac{3}{2}})_{m+r},
    \nonu \\
   && \comm{(W^{\la,\hat{A}}_{B,\,h_1})_m}{(Q^{\la}_{h_2+\frac{1}{2}})_r} =
    \comm{(W^{\la}_{B,\,h_1})_m}{(Q^{\la,\hat{A}}_{h_2+\frac{1}{2}})_r},
    \nonu \\
   && \comm{(W^{\la,\hat{A}}_{B,\,h_1})_m}{(Q^{\la,\hat{B}}_{h_2+\frac{1}{2}})_r}
    =
    q^2\,\frac{1}{4}\,\delta^{\hat{A}\hat{B}}\,\frac{1}{K}\,(Q^{\la}_{h_1+h_2-\frac{1}{2}})_{m+r}
    \nonu    \\
        && + q^0\,\frac{1}{8}\, ( d^{\hat{A}\hat{B}\hat{C}} - i\,
     f^{\hat{A}\hat{B}\hat{C}}   )\,(Q^{\la,\hat{C}}_{h_1+h_2-\frac{1}{2}})_{m+r}
         + \cdots,
       \nonu  \\
 && \comm{(W_{F,\,h_1}^{\la})_m}{(\bar{Q}^{\la}_{h_2+\frac{1}{2}})_r}
  = \frac{1}{4q^2}\,(\bar{Q}^{\la}_{h_1+h_2-\frac{1}{2}})_{m+r}
+ \cdots    \nonu \\
   && +  q^0
    \begin{cases}
      \frac{h_2(h_2-1+2\la)}{(2h_2-1)}\, m,   \,\,
      \text{If}\,h_1=1
       \nonu \\
    \frac{(2h_2^2-h_1-4h_2+2h_1 h_2+1+2\la)((2h_2-1)m-2(h_1-1)r)}{2(2h_2-1)(2h_1+2h_2-3)},
     \,\, \text{If}\,h_1>1
    \end{cases}
    \Bigg\}(\bar{Q}^{\la}_{h_1+h_2-\frac{3}{2}})_{m+r},
  \nonu \\
 && \comm{(W_{F,\,h_1}^{\la})_m}{(\bar{Q}^{\la,\hat{A}}_{h_2+\frac{1}{2}})_r}
 %
= \frac{1}{4q^2}\,(\bar{Q}^{\la,\hat{A}}_{h_1+h_2-\frac{1}{2}})_{m+r}
  + \cdots  \nonu \\
  && +  q^0
    \begin{cases}
      \frac{h_2(h_2-1+2\la)}{(2h_2-1)}\, m, \,\,
      \text{If}\,h_1=1
        \nonu \\
    \frac{(2h_2^2-h_1-4h_2+2h_1 h_2+1+2\la)((2h_2-1)m-2(h_1-1)r)}{2(2h_2-1)(2h_1+2h_2-3)},
     \,\, \text{If}\,h_1>1
    \end{cases}
    \Bigg\}(\bar{Q}^{\la,\hat{A}}_{h_1+h_2-\frac{3}{2}})_{m+r},
    \nonu \\
  &&  \comm{(W_{F,\,h_1}^{\la,\hat{A}})_m}{(\bar{Q}^{\la}_{h_2+\frac{1}{2}})_r}  =
  \comm{(W_{F,\,h_1}^{\la})_m}{(\bar{Q}^{\la,\hat{A}}_{h_2+\frac{1}{2}})_r},   
\nonu \\
&& \comm{(W_{F,\,h_1}^{\la,\hat{A}})_m}{(\bar{Q}^{\la,\hat{B}}_{h_2+\frac{1}{2}})_r}
 =
 q^2\,\frac{1}{4}\,\delta^{\hat{A}\hat{B}}\,\frac{1}{K}\,(\bar{Q}^{\la}_{h_1+h_2-\frac{1}{2}})_{m+r}
 \nonu   \\
    &&+ q^0\,\frac{1}{8}\, (d^{\hat{A}\hat{B}\hat{C}} - i \,
f^{\hat{A}\hat{B}\hat{C}}    )\,
    (\bar{Q}^{\la,\hat{C}}_{h_1+h_2-\frac{1}{2}})_{m+r}
       + \cdots, 
  \nonu \\
 && \comm{(W_{B,\,h_1}^{\la})_m}{(\bar{Q}^{\la}_{h_2+\frac{1}{2}})_r}
 = -\frac{1}{4q^2}\,(\bar{Q}^{\la}_{h_1+h_2-\frac{1}{2}})_{m+r}
+ \cdots    \nonu \\
 && +  q^0
    \begin{cases}
      \frac{h_2(h_2-2\la)}{(2h_2-1)}\, m, \,\,
      \text{If}\,h_1=1
       \nonu  \\
    \frac{(2h_2^2-h_1-4h_2+2h_1 h_2+2-2\la)((2h_2-1)m-2(h_1-1)r)}{2(2h_2-1)(2h_1+2h_2-3)},
     \,\, \text{If}\,h_1>1
    \end{cases}
    \Bigg\}(\bar{Q}^{\la}_{h_1+h_2-\frac{3}{2}})_{m+r},
  \nonu \\
 && \comm{(W_{B,\,h_1}^{\la})_m}{(\bar{Q}^{\la,\hat{A}}_{h_2+\frac{1}{2}})_r}
   =
  -\frac{1}{4q^2}\,(\bar{Q}^{\la,\hat{A}}_{h_1+h_2-\frac{1}{2}})_{m+r}
+ \cdots    \nonu \\
 && +  q^0
    \begin{cases}
       \frac{h_2(h_2-2\la)}{(2h_2-1)}\, m, \,\,   \text{If}\,h_1=1
       \nonu \\
    \frac{(2h_2^2-h_1-4h_2+2h_1 h_2+2-2\la)((2h_2-1)m-2(h_1-1)r)}{2(2h_2-1)(2h_1+2h_2-3)},
     \,\, \text{If}\,h_1>1
    \end{cases}
    \Bigg\}(\bar{Q}^{\la,\hat{A}}_{h_1+h_2-\frac{3}{2}})_{m+r},
    \nonu \\
&&    \comm{(W_{B,\,h_1}^{\la,\hat{A}})_m}{(\bar{Q}^{\la}_{h_2+\frac{1}{2}})_r} =
    \comm{(W_{B,\,h_1}^{\la})_m}{(\bar{Q}^{\la,\hat{A}}_{h_2+\frac{1}{2}})_r},
    \nonu \\
 && \comm{(W_{B,\,h_1}^{\la,\hat{A}})_m}{(\bar{Q}^{\la,\hat{B}}_{h_2+\frac{1}{2}})_r}
 =
 -q^2\,\frac{1}{4}\,\delta^{\hat{A}\hat{B}}\,\frac{1}{K}\,(\bar{Q}^{\la}_{h_1+h_2-\frac{1}{2}})_{m+r}
 \nonu  \\
   && -  q^0\,\frac{1}{8}\, (d^{\hat{A}\hat{B}\hat{C}} + i \,
f^{\hat{A}\hat{B}\hat{C}}   )\,
    (\bar{Q}^{\la,\hat{C}}_{h_1+h_2-\frac{1}{2}})_{m+r}
+ \cdots,
       \nonu \\
%
       &&   \acomm{(Q^{\la}_{h_1+\frac{1}{2}})_r}{(\bar{Q}^{\la}_{h_2+\frac{1}{2}})_s}
       =
        q^{2(h_1+h_2-2)}\,K\,N\, c^{h_1,h_2}_{Q}(\la)\,[r+h_1-\tfrac{1}{2}]_{h_1+h_2}
\nonu \\
&& + q^0 \,2\,\Big( (W^{\la}_{F,\,h_1+h_2})_{r+s}+(W^{\la}_{B,\,h_1+h_2})_{r+s} \Big)
  + \cdots   \,,
   \nonu \\
  &&  \acomm{(Q^{\la}_{h_1+\frac{1}{2}})_r}{(\bar{Q}^{\la,\hat{{A}}}_{h_2+\frac{1}{2}})_s}
 =     q^0 \,2\,\Big( (W^{\la,\hat{{A}}}_{F,\,h_1+h_2})_{r+s}
      +(W^{\la,\hat{{A}}}_{B,\,h_1+h_2})_{r+s}\Big)+ \cdots \,,
    \nonu  \\
 &&   \acomm{(Q^{\la,\hat{{A}}}_{h_1+\frac{1}{2}})_r}{(\bar{Q}^{\la}_{h_2+\frac{1}{2}})_s}  =
    \acomm{(Q^{\la}_{h_1+\frac{1}{2}})_r}{(\bar{Q}^{\la,\hat{{A}}}_{h_2+\frac{1}{2}})_s},
    \nonu  \\
    &&  \acomm{(Q^{\la,\hat{{A}}}_{h_1+\frac{1}{2}})_r}{(\bar{Q}^{\la,\hat{{B}}}_{h_2+\frac{1}{2}})_s}= q^{2(h_1+h_2)}\,
      \delta^{\hat{A}\hat{B}}\,N\, c^{h_1,h_2}_{Q}(\la)\,[r+h_1-\tfrac{1}{2}]_{h_1+h_2}
      \nonu  \\ & &+
      q^4\,\delta^{\hat{A}\hat{B}}\,\frac{2}{K}\,
      \Big( (W^{\la}_{F,\,h_1+h_2})_{r+s}
      +(W^{\la}_{B,\,h_1+h_2})_{r+s} \Big)
    \nonu  \\&& + q^2\,d^{\hat{A}\hat{B}\hat{C}}\,
     \Big( (W^{\la,\hat{C}}_{F,\,h_1+h_2})_{r+s}
     +(W^{\la,\hat{C}}_{B,\,h_1+h_2})_{r+s} \Big)
    \nonu  \\&& + q^2\,f^{\hat{A}\hat{B}\hat{C}}\,
     \Big( (W^{\la,\hat{C}}_{F,\,h_1+h_2})_{r+s}
     -(W^{\la,\hat{C}}_{B,\,h_1+h_2})_{r+s} \Big)  + \cdots.
\label{appd}
\eea
Note that the commutators with some modes between
the bosonic currents and the fermionic 
currents contain $\frac{1}{q^2}$ factors.
We need to find out the particular combinations which
will remove these divergent factors for the $q \rightarrow 0$ limit. 
The last
anticommutator of Appendix (\ref{appd}) vanishes
under the $q \rightarrow 0$ limit.

\section{ The (anti)commutators in the section $4$ }


We present the complete (anti)commutators in section $4$ as follows:
\bea
&& \comm{(\mathcal{L}^{\alpha}_{\,\,\,\beta,\,h_1})_m}{(\mathcal{L}^{\gamma}_{\,\,\,\delta,\,h_2})_n}  =  
q^{2(h_1-2)}\,\delta_{h_1,h_2}(\tfrac{1}{2}\delta^{\alpha}_{\beta}\delta^{\gamma}_{\delta}-\delta^{\alpha}_{\delta}\delta^{\gamma}_{\beta})\,c_{W_F,h}(m)\,\delta_{m+n}
\nonu \\
&& + \frac{1}{2} \sum_{h=-1}^{h_1+h_2-3}
q^h\,p_B^{h_1,h_2,h}(m,n,\tfrac{1}{2})\,
\Bigg[
\delta^{\alpha}_{\delta}\,(\mathcal{L}^{\gamma}_{\,\,\,\beta,\,h_1+h_2-h-2})_{m+n}
+
(-1)^h\delta^{\gamma}_{\beta}\,(\mathcal{L}^{\alpha}_{\,\,\,\delta,\,h_1+h_2-h-2})_{m+n}\Bigg]
\nonu \\
&& - \frac{1}{2}
\sum_{h=0}^{(h_1+h_2-3)/2}
q^{2h}\,p_B^{h_1,h_2,2h}(m,n,\tfrac{1}{2})\,
\Bigg[
\delta^{\alpha}_{\beta}
\,(\mathcal{L}^{\gamma}_{\,\,\,\delta,\,h_1+h_2-2h-2})_{m+n}
+
\delta^{\gamma}_{\delta}\,
(\mathcal{L}^{\alpha}_{\,\,\,\beta,\,h_1+h_2-2h-2})_{m+n}
\nonu \\
&&
-
\big(\delta^{\beta}_{\gamma}\delta^{\alpha}_{\delta}-\tfrac{1}{2}\delta^{\alpha}_{\beta}\delta^{\gamma}_{\delta}\big)
(\mathcal{U}_{h_1+h_2-2h-2})_{m+n}
\Bigg]\,,
\nonu     \\
&& \comm{(\mathcal{L}^{\alpha}_{\,\,\,\beta,\,h_1})_m}{(\mathcal{Q}^{a}_{\,\,\,\gamma,\,h_2})_n}  =  
\nonu \\
& & \frac{1}{2} \sum_{h=-1}^{h_1+h_2-3}
q^h\,p_B^{h_1,h_2,h}(m,n,\tfrac{1}{2})\,
\Bigg[
\delta^{\alpha}_{\gamma}\,(\mathcal{Q}^{a}_{\,\,\,\beta,\,h_1+h_2-h-2})_{m+n}
-\frac{1}{2}
\delta^{\alpha}_{\beta}\,(\mathcal{Q}^{a}_{\,\,\,\gamma,\,h_1+h_2-h-2})_{m+n}
\Bigg]\,,
\nonu \\
& &
\comm{(\mathcal{L}^{\alpha}_{\,\,\,\beta,\,h_1})_m}{(\mathcal{P}^{\dot{\alpha}}_{\,\,\,\gamma,\,h_2})_n} =
\nonu \\
& & \frac{1}{2} \sum_{h=-1}^{h_1+h_2-3}
q^h\,p_B^{h_1,h_2,h}(m,n,\tfrac{1}{2})\,
\Bigg[
\delta^{\alpha}_{\gamma}\,(\mathcal{P}^{\dot{\alpha}}_{\,\,\,\beta,\,h_1+h_2-h-2})_{m+n}
-\frac{1}{2}
\delta^{\alpha}_{\beta}\,(\mathcal{P}^{\dot{\alpha}}_{\,\,\,\gamma,\,h_1+h_2-h-2})_{m+n}\Bigg]\,,
\nonu \\
& &
\comm{(\mathcal{L}^{\alpha}_{\,\,\,\beta,\,h_1})_m}{(\mathcal{S}^{\gamma}_{\,\,\,a,\,h_2})_n} =
\nonu \\
& & \frac{1}{2} \sum_{h=-1}^{h_1+h_2-3}
(-q)^h\,p_B^{h_1,h_2,h}(m,n,\tfrac{1}{2})\,
\Bigg[
\delta^{\gamma}_{\beta}\,(\mathcal{S}^{\alpha}_{\,\,\,a,\,h_1+h_2-h-2})_{m+n}
-\frac{1}{2}
\delta^{\alpha}_{\beta}\,(\mathcal{S}^{\gamma}_{\,\,\,a,\,h_1+h_2-h-2})_{m+n}
\Bigg]\,,
\nonu \\
&&  \comm{(\mathcal{L}^{\alpha}_{\,\,\,\beta,\,h_1})_m}{(\mathcal{K}^{\gamma}_{\,\,\,\dot{\beta},\,h_2})_n} = 
\nonu \\
&& \frac{1}{2} \sum_{h=-1}^{h_1+h_2-3}
(-q)^h\,p_B^{h_1,h_2,h}(m,n,\tfrac{1}{2})\,
\Bigg[
\delta^{\gamma}_{\beta}\,(\mathcal{K}^{\alpha}_{\,\,\,\dot{\beta},\,h_1+h_2-h-2})_{m+n}
-\frac{1}{2}
\delta^{\alpha}_{\beta}\,(\mathcal{K}^{\gamma}_{\,\,\,\dot{\beta},\,h_1+h_2-h-2})_{m+n}\Bigg]\,,
\nonu \\
&&
\comm{(\dot{\mathcal{L}}^{\dot{\alpha}}_{\,\,\,\dot{\beta},\,h_1})_m}{(\dot{\mathcal{L}}^{\dot{\gamma}}_{\,\,\,\dot{\delta},\,h_2})_n}= 
q^{2(h_1-2)}\,\delta_{h_1,h_2}(\tfrac{1}{2}\delta^{\dot{\alpha}}_{\dot{\beta}}\delta^{\dot{\gamma}}_{\dot{\delta}}-\delta^{\dot{\alpha}}_{\dot{\delta}}\delta^{\dot{\gamma}}_{\dot{\beta}})\,c_{W_F,h}(m)\,\delta_{m+n}
\nonu \\
&& +\frac{1}{2} \sum_{h=-1}^{h_1+h_2-3}
q^h\,p_B^{h_1,h_2,h}(m,n,\tfrac{1}{2})\,
\Bigg[
\delta^{\dot{\alpha}}_{\dot{\delta}}\,(\dot{\mathcal{L}}^{\dot{\gamma}}_{\,\,\,\dot{\beta},\,h_1+h_2-h-2})_{m+n}
+(-1)^h\delta^{\dot{\gamma}}_{\dot{\beta}}\,(\dot{\mathcal{L}}^{\dot{\alpha}}_{\,\,\,\dot{\delta},\,h_1+h_2-h-2})_{m+n}
\Bigg]
\nonu \\
&& -\frac{1}{2}
\sum_{h=0}^{(h_1+h_2-3)/2}
q^{2h}\,p_B^{h_1,h_2,2h}(m,n,\tfrac{1}{2})\,
\Bigg[
\delta^{\dot{\alpha}}_{\dot{\beta}}
\,(\dot{\mathcal{L}}^{\dot{\gamma}}_{\,\,\,\dot{\delta},\,h_1+h_2-2h-2})_{m+n}
+
\delta^{\dot{\gamma}}_{\dot{\delta}}\,
(\mathcal{L}^{\dot{\alpha}}_{\,\,\,\dot{\beta},\,h_1+h_2-2h-2})_{m+n}
\nonu \\
& &
-
\big(\delta^{\dot{\beta}}_{\dot{\gamma}}\delta^{\dot{\alpha}}_{\dot{\delta}}-\tfrac{1}{2}\delta^{\dot{\alpha}}_{\dot{\beta}}\delta^{\dot{\gamma}}_{\dot{\delta}}\big)
(\dot{\mathcal{U}}_{h_1+h_2-2h-2})_{m+n}
\Bigg]\, ,
\nonu   \\
&&
\comm{(\dot{\mathcal{L}}^{\dot{\alpha}}_{\,\,\,\dot{\beta},\,h_1})_m}{(\dot{\mathcal{Q}}^{\dot{\gamma}}_{\,\,\,a,\,h_2})_n} =  
\nonu \\
& & \frac{1}{2} \sum_{h=-1}^{h_1+h_2-3}
(-q)^h\,p_B^{h_1,h_2,h}(m,n,\tfrac{1}{2})\,
\Bigg[
\delta^{\dot{\gamma}}_{\dot{\beta}}\,(\dot{\mathcal{Q}}^{\dot{\alpha}}_{\,\,\,a,\,h_1+h_2-h-2})_{m+n}
-
\frac{1}{2}\delta^{\dot{\alpha}}_{\dot{\beta}}\,(\dot{\mathcal{L}}^{\dot{\gamma}}_{\,\,\,a,\,h_1+h_2-h-2})_{m+n}\Bigg]\,,
\nonu   \\
&&  \comm{(\dot{\mathcal{L}}^{\dot{\alpha}}_{\,\,\,\dot{\beta},\,h_1})_m}{(\mathcal{P}^{\dot{\gamma}}_{\,\,\,\delta,\,h_2})_n} =  
\nonu \\
& & \frac{1}{2} \sum_{h=-1}^{h_1+h_2-3}
(-q)^h\,p_B^{h_1,h_2,h}(m,n,\tfrac{1}{2})\,
\Bigg[
\delta^{\dot{\gamma}}_{\dot{\beta}}\,(\mathcal{P}^{\dot{\alpha}}_{\,\,\,\delta,\,h_1+h_2-h-2})_{m+n}
-\frac{1}{2}
\delta^{\dot{\alpha}}_{\dot{\beta}}\,(\mathcal{P}^{\dot{\gamma}}_{\,\,\,\delta,\,h_1+h_2-h-2})_{m+n}\Bigg]\,,
\nonu  \\
&&
\comm{(\dot{\mathcal{L}}^{\dot{\alpha}}_{\,\,\,\dot{\beta},\,h_1})_m}{(\dot{\mathcal{S}}^{a}_{\,\,\,\dot{\gamma},\,h_2})_n} =  
\nonu \\
& & \frac{1}{2} \sum_{h=-1}^{h_1+h_2-3}
q^h\,p_B^{h_1,h_2,h}(m,n,\tfrac{1}{2})\,
\Bigg[
\delta^{\dot{\alpha}}_{\dot{\gamma}}\,(\dot{\mathcal{S}}^{a}_{\,\,\,\dot{\beta},\,h_1+h_2-h-2})_{m+n}
-\frac{1}{2}
\delta^{\dot{\alpha}}_{\dot{\beta}}\,(\dot{\mathcal{S}}^{a}_{\,\,\,\dot{\gamma},\,h_1+h_2-h-2})_{m+n}\Bigg]\,,
\nonu \\
&&
\comm{(\dot{\mathcal{L}}^{\dot{\alpha}}_{\,\,\,\dot{\beta},\,h_1})_m}{(\mathcal{K}^{\gamma}_{\,\,\,\dot{\delta},\,h_2})_n} =  
\nonu \\
& & \frac{1}{2} \sum_{h=-1}^{h_1+h_2-3}
q^h\,p_B^{h_1,h_2,h}(m,n,\tfrac{1}{2})\,
\Bigg[
\delta^{\dot{\alpha}}_{\dot{\delta}}\,(\mathcal{K}^{\gamma}_{\,\,\,\dot{\beta},\,h_1+h_2-h-2})_{m+n}
-\frac{1}{2}
\delta^{\dot{\alpha}}_{\dot{\beta}}\,(\mathcal{K}^{\gamma}_{\,\,\,\dot{\delta},\,h_1+h_2-h-2})_{m+n}\Bigg]\,,
\nonu \\
&&
\comm{(\mathcal{R}^{a}_{\,\,\,b,\,h_1})_m}{(\mathcal{R}^{c}_{\,\,\,d,\,h_2})_n}
= 
q^{2(h_1-2)}\,\delta_{h_1,h_2}(-\tfrac{1}{4}\delta^{a}_{b}\delta^{c}_{d}+\delta^{a}_{d}\delta^{c}_{b}
)\,c_{W_F,h}(m)\,\delta_{m+n}
\nonu \\
& & +\frac{1}{2} \sum_{h=-1}^{h_1+h_2-3}
q^h\,p_B^{h_1,h_2,h}(m,n,\tfrac{1}{2})\,
\Bigg[
\delta^{a}_{d}\,(\mathcal{R}^{c}_{\,\,\,b,\,h_1+h_2-h-2})_{m+n}
+(-1)^h\delta^{c}_{b}\,(\mathcal{R}^{a}_{\,\,\,d,\,h_1+h_2-h-2})_{m+n}\Bigg]
\nonu \\
& & -\frac{1}{4}
\sum_{h=0}^{(h_1+h_2-3)/2}
q^{2h}\,p_B^{h_1,h_2,2h}(m,n,\tfrac{1}{2})\,
\Bigg[
\delta^{a}_{b}
\,(\mathcal{R}^{c}_{\,\,\,d,\,h_1+h_2-2h-2})_{m+n}
+
\delta^{c}_{d}\,
(\mathcal{R}^{a}_{\,\,\,b,\,h_1+h_2-2h-2})_{m+n}
\nonu \\
&&
-
\big(\delta^{c}_{b}\delta^{a}_{d}-\tfrac{1}{4}\delta^{a}_{b}\delta^{c}_{d}\big)
(\mathcal{V}_{h_1+h_2-2h-2})_{m+n}
\Bigg]\,,
\nonu \\
&&
\comm{(\mathcal{R}^{a}_{\,\,\,b,\,h_1})_m}{(\mathcal{Q}^{c}_{\,\,\,\alpha,\,h_2})_n} =
\nonu \\
&&
\frac{1}{2} \sum_{h=-1}^{h_1+h_2-3}
(-q)^h\,p_B^{h_1,h_2,h}(m,n,\tfrac{1}{2})\,
\Bigg[
\delta^{c}_{b}\,(\mathcal{Q}^{a}_{\,\,\,\alpha,\,h_1+h_2-h-2})_{m+n}
-\frac{1}{4}
\delta^{a}_{b}\,(\mathcal{Q}^{c}_{\,\,\,\alpha,\,h_1+h_2-h-2})_{m+n}\Bigg]\,,
\nonu \\
&&
\comm{(\mathcal{R}^{a}_{\,\,\,b,\,h_1})_m}{(\dot{\mathcal{Q}}^{\dot{\alpha}}_{\,\,\,c,\,h_2})_n} =
\nonu \\
&& \frac{1}{2} \sum_{h=-1}^{h_1+h_2-3}
q^h\,p_B^{h_1,h_2,h}(m,n,\tfrac{1}{2})\,
\Bigg[
\delta^{a}_{c}\,(\dot{\mathcal{Q}}^{\dot{\alpha}}_{\,\,\,b,\,h_1+h_2-h-2})_{m+n}
-\frac{1}{4}
\delta^{a}_{b}\,(\dot{\mathcal{Q}}^{\dot{\alpha}}_{\,\,\,c,\,h_1+h_2-h-2})_{m+n}
\Bigg]\,,
\nonu  \\
&&
\comm{(\mathcal{R}^{a}_{\,\,\,b,\,h_1})_m}{(\mathcal{S}^{\alpha}_{\,\,\,c,\,h_2})_n} =
\nonu \\
&& \frac{1}{2} \sum_{h=-1}^{h_1+h_2-3}
q^h\,p_B^{h_1,h_2,h}(m,n,\tfrac{1}{2})\,
\Bigg[
\delta^{a}_{c}\,(\mathcal{S}^{\alpha}_{\,\,\,b,\,h_1+h_2-h-2})_{m+n}
-\frac{1}{4}
\delta^{a}_{b}\,(\mathcal{S}^{\alpha}_{\,\,\,c,\,h_1+h_2-h-2})_{m+n}\Bigg]\,,
\nonu \\
&&
\comm{(\mathcal{R}^{a}_{\,\,\,b,\,h_1})_m}{(\dot{\mathcal{S}}^{c}_{\,\,\,\dot{\alpha},\,h_2})_n} =
\nonu \\
&& \frac{1}{2} \sum_{h=-1}^{h_1+h_2-3}
(-q)^h\,p_B^{h_1,h_2,h}(m,n,\tfrac{1}{2})\,
\Bigg[
\delta^{c}_{b}\,(\dot{\mathcal{S}}^{a}_{\,\,\,\dot{\alpha},\,h_1+h_2-h-2})_{m+n}
-\frac{1}{4}
\delta^{a}_{b}\,(\dot{\mathcal{S}}^{c}_{\,\,\,\dot{\alpha},\,h_1+h_2-h-2})_{m+n}
\Bigg]\,,
\nonu \\
&&
\comm{(\mathcal{B}_{h_1})_m}{(\mathcal{B}_{\,h_2})_n} =
\nonu \\
&&
-q^{2(h_1-2)}\,\delta_{h_1,h_2}\,c_{W_F,h_1}(m)\,\delta_{m+n}
+\frac{1}{2} \sum_{h=0}^{(h_1+h_2-3)/2}
q^{2h}\,p_B^{h_1,h_2,2h}(m,n,\tfrac{1}{2})\,
(\mathcal{B}_{h_1+h_2-2h-2})_{m+n}\,,
\nonu \\
&& \comm{(\mathcal{B}_{h_1})_m}{(\mathcal{C}_{\,h_2})_n} =
\comm{(\mathcal{B}_{h_1})_m}{(\mathcal{B}_{\,h_2})_n}\,,
\nonu \\
&&  \comm{(\mathcal{B}_{h_1})_m}{(\mathcal{Q}^{a}_{\,\,\,\alpha,\,h_2})_n} =
 \frac{1}{4} \sum_{h=-1}^{h_1+h_2-3}
q^h\,p_B^{h_1,h_2,h}(m,n,\tfrac{1}{2})\,
(\mathcal{Q}^{a}_{\,\,\,\alpha,\,h_1+h_2-h-2})_{m+n}\,, 
\nonu \\
&&  \comm{(\mathcal{B}_{h_1})_m}{(\dot{\mathcal{Q}}^{\dot{\alpha}}_{\,\,\,a,\,h_2})_n}
= \frac{1}{4} \sum_{h=-1}^{h_1+h_2-3}
(-q)^h\,p_B^{h_1,h_2,h}(m,n,\tfrac{1}{2})\,
(\dot{\mathcal{Q}}^{\dot{\alpha}}_{\,\,\,a,\,h_1+h_2-h-2})_{m+n}\,, 
\nonu \\
&&  \comm{(\mathcal{B}_{\,h_1})_m}{(\mathcal{S}^{\alpha}_{\,\,\,a,\,h_2})_n} =
\frac{1}{4} \sum_{h=-1}^{h_1+h_2-3}
(-q)^{h}\,p_B^{h_1,h_2,h}(m,n,\tfrac{1}{2})\,
(\mathcal{S}^{\alpha}_{\,\,\,a,\,h_1+h_2-h-2})_{m+n}\,, 
\nonu \\
&&  \comm{(\mathcal{B}_{\,h_1})_m}{(\dot{\mathcal{S}}^{a}_{\,\,\,\dot{\alpha},\,h_2})_n} =
\frac{1}{4} \sum_{h=-1}^{h_1+h_2-3}
q^{h}\,p_B^{h_1,h_2,h}(m,n,\tfrac{1}{2})\,
(\dot{\mathcal{S}}^{a}_{\,\,\,\dot{\alpha},\,h_1+h_2-h-2})_{m+n}\,,
\nonu \\
&& \comm{(\mathcal{D}_{h_1})_m}{(\mathcal{D}_{\,h_2})_n} = 
\comm{(\mathcal{B}_{h_1})_m}{(\mathcal{B}_{\,h_2})_n} \, ,
\nonu \\
&& \comm{(\mathcal{D}_{h_1})_m}{(\mathcal{Q}^{a}_{\,\,\,\alpha,\,h_2})_n}=
\comm{(\mathcal{B}_{h_1})_m}{(\mathcal{Q}^{a}_{\,\,\,\alpha,\,h_2})_n} \, ,
\nonu \\
&& \comm{(\mathcal{D}_{h_1})_m} {(\dot{\mathcal{Q}}^{\dot{\alpha}}_{\,\,\,a,\,h_2})_n}= -
 \comm{(\mathcal{B}_{h_1})_m}{(\dot{\mathcal{Q}}^{\dot{\alpha}}_{\,\,\,a,\,h_2})_n}\, ,
\nonu \\
&&  \comm{(\mathcal{D}_{h_1})_m}{(\mathcal{P}^{\dot{\alpha}}_{\,\,\,\beta,\,h_2})_n} =
\frac{1}{2} \sum_{h=-1}^{(h_1+h_2-4)/2}
q^{2h+1}\,p_B^{h_1,h_2,2h+1}(m,n,\tfrac{1}{2})\,
(\mathcal{P}^{\dot{\alpha}}_{\,\,\,\beta,\,h_1+h_2-2h-3})_{m+n}\,,
\nonu \\
&& \comm{(\mathcal{D}_{\,h_1})_m}{(\mathcal{S}^{\alpha}_{\,\,\,a,\,h_2})_n}=
 \comm{(\mathcal{B}_{\,h_1})_m}{(\mathcal{S}^{\alpha}_{\,\,\,a,\,h_2})_n}\, ,
\nonu \\
&&
\comm{(\mathcal{D}_{\,h_1})_m}{(\dot{\mathcal{S}}^{a}_{\,\,\,\dot{\alpha},\,h_2})_n} = - \comm{(\mathcal{B}_{\,h_1})_m}{(\dot{\mathcal{S}}^{a}_{\,\,\,\dot{\alpha},\,h_2})_n}\, ,
\nonu \\
& &
\comm{(\mathcal{D}_{\,h_1})_m} {(\mathcal{K}^{\alpha}_{\,\,\,\dot{\beta},\,h_2})_n}=
\frac{1}{2} \sum_{h=-1}^{(h_1+h_2-4)/2}
(-q)^{2h+1}\,p_B^{h_1,h_2,2h+1}(m,n,\tfrac{1}{2})\,
(\mathcal{K}^{\alpha}_{\,\,\,\dot{\beta},\,h_1+h_2-2h-3})_{m+n}\,,
\nonu \\
&&  \acomm{(\mathcal{Q}^{a}_{\,\,\,\beta,\,h_1})_m}{(\dot{\mathcal{Q}}^{\dot{\alpha}}_{\,\,\,b,\,h_2})_n} =
\frac{1}{2} \sum_{h=-1}^{h_1+h_2-3}
q^h\,p_B^{h_1,h_2,h}(m,n,\tfrac{1}{2})\,\delta^{a}_{b}\,
(\mathcal{P}^{\dot{\alpha}}_{\,\,\,\beta,\,h_1+h_2-h-2})_{m+n}
\,,
\nonu \\
&&  \acomm{(\mathcal{Q}^{a}_{\,\,\,\alpha,\,h_1})_m}{(\mathcal{S}^{\beta}_{\,\,\,b,\,h_2})_n} =
-q^{2(h_1-2)}\,\delta_{h_1,h_2}\,\delta^{a}_{b}\delta^{\alpha}_{\beta}\,c_{W_F,h}(m)\,\delta_{m+n}
\nonu \\
&&
+\frac{1}{2} \sum_{h=-1}^{h_1+h_2-3}
q^h\,p_B^{h_1,h_2,h}(m,n,\tfrac{1}{2})
\Bigg[
(-1)^{h+1}\,\delta^{\beta}_{\alpha}\,(\mathcal{R}^{a}_{\,\,\,b,\,h_1+h_2-h-2})_{m+n}
+\delta^{a}_{b}\,(\mathcal{L}^{\beta}_{\,\,\,\alpha,\,h_1+h_2-h-2})_{m+n}
\nonu \\
&&
+\frac{1}{2}\,\delta^{a}_{b} \delta^{\beta}_{\alpha}\,\bigg(
(\mathcal{U}_{h_1+h_2-h-2})_{m+n}
-\frac{(-1)^h}{2}\,
(\mathcal{V}_{h_1+h_2-h-2})_{m+n}\bigg)
\Bigg]
\,,
\nonu \\
&&  \comm{(\mathcal{Q}^{a}_{\,\,\,\alpha,\,h_1})_m}{(\mathcal{K}^{\beta}_{\,\,\,\dot{\gamma},\,h_2})_n} =
\frac{1}{2} \sum_{h=-1}^{h_1+h_2-3}
(-q)^h\,p_B^{h_1,h_2,h}(m,n,\tfrac{1}{2})\,\delta^{\beta}_{\alpha}\,
(\dot{\mathcal{S}}^{a}_{\,\,\,\dot{\gamma},\,h_1+h_2-h-2})_{m+n}\,,
\nonu \\
&&  \acomm{(\dot{\mathcal{Q}}^{\dot{\alpha}}_{\,\,\,a,\,h_1})_m}{(\dot{\mathcal{S}}^{b}_{\,\,\,\dot{\beta},\,h_2})_n} =
q^{2(h_1-2)}\,\delta_{h_1,h_2}\,\delta^{b}_{a}\delta^{\dot{\alpha}}_{\dot{\beta}}\,c_{W_F,h}(m)\,\delta_{m+n}
\nonu \\
&&
+\frac{1}{2} \sum_{h=-1}^{h_1+h_2-3}
q^h\,p_B^{h_1,h_2,h}(m,n,\tfrac{1}{2})
\Bigg[\,
\delta^{\dot{\alpha}}_{\dot{\beta}}\,(\mathcal{R}^{b}_{\,\,\,a,\,h_1+h_2-h-2})_{m+n}-(-1)^h\,\delta^{b}_{a}\,(\dot{\mathcal{L}}^{\dot{\alpha}}_{\,\,\,\dot{\beta},\,h_1+h_2-h-2})_{m+n}
\nonu \\
&&
-\delta^{b}_{a} \delta^{\dot{\alpha}}_{\dot{\beta}}\,\bigg(\frac{(-1)^h}{2}\,
(\dot{\mathcal{U}}_{h_1+h_2-h-2})_{m+n}
-\frac{1}{4}\,
(\mathcal{V}_{h_1+h_2-h-2})_{m+n}\bigg)
\Bigg]
\,,
\nonu \\
&&
\comm{(\dot{\mathcal{Q}}^{\dot{\alpha}}_{\,\,\,a,\,h_1})_m}{(\mathcal{K}^{\beta}_{\,\,\,\dot{\gamma},\,h_2})_n} =
\frac{1}{2} \sum_{h=-1}^{h_1+h_2-3}
q^h\,p_B^{h_1,h_2,h}(m,n,\tfrac{1}{2})\,\delta^{\dot{\alpha}}_{\dot{\gamma}}\,
(\mathcal{S}^{\beta}_{\,\,\,a,\,h_1+h_2-h-2})_{m+n}\,,
\nonu \\
&&
\comm{(\mathcal{P}^{\dot{\alpha}}_{\,\,\,\beta,\,h_1})_m}{(\mathcal{S}^{\gamma}_{\,\,\,a,\,h_2})_n} =
\frac{1}{2} \sum_{h=-1}^{h_1+h_2-3}
(-q)^{h}\,p_B^{h_1,h_2,h}(m,n,\tfrac{1}{2})\,\delta^{\gamma}_{\beta}\,
(\dot{\mathcal{Q}}^{\dot{\alpha}}_{\,\,\,a,\,h_1+h_2-h-2})_{m+n}\,, 
\nonu \\
&&  \comm{(\mathcal{P}^{\dot{\alpha}}_{\,\,\,\beta,\,h_1})_m}{(\dot{\mathcal{S}}^{a}_{\,\,\,\dot{\gamma},\,h_2})_n} =
\frac{1}{2} \sum_{h=-1}^{h_1+h_2-3}
q^{h}\,p_B^{h_1,h_2,h}(m,n,\tfrac{1}{2})\,\delta^{\dot{\alpha}}_{\dot{\gamma}}\,
(\mathcal{Q}^{a}_{\,\,\,\beta,\,h_1+h_2-h-2})_{m+n}\,, 
\nonu \\
&&  \comm{(\mathcal{P}^{\dot{\alpha}}_{\,\,\,\beta,\,h_1})_m}{(\mathcal{K}^{\gamma}_{\,\,\,\dot{\delta},\,h_2})_n} =
-q^{2(h_1-2)}\,
\delta_{h_1,h_2}\,\delta^{\gamma}_{\beta}\,\delta^{\dot{\alpha}}_{\dot{\delta}}\,c_{W_F,h_1}(m)\,\delta_{m+n}
\nonu \\
&& +\frac{1}{2} \sum_{h=-1}^{h_1+h_2-3}
q^h\,p_B^{h_1,h_2,h}(m,n,\tfrac{1}{2})
\Bigg[
(-1)^h\,\delta^{\gamma}_{\beta}\,(\dot{\mathcal{L}}^{\dot{\alpha}}_{\,\,\,\dot{\delta},\,h_1+h_2-h-2})_{m+n}
+\delta^{\dot{\alpha}}_{\dot{\delta}}\,(\mathcal{L}^{\gamma}_{\,\,\,\beta,\,h_1+h_2-h-2})_{m+n}
\nonu \\
&&
+\frac{1}{2}\,\delta^{\dot{\alpha}}_{\dot{\delta}}\delta^{\gamma}_{\beta} \,
\bigg(
(\mathcal{U}_{h_1+h_2-h-2})_{m+n}
+(-1)^h\,
(\dot{\mathcal{U}}_{h_1+h_2-h-2})_{m+n}\bigg)
\Bigg]
\,,
\nonu \\
&&
\acomm{(\mathcal{S}^{\alpha}_{\,\,\,a,\,h_1})_m}{(\dot{\mathcal{S}}^{b}_{\,\,\,\dot{\beta},\,h_2})_n} =
-\frac{1}{2} \sum_{h=-1}^{h_1+h_2-3}
(-q)^{h}\,p_B^{h_1,h_2,h}(m,n,\tfrac{1}{2})\,\delta^{b}_{a}\,
(\mathcal{K}^{\alpha}_{\,\,\,\dot{\beta},\,h_1+h_2-h-2})_{m+n}\,.
 \label{generalexp}
\eea
Note that the twelfth, the thirteenth, the fourteenth and the
fifteenth relations for $b=a$ in (\ref{generalexp}) vanish
for ${\cal N}=4$ supersymmetry. For different ${\cal N}$
with ${\cal N}=1,2$ or ${\cal N}=3$ supersymmetries,
they do not vanish.
According to the rescalings in the subsection
\ref{qlimit} with the footnote
\ref{pvalues}, the commutators having the
group indices contain the $\frac{1}{q^2}$ factors
on the right hand sides while the anticommutators do not have
the divergent factors for the $q \rightarrow 0$ limit.

There are also additional ones which vanish for $h_1=h_2=1$
\bea
\comm{(\mathcal{L}^{\alpha}_{\,\,\,\beta,\,h_1})_m}{(\mathcal{B}_{h_2})_n} & = &
\frac{1}{2} \sum_{h=0}^{[(h_1+h_2-3)/2]}
q^{2h}\,p_B^{h_1,h_2,2h}(m,n,\tfrac{1}{2})\,(\mathcal{L}^{\alpha}_{\,\,\,\beta,\,h_1+h_2-2h-2})_{m+n}\,,
\nonu \\
\comm{(\mathcal{L}^{\alpha}_{\,\,\,\beta,\,h_1})_m}{(\mathcal{C}_{h_2})_n}
& = &
\comm{(\mathcal{L}^{\alpha}_{\,\,\,\beta,\,h_1})_m}{(\mathcal{B}_{h_2})_n},
\nonu \\
\comm{(\mathcal{L}^{\alpha}_{\,\,\,\beta,\,h_1})_m}{(\mathcal{D}_{h_2})_n}
& = &
\comm{(\mathcal{L}^{\alpha}_{\,\,\,\beta,\,h_1})_m}{(\mathcal{B}_{h_2})_n},
\nonu \\
\comm{(\dot{\mathcal{L}}^{\dot{\alpha}}_{\,\,\,\dot{\beta},\,h_1})_m}{(\mathcal{B}_{h_2})_n} & = &
\frac{1}{2} \sum_{h=0}^{[(h_1+h_2-3)/2]}
q^{2h}\,p_B^{h_1,h_2,2h}(m,n,\tfrac{1}{2})\,(\dot{\mathcal{L}}^{\dot{\alpha}}_{\,\,\,\dot{\beta},\,h_1+h_2-2h-2})_{m+n}
\,,
\nonu \\
\comm{(\dot{\mathcal{L}}^{\dot{\alpha}}_{\,\,\,\dot{\beta},\,h_1})_m}{(\mathcal{C}_{h_2})_n} & = &
\comm{(\dot{\mathcal{L}}^{\dot{\alpha}}_{\,\,\,\dot{\beta},\,h_1})_m}{(\mathcal{B}_{h_2})_n},
\nonu \\
\comm{(\dot{\mathcal{L}}^{\dot{\alpha}}_{\,\,\,\dot{\beta},\,h_1})_m}{
  (\mathcal{D}_{h_2})_n} & = &
-\comm{(\dot{\mathcal{L}}^{\dot{\alpha}}_{\,\,\,\dot{\beta},\,h_1})_m}{(\mathcal{B}_{h_2})_n},
\nonu \\
\comm{(\mathcal{R}^{a}_{\,\,\,b,\,h_1})_m}{(\mathcal{C}_{h_2})_n} & = &
\frac{1}{2} \sum_{h=0}^{[(h_1+h_2-3)/2]}
q^{2h}\,p_B^{h_1,h_2,2h}(m,n,\tfrac{1}{2})\,(\mathcal{R}^{a}_{\,\,\,b,\,h_1+h_2-2h-2})_{m+n}
\,,
\nonu \\
\comm{(\mathcal{Q}^{a}_{\,\,\,\alpha,\,h_1})_m}{(\mathcal{C}_{h_1})_n}
& = & \frac{1}{2} \sum_{h=0}^{[(h_1+h_2-3)/2]}
q^{2h}\,p_B^{h_1,h_2,2h}(m,n,\tfrac{1}{2})\,
(\mathcal{Q}^{a}_{\,\,\,\alpha,\,h_1+h_2-2h-2})_{m+n}\,,
\nonu \\
\comm{(\dot{\mathcal{Q}}^{\dot{\alpha}}_{\,\,\,a,\,h_1})_m}{(\mathcal{C}_{h_2})_n} & = &
\frac{1}{2} \sum_{h=0}^{[(h_1+h_2-3)/2]}
q^{2h}\,p_B^{h_1,h_2,2h}(m,n,\tfrac{1}{2})\,
(\dot{\mathcal{Q}}^{\dot{\alpha}}_{\,\,\,a,\,h_1+h_2-2h-2})_{m+n}\,,
\nonu \\
\comm{(\mathcal{P}^{\dot{\alpha}}_{\,\,\,\beta,\,h_1})_m}{(\mathcal{B}_{h_2})_n}
& = &
\frac{1}{2} \sum_{h=0}^{[(h_1+h_2-3)/2]}
q^{2h}\,p_B^{h_1,h_2,2h}(m,n,\tfrac{1}{2})\,
(\mathcal{P}^{\dot{\alpha}}_{\,\,\,\beta,\,h_1+h_2-2h-2})_{m+n}\,, 
\nonu \\
\comm{(\mathcal{P}^{\dot{\alpha}}_{\,\,\,\beta,\,h_1})_m}{(\mathcal{C}_{h_2})_n}
& = &
 \comm{(\mathcal{P}^{\dot{\alpha}}_{\,\,\,\beta,\,h_1})_m}{(\mathcal{B}_{h_2})_n},
\nonu \\
\comm{(\mathcal{S}^{\alpha}_{\,\,\,a,\,h_1})_m}{(\mathcal{C}_{\,h_2})_n} & = &
\frac{1}{2} \sum_{h=0}^{[(h_1+h_2-3)/2]}
q^{2h}\,p_B^{h_1,h_2,2h}(m,n,\tfrac{1}{2})\,
(\mathcal{S}^{\alpha}_{\,\,\,a,\,h_1+h_2-2h-2})_{m+n}\,,
\nonu \\
\comm{(\dot{\mathcal{S}}^{a}_{\,\,\,\dot{\alpha},\,h_1})_m}{(\mathcal{C}_{\,h_2})_n}& = &
\frac{1}{2} \sum_{h=0}^{[(h_1+h_2-3)/2]}
q^{2h}\,p_B^{h_1,h_2,2h}(m,n,\tfrac{1}{2})\,
(\dot{\mathcal{S}}^{a}_{\,\,\,\dot{\alpha},\,h_1+h_2-2h-2})_{m+n}\,,
\nonu \\
\comm{(\mathcal{K}^{\alpha}_{\,\,\,\dot{\beta},\,h_1})_m}{(\mathcal{B}_{\,h_2})_n}
& = &
\frac{1}{2} \sum_{h=0}^{[(h_1+h_2-3)/2]}
q^{2h}\,p_B^{h_1,h_2,2h}(m,n,\tfrac{1}{2})\,
(\mathcal{K}^{\alpha}_{\,\,\,\dot{\beta},\,h_1+h_2-2h-2})_{m+n}\,,
\nonu \\
\comm{(\mathcal{K}^{\alpha}_{\,\,\,\dot{\beta},\,h_1})_m}{(\mathcal{C}_{\,h_2})_n}
& = &
\comm{(\mathcal{K}^{\alpha}_{\,\,\,\dot{\beta},\,h_1})_m}{(\mathcal{B}_{\,h_2})_n},
\nonu \\
\comm{(\mathcal{B}_{h_1})_m}{(\mathcal{D}_{\,h_2})_n} & = &
\frac{1}{2} \sum_{h=0}^{[(h_1+h_2-3)/2]}
q^{2h}\,p_B^{h_1,h_2,2h}(m,n,\tfrac{1}{2})\,
(\mathcal{D}_{h_1+h_2-2h-2})_{m+n}\,,
\nonu \\
\comm{(\mathcal{C}_{h_1})_m}{(\mathcal{C}_{\,h_2})_n} & = &
\frac{1}{2} \sum_{h=0}^{[(h_1+h_2-3)/2]}
q^{2h}\,p_B^{h_1,h_2,2h}(m,n,\tfrac{1}{2})\,
(\mathcal{C}_{h_1+h_2-2h-2})_{m+n}\,,
\nonu \\
\comm{(\mathcal{C}_{h_1})_m}{(\mathcal{D}_{\,h_2})_n}  & = &
\frac{1}{2} \sum_{h=0}^{[(h_1+h_2-3)/2]}
q^{2h}\,p_B^{h_1,h_2,2h}(m,n,\tfrac{1}{2})\,
(\mathcal{D}_{h_1+h_2-2h-2})_{m+n}\,.
\label{finalresult}
\eea
In Appendix
(\ref{finalresult}), there are no divergent factors when we take
$q \rightarrow 0$ limit after a rescaling as done in the subsection
\ref{qlimit} \footnote{
  The notation $[x]$ appearing in the summation of
  Appendix (\ref{finalresult})
  is the greatest integer less than or equal to $x$. }.
Note that there exists nonzero commutator relation
from the seventh equation of Appendix (\ref{finalresult})
which is identically zero for $h_1=1=h_2$.



\end{document}